\documentclass[reprint,twocolumn,superscriptaddress,numerical,prl,floats]{revtex4-2}

\usepackage{bbold}
\usepackage{amsmath}
\usepackage{bm}
\usepackage{physics}
\usepackage{calrsfs}
\usepackage{graphicx}
\usepackage{bbm}
\usepackage{xcolor}
\usepackage{hyperref}
\DeclareMathAlphabet{\pazocal}{OMS}{zplm}{m}{n}

\usepackage[protrusion=true, expansion=true,
tracking=true, kerning=true, spacing=true, 
final]{microtype}
\usepackage{hyphenat}
\usepackage{tikz}
\usetikzlibrary{calc,arrows.meta}
\usepackage{pgfplots} 
\pgfplotsset{compat=1.17}
\usetikzlibrary{decorations.markings}
\tikzset{
    set arrow inside/.code={\pgfqkeys{/tikz/arrow inside}{#1}},
    set arrow inside={end/.initial=stealth, opt/.initial=},
    /pgf/decoration/Mark/.style={
        mark/.expanded=at position #1 with
        {
            \noexpand\arrow[\pgfkeysvalueof{/tikz/arrow inside/opt}]{\pgfkeysvalueof{/tikz/arrow inside/end}}
        }
    },
    arrow inside/.style 2 args={
        set arrow inside={#1},
        postaction={
            decorate,decoration={
                markings,Mark/.list={#2}
            }
        }
    },
}

\begin{document}

\title{Variational approach to open quantum systems with long-range competing interactions}

\author{Dawid A. Hryniuk}
\email{d.hryniuk@ucl.ac.uk}
\affiliation{Department of Physics and Astronomy, University College London,
Gower Street, London, WC1E 6BT, United Kingdom}
\affiliation{London Centre for Nanotechnology, University College London, Gordon Street, London WC1H 0AH, United Kingdom}
\author{Marzena H. Szymańska}
\affiliation{Department of Physics and Astronomy, University College London,
Gower Street, London, WC1E 6BT, United Kingdom}

\date{\today}

\begin{abstract}
    Competition between short- and long-range interactions underpins many emergent phenomena in nature.
    Despite rapid progress in their experimental control, computational methods capable of accurately simulating open quantum many-body systems with complex long-ranged interactions at scale remain scarce. 
    Here, we address this limitation by introducing an efficient and scalable approach to dissipative quantum lattices in one and two dimensions, combining matrix product operators and time-dependent variational Monte Carlo. 
    We showcase the versatility, effectiveness, and unique methodological advantages of our algorithm by simulating the non-equilibrium dynamics and steady states of spin-$\frac{1}{2}$ lattices with competing algebraically-decaying interactions for as many as $N=200$ sites, revealing the emergence of spatially-modulated magnetic order far from equilibrium.
    This approach offers promising prospects for advancing our understanding of the complex non-equilibrium properties of a diverse variety of experimentally-realizable quantum systems with long-ranged interactions, including Rydberg atoms, ultracold dipolar molecules, and trapped ions. 
\end{abstract}

\maketitle

\begin{center}\noindent\textbf{INTRODUCTION}\end{center}

The intricate interplay of short- and long-distance interactions gives rise to 
emergent behaviors in a diverse range of in- and out-of-equilibrium systems found in nature, including the collective motion of bacterial swarms and animal flocks \cite{SANTORE2022102665, doi:10.1098/rspb.2019.0865, B812146J}, transcription and replication in RNA viruses \cite{ZIV20201067, Huber2019, Nicholson2014}
self-organization of polymers and liquid crystals \cite{doi:10.1126/science.277.5330.1225, Desai_Kapral_2009, doi:10.1021/jp807770n}, formation of striped domains in thin magnetic films and molecular monolayers at liquid-air interfaces \cite{PhysRevLett.104.077203, PhysRevB.51.1023, PhysRevLett.82.1602}, and onset of complex quantum phases in spin liquids and ensembles of ultracold atoms \cite{Tokiwa2014, Landig2016, Su2023}.
In the quantum condensed matter community, modern experimental platforms involving e.g. dipolar atoms and molecules in optical lattices \cite{Landig2016, Moses2017, Su2023, Weimer2010, doi:10.1126/science.abg2530, PhysRevLett.113.195302}, trapped ions \cite{annurev:/content/journals/10.1146/annurev-conmatphys-032822-045619, RevModPhys.93.025001, doi:10.1126/science.abk2400}, and nitrogen-vacancy centers \cite{Davis2023, PhysRevLett.118.093601, Choi2017} have enabled the realization of a rich family of long-ranged interactions and detailed study of the resultant exotic quantum phases, such as "time crystals" \cite{doi:10.1126/science.abg2530, Choi2017, Zhang2017} 
that emerge in systems with non-local couplings and away from equilibrium \cite{PhysRevB.106.224308, PhysRevLett.117.090402}.
In turn, these experimental platforms are now used to facilitate quantum simulation of complex quantum many-body phenomena and chemical reactions \cite{Weimer2010, Su2023, Moses2017, PRXQuantum.5.020358, RevModPhys.93.025001, Choi2017, doi:10.1126/science.abk2400}.
Additionally, long-ranged interactions are frequently resorted to in the design of innovative quantum technologies: non-local couplings are utilized when implementing high-fidelity multi-qubit gates \cite{Weimer2010, PhysRevX.10.021054, Evered2023, C8SC02355G}, while the frustration induced by the competition between long-ranged interactions may enhance the robustness to decoherence of quantum batteries \cite{PRXQuantum.5.030319, PhysRevLett.120.117702}.

\begin{figure}[b]
    \includegraphics[width=\columnwidth]{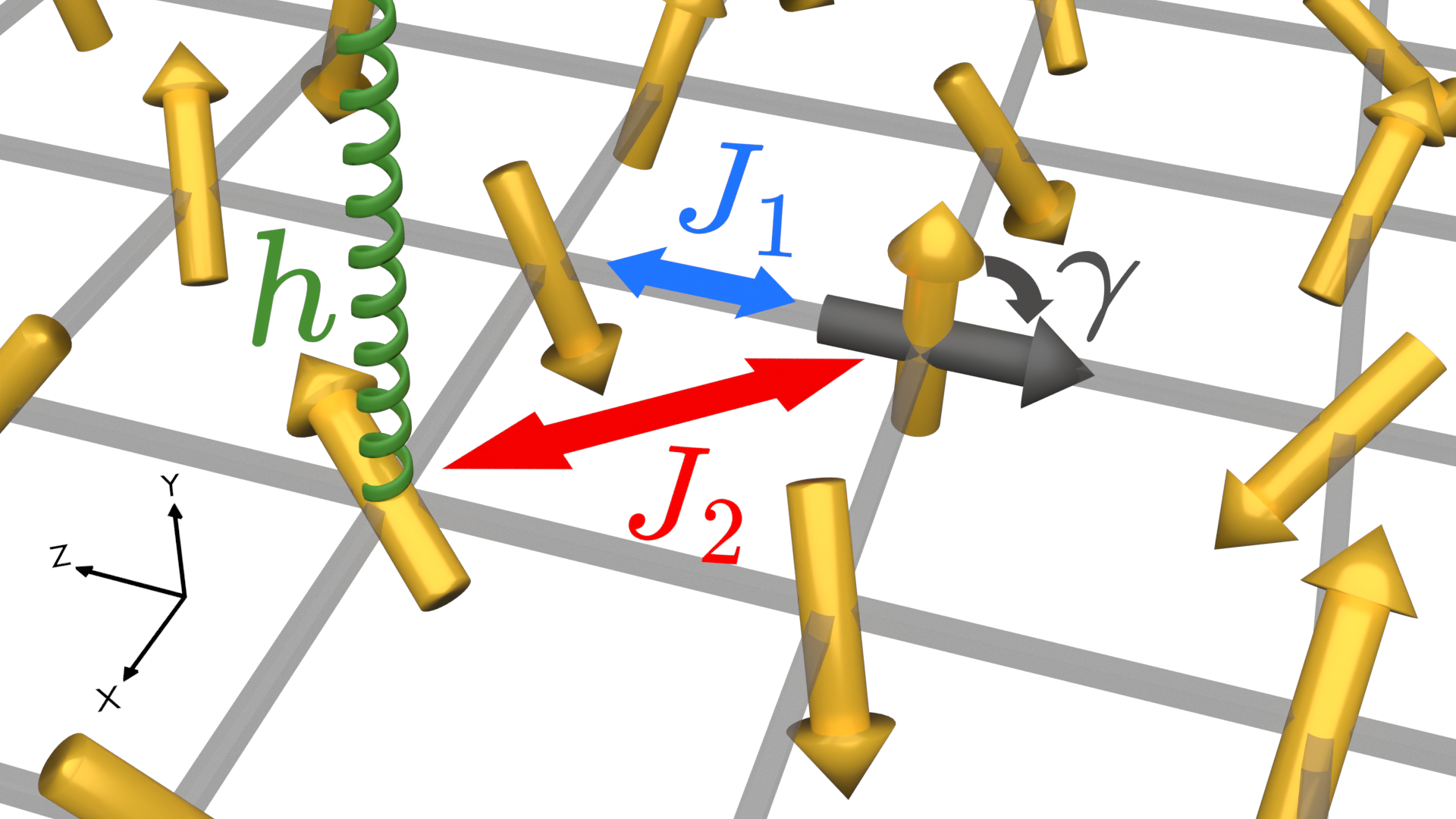}
    \caption{\textbf{Interacting spins on a driven-dissipative square lattice.} The system is driven by a field $h$ and coupled to an external environment with dissipation rate $\gamma$. Spins interact non-locally through example couplings $J_1$ (nearest-neighbor) and $J_2$ (next-nearest-neighbor) illustrated.}    \label{fig: blender}
\end{figure}

In view of the rapid experimental progress, analytical and computational methods capable of efficiently solving the relevant quantum many-body problem 
are becoming increasingly indispensable  in bridging theory with experimental observations \cite{Reh}.
Nevertheless, with their scalability limited by short coherence times due to high sensitivity to external perturbations \cite{PhysRevLett.113.195302}, the aforementioned platforms should be regarded as open quantum systems, i.e. where the external environment retains crucial influence over the behavior of the system. 
Therefore, to accurately model the long-range interacting system, it is necessary to account for the impact of the external environment.
Despite that, given the high inherent difficulty of describing open many-body quantum systems, much of the theoretical research literature resorted to replacing long-range interactions with more tractable effective short-range interactions \cite{PhysRevLett.130.163601}. Simultaneously, the inclusion of long-ranged interactions in leading numerical approaches, such as tensor network algorithms which can quantify correlations to a high degree of accuracy for low-dimensional lattices with limited entanglement \cite{RevModPhys.93.015008, fazio2024manybodyopenquantumsystems}, remains challenging, although some progress has been possible very recently with tree tensor networks \cite{PhysRevA.109.022420}. 
Effective treatment of multiple competing long-ranged interactions, due to the frustration they induce in the many-body system, is expected to be even more out of reach.
New computational techniques capable of accurately characterizing the combined effects of drive, dissipation, and complex long-ranged interactions are therefore becoming urgently needed tools for probing fundamental questions on the non-equilibrium physics of an increasingly diverse range of experimentally-realizable many-body systems, as well as for facilitating the practical implementation of near-term quantum technologies.

In this Letter, we propose a time-dependent variational Monte Carlo approach utilizing a matrix product operator (MPO) tensor network ansatz to simulating open quantum lattice systems.
Related hybrid approaches have been proposed before for closed quantum systems \cite{PhysRevLett.99.220602, PhysRevB.83.134421, PhysRevB.111.075102} and have proved capable of advancing our understanding of the physics of e.g. frustrated spin \cite{PhysRevB.103.235155} and strongly-correlated electrons in two dimensions \cite{r4q9-4yvj}, or 2+1 dimensional lattice gauge theories \cite{PhysRevD.107.014505}.
Our method relies on efficiently solving the variational equations of motion for the Lindblad quantum master equation by leveraging a compact tensor network representation of the many-body density matrix in combination with Monte Carlo sampling.
We provide a publicly-accessible implementation that permits efficient simulation of large-scale many-body dissipative interacting quantum lattices while offering unique technical advantages over comparable state-of-the-art tensor and neural network approaches.
Importantly, it enables accurate simulation of open quantum systems with hitherto intractable long-ranged couplings, including arbitrary combinations of spatially-decaying non-local interactions, which we showcase by investigating the dynamics and steady states of driven-dissipative one- and two-dimensional lattices with long-range competing Ising and XYZ interactions, as pictured in Fig. \ref{fig: blender} for a square lattice. In particular, we report on the emergence of a spatially-modulated magnetic ordering in the non-equilibrium steady state.

\begin{center}\noindent\textbf{RESULTS}\end{center}

\begin{center}\noindent\textbf{{Time-dependent variational Monte Carlo with matrix product operators}}\end{center}

We consider open quantum many-body systems whose time evolution is described by a Markovian Lindblad quantum master equation (setting $\hbar=1$)
\begin{equation}
    \frac{\dd}{\dd t} \rho = \pazocal{L}\rho := -i[H,\rho] + \sum_k \left(\Gamma_k \rho \Gamma_k^\dagger - \frac{1}{2}\{\Gamma_k^\dagger \Gamma_k,\rho\}\right),
    \label{eq: lindblad master equation}
\end{equation}
with the Hamiltonian $H$ and Lindblad jump operators $\{\Gamma_k\}$ governing the coherent and dissipative dynamics of the system, respectively.
The superoperator $\pazocal{L}$ is called the Lindbladian.
We express the density matrix in terms of a set of variational parameters $\bm{a} = \{a_i\}_{i=1}^{N_\text{param}}$ such that $\rho(t) := \rho(\bm{a}(t))$.
Specifically, we represent the density matrix as a matrix product operator (MPO) tensor network, $\rho(t) = \sum_{\{\sigma_i\},\{\sigma'_i\}}\tr A_1^{\sigma_1\sigma'_1}\dots A_N^{\sigma_N\sigma'_N} \ket{\bm{\sigma}} \bra{\bm{\sigma'}}$, where $\bm{\sigma}$ and $\bm{\sigma'}$ denote many-body configurations in the computational basis. The MPO can be vectorized into a matrix product state (MPS), $\ket{\rho(t)} := \sum_{\{x_i\}}\tr A_1^{x_1}\dots A_N^{x_N} \ket{\bm{x}}$, where $x_i := (\sigma_i,\sigma'_i)$.
The variational parameters are then the elements of the matrices $\{A^s_i\}_{i=1, s=1}^{N, d^2}$ of the MPS (where $d$ denotes the local Hilbert space dimension). We take the matrices to be square with equal (bond) dimension $\chi$.

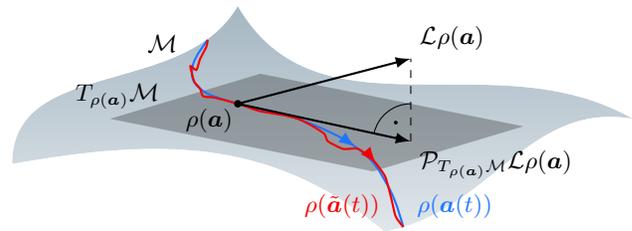
\begin{figure}
    \begin{tikzpicture}
    
    \definecolor{mylightgreen}{RGB}{179, 194, 204}
    \definecolor{mydarkgreen}{rgb}{40,52,60}
    \definecolor{myblue}{RGB}{31,117,255}
    \definecolor{myred}{RGB}{245,0,0}
    
    \shade[thin, top color=mylightgreen!120, bottom color=mylightgreen!40, 
    shift={(0.2, 0.7)},scale=1.5]
    (0, 0.8) to[out=10, in=140] 
    (3.3, 0.2) to [out=10, in=-190] 
    (5.5, 1.2) to [out=150, in=-60] 
    (2, 2.2) to[out=-120, in=90] cycle;
    
    \coordinate (A) at (1.5,2.5);
    \coordinate (B) at (5.0,1.8);
    \coordinate (C) at (7.0,2.4);
    \coordinate (D) at (3.5,3.1);
    
    \fill[gray!180, fill opacity=0.3] (A) -- (B) -- (C) -- (D) -- cycle;
    
    \draw [thick, myblue] plot [smooth, tension=0.7] coordinates {(2.8,3.55) (2.6,3.2) (2.7,2.9) (3.2,2.7) (4.0,2.5) (4.8,2.1) (5.2,1.6) (5.4,1.06)} [arrow inside={end=latex,opt={myblue,scale=1.2}}{0.7}];

    \draw [thick, myred] plot [smooth, tension=0.4] coordinates {
        (2.8,3.55) 
        (2.78,3.45)
        (2.7,3.3) 
        (2.65,3.15) 
        (2.57,3.2) 
        (2.62,3.0)
        (2.7,2.9) 
        (2.74,2.8)
        (2.9,2.78)
        (3.2,2.7) 
        (3.35,2.65)
        (3.5,2.6)
        (3.8,2.55)
        (4.0,2.5) 
        (4.2,2.4)
        (4.35,2.3)
        (4.4,2.2) 
        (4.6,2.1)
        (4.8,2.1) 
        (5.0,1.95) 
        (5.15,1.75)
        (5.2,1.6) 
        (5.25,1.4)
        (5.3,1.2)
        (5.4,1.06)
    } [arrow inside={end=latex,opt={myred,scale=1.2}}{0.79}];
    
    \coordinate (center) at (3.2,2.7);
    \coordinate (exact) at (5.5,3.3);
    \coordinate (projected) at (5.5,2.2);

    \draw[-{Latex[length=2mm]}, thick] (center) -- (exact);
    
    \draw[-{Latex[length=2mm]}, thick] (center) -- (projected);
    
    \draw[dashed] (projected) -- (exact);
    
    \node at (center)[circle,fill,inner sep=1.0pt]{};
    
    \node at (5.3,2.45)[circle,fill,inner sep=0.5pt]{};
    
    \usetikzlibrary {arrows.meta}
    \usetikzlibrary{angles, quotes}
    
    \node[below left] at (3.25,2.75) {$\rho(\boldsymbol{a})$};
    \node[above right] at (exact) {$\pazocal{L}\rho(\boldsymbol{a})$};
    \node[below right] at (projected) {$\pazocal{P}_{T_{\rho(\boldsymbol{a})}\pazocal{M}}\pazocal{L}\rho(\boldsymbol{a})$};

    \pic [draw, angle radius=5mm, angle eccentricity=1.2, ] {angle = exact--projected--center};
    \node at (5.32,2.25) {.};
    
    \node at (2.2,3.5) {$\pazocal{M}$};
    
    \node at (1.6,2.8) {$T_{\rho(\boldsymbol{a})}\pazocal{M}$};
    
    \node[above left, text=myred] at (5.2,1.06) {$\rho(\tilde{\boldsymbol{a}}(t))$};
    \node[above left, text=myblue] at (6.7,1.06) {$\rho(\boldsymbol{a}(t))$};
    
    \end{tikzpicture}
    \caption{\textbf{Illustration of the Dirac-Frenkel variational principle applied to the Lindblad master equation \eqref{eq: lindblad master equation}.} Given a set of variational parameter values $\boldsymbol{a}$, spanning a variational submanifold $\pazocal{M}$ of the projective Hilbert space, the exact time-evolution vector $\pazocal{L}\rho(\boldsymbol{a})$, which generally lies outside of $\pazocal{M}$, is orthogonally projected onto the tangent space $T_{\rho(\boldsymbol{a})}\pazocal{M}$ of $\pazocal{M}$ at $\rho(\boldsymbol{a})$ by means of the projection operator $\pazocal{P}_{T_{\rho(\boldsymbol{a})}\pazocal{M}}$. The blue path $\rho(\boldsymbol{a}(t))$ on $\pazocal{M}$, resultant from a succession of projections, is the best approximation to the exact dynamics \cite{PhysRevLett.107.070601}. The red path $\rho(\tilde{\boldsymbol{a}}(t))$ is a stochastic solution to the variational equations of motion.}
    \label{fig: variational principle}
\end{figure}

In the time-dependent variational approach, one seeks to minimize the distance between the variationally and exactly time-evolved density matrices, i.e. solve the optimization problem
\begin{equation}
    \bm{a}(t) = \text{arg\,min} \Bigg|\Bigg| \sum_j \dot{a}_j \partial_j \rho(t) - \pazocal{L}\rho(t) \Bigg|\Bigg|^2,
\end{equation}
where $\partial_j := \frac{\partial}{\partial a_j}$.
A formal solution can be obtained via a projection argument of Dirac and Frenkel \cite{Dirac_1930, frenkel1934wave}, whereby a dynamics restricted to the variational manifold is solved instead of the exact Lindblad master equation \eqref{eq: lindblad master equation} (see Fig. \ref{fig: variational principle}).
With a Hilbert-Schmidt inner product and norm assumed, the resultant equations of motion for the variational parameters become \cite{Yuan2019theoryofvariational}
\begin{equation}
    \sum_j S_{ij} \dot{a}_j = f_i,
    \label{eq: variational equations of motion}
\end{equation}
where, expressed in Liouville-space notation \cite{Hryniuk2024tensornetworkbased, Gyamfi2020}, we defined the metric tensor (or normalized Gram matrix) 
$
S_{ij} = \bra{\partial_i\rho(t)} \ket{\partial_j\rho(t)}/Z
$
and variational forces
$
f_i = \bra{\partial_i\rho(t)} \ket{\pazocal{L} \rho(t) }/Z,
$
with the normalization constant $Z = \bra{\rho(t)} \ket{\rho(t)}$. A solution to Eq. \ref{eq: variational equations of motion} yields the non-equilibrium dynamics of the open quantum system.

The above expectation values are difficult to compute exactly for larger many-body systems due to the exponential growth in size of the Hilbert space. 
Instead, let us express them as ensemble averages over many-body configurations $\{\bm{x}\}$, distributed according to the probability density $p(\bm{x}) = |\langle \bm{x}|\rho (t) \rangle|^2/Z$:
$
    S_{ij} = \mathbbm{E}_{\bm{x}\sim p(\bm{x})}[  \Delta_i^\dagger (\bm{x}) \Delta_j (\bm{x}) ]
$
and
$
    f_i = \mathbbm{E}_{\bm{x}\sim p(\bm{x})}[ \Delta_i^\dagger (\bm{x}) \pazocal{L}_\text{loc}(\bm{x}) ],
$
where we defined the diagonal log-derivative superoperator,
$
    \Delta_i = \sum_{\bm{x}} \partial_{i} \ln{\bra{\bm{x}}\ket{\rho(t)}} \ket{\bm{x}}\bra{\bm{x}}
$
(satisfying
$
    \Delta_i \ket{\rho} = \partial_i \ket{\rho}
$
and
$
    \Delta_i(\bm{x}) = \bra{\bm{x}} \Delta_i \ket{\bm{x}}),
$
and the local estimator of the Lindbladian,
\begin{equation}
    \pazocal{L}_\text{loc}(\bm{x}) = \frac{\bra{\bm{x}}\pazocal{L}\ket{\rho(t)}}{\bra{\bm{x}}\ket{\rho(t)}}.
    \label{eq: local estimator}
\end{equation}
One can thus approximate the metric tensor and variational forces stochastically by averaging $S_{ij}$ and $f_i$ over a subset of representative many-body configurations $\{\bm{x}\}$.
In our implementation, the samples are drawn via a sequential Metropolis Markov-chain Monte Carlo algorithm, where partially contracted matrix products are saved to memory and reused during subsequent lattice sweeps. Besides producing independent samples efficiently, this algorithm minimizes costly tensor contractions in computing the probability amplitudes \cite{Hryniuk2024tensornetworkbased}. For a detailed description of the sampling algorithm, see Supplementary Note 1 A.

Let us consider more closely the local estimator of the Lindbladian in Eq. \eqref{eq: local estimator}. Unlike in comparable variational Monte Carlo (VMC) algorithms reliant on non-linear neural-network architectures \cite{Vicentini2019,Nagy2019,PhysRevLett.122.250502,PhysRevB.99.214306,Mellak2022,mellak2024deep, vicentini2022positivedefiniteparametrizationmixedquantum}, where an explicit summation over auxiliary many-body configurations must be introduced to evaluate the local estimator, here we can contract directly the underlying tensors, reducing computational cost.
Specifically, we decompose the Lindbladian superoperator into a sum of distinct $n$-local superoperators, $\pazocal{L} = \sum_{\{n\}} \sum_{j=1}^N l_j^{[n]}$, which allows for the numerator in the local estimator to be efficiently contracted for systems with quasi-local terms (i.e. with small $n$), or with arbitrary non-local terms with purely-diagonal $\{ l_j^{[n]} \}$ in the computational basis; we leverage the latter observation to simulate spin lattices with otherwise prohibitively-difficult competing long-ranged Ising interactions in a subsequent part of this Letter. For a detailed discussion on the contraction of the local estimator, see Supplementary Note 1 B.

Our approach enjoys further advantages over comparable methods.
In contrast to other common tensor network time-evolution algorithms \cite{PAECKEL2019167998, SCHOLLWOCK201196}, the variational approach does not rely on a Suzuki-Trotter decomposition of the propagator $e^{\pazocal{L} t}$, and as such is free of a Trotter error \cite{PhysRevLett.107.070601}. This may be relevant for reducing the error in steady-state measurements extracted at long-times, where the repeated projections from the TDVP may suppress time-discretization errors acquired during time-propagation when approaching the fixed point. In addition, our method does not rely on expressing the Lindbladian superoperator in MPO form, which may require a large bond dimension that impacts the computational complexity \cite{PhysRevA.92.022116}. The present approach is also readily applicable to finite periodic systems
(with the MPO density matrix forming a closed loop), which are not efficiently treatable with conventional tensor network methods \cite{SCHOLLWOCK201196, 10.21468/SciPostPhysLectNotes.8}.
In contrast again to current neural network VMC approaches \cite{Reh, mellak2024deep, Mellak2022, Vicentini2019, Nagy2019, PhysRevLett.122.250502, PhysRevB.99.214306, vicentini2022positivedefiniteparametrizationmixedquantum, PhysRevLett.128.090501},
relevant observables such as magnetization and correlation functions (and some non-linear functions of the density matrix such as the purity or Rényi-2 entropy) can be evaluated exactly via an appropriate tensor contraction, and so are free of statistical errors in the measurements.

\begin{figure}
    \includegraphics[width=\columnwidth]{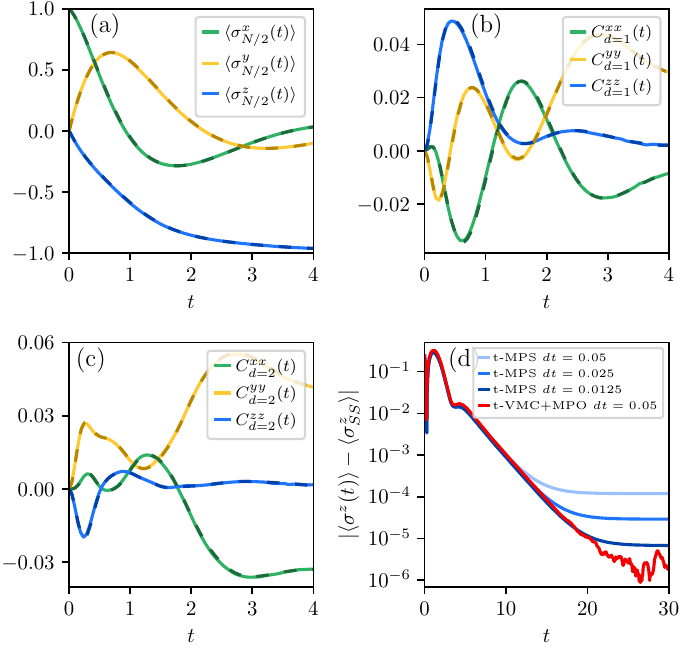}
    \caption{\textbf{Non-equilibrium relaxation dynamics of spin chains obtained with t-VMC+MPO and t-MPS.} Dynamics of the bulk (a) magnetizations, (b) nereast-neighbor, and (c) next-nearest-neighbor spin-spin correlation functions for the dissipative anisotropic antiferromagnetic Heisenberg spin chain with $(J_x, J_y, J_z)=(-1.0,-0.9,-1.2)$ and $N=200$ sites at $\gamma=-h=1$, starting from the product state $\expval{\sigma^x}=1$. 
    Solid and darker dashed lines are respectively t-VMC+MPO and t-MPS results, each with $\chi=20$. 
    (d) Long-time relaxation of the difference of bulk spin magnetization and corresponding exact steady-state expectation value for a dissipative transverse-field Ising chain with $N=10$, $J_z=-0.5$, and $\gamma=-h=1$. Compared are t-VMC+MPO and t-MPS dynamics (for multiple time-step sizes), each with $\chi=20$.}
    \label{fig: XYZ}
\end{figure}

\begin{center}\noindent\textbf{{Numerical simulations}}\end{center}

To benchmark our method, we first simulate the dynamics of a dissipative quantum spin chain defined by the anisotropic Heisenberg Hamiltonian
\begin{equation}
    H = -\sum_{i} J^x \sigma_i^x \sigma_{i+1}^x + J^y \sigma_i^y \sigma_{i+1}^y + J^z \sigma_i^z \sigma_{i+1}^z - h \sum_i \sigma_i^z,
\end{equation}
together with the spin decay Lindblad jump operators $\Gamma_k = \frac{\sqrt{\gamma}}{2}(\sigma^x_k - i\sigma^y_k)$ for $k=1,\dots,N$, where $\{\sigma^\alpha\}_{\alpha=x,y,z}$ denote Pauli spin matrices.
In Fig. \ref{fig: XYZ} (a-c), we study the dynamics of the magnetizations (defined as $ \langle \sigma^\alpha (t) \rangle = \tr \{\sigma^\alpha \rho (t)\}$), nearest-neighbor, and next-nearest-neighbor spin-spin correlation functions (defined as $C_d^{\alpha\alpha}(t) = \langle \sigma^\alpha_i \sigma^\alpha_{i+d} (t) \rangle - \langle \sigma^\alpha_i (t)\rangle \langle \sigma^\alpha_{i+d} (t) \rangle$) for the above model with antiferromagnetic interactions and $N=200$ sites, comparing our t-VMC+MPO results to that of t-MPS with second order Suzuki-Trotter decomposition \cite{SCHOLLWOCK201196}.
To ensure efficient integration of the variational equations of motion while minimizing local truncation errors, we utilize a second-order Heun's method with adaptive step sizing, together with a signal-to-noise ratio regularization scheme as proposed in \cite{PhysRevLett.125.100503}. 
Excellent agreement throughout between all considered quantities can be observed.

Real-time evolution algorithms are routinely used to also study the non-equilibrium steady state, since "ground-state search" algorithms that specifically target the steady state \cite{Cui, PhysRevLett.119.010501, CASAGRANDE2021108060, PhysRevE.103.013309, PhysRevB.105.195152} can be inefficient due to functionals that are non-linear in the Lindbladian superoperator, and therefore highly non-local \cite{Weimer2015, Kshetrimayum_2017}. However, many common time evolution algorithms can suffer significant errors due to the Trotterization of the dynamics. In Fig. \ref{fig: XYZ} (d), we consider the relaxation towards the steady state for a chain with antiferromagnetic Ising interactions in a transverse external field. Our stochastic dynamics, obtained via a simple forward Euler integration with fixed step size, but which are free of a compounding Trotter error, attain a smaller error in the estimate of the steady-state magnetization than t-MPS with second-order Suzuki-Trotter decomposition and a fourfold smaller step size.

\begin{center}\noindent\textbf{{Long-range competing interactions}}\end{center}

The competition between multiple interactions at separate length scales is known to give rise to non-trivial phases across different classical and closed quantum systems \cite{RevModPhys.95.035002}, but remains unexplored in the context of open quantum systems.
As a minimal example of an open quantum many-body system with long-range competing interactions, we consider a spin lattice with long-range power-law Ising interactions in a transverse external field, described by the time-independent Hamiltonian
\begin{equation}
    H = - \sum_{n}  J_n \sum_{i,j} d_{ij}^{-\alpha_n} \sigma^z_i \sigma^z_j - h\sum_i \sigma^x_i,
    \label{eq: competing interactions Hamiltonian}
\end{equation}
where $d_{ij}$ denotes the Euclidean distance between sites labelled with indices $i$ and $j$, while $\alpha$ is an exponent characterizing the decline with distance of the interaction strength.
We first model dissipation by the one-body spin-decay Lindblad jump operators $\Gamma_k = \frac{\sqrt{\gamma}}{2}(\sigma_k^x - i\sigma_k^y)$ for $k=1,\dots,N$. This model for a square two-dimensional lattice is illustrated in Fig. \ref{fig: blender}.

\begin{figure}[t]
    \includegraphics[width=\columnwidth]{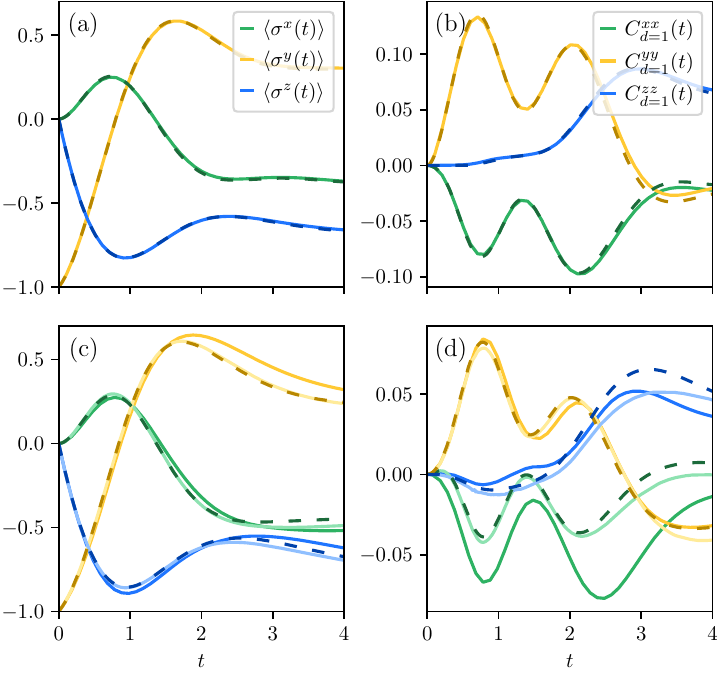}
    \caption{\textbf{Non-equilibrium relaxation dynamics of 1D and 2D spin lattices with long-range competing Ising interactions.} Dynamics of (a) magnetizations and (b) nearest-neighbor correlation functions for a dissipative spin chain with $N=200$ sites and competing long-ranged dipolar Ising interactions ($J_1=-1/2$, $\alpha_1=3$ and $J_2=1$, $\alpha_2=6$), starting from the product state $\langle \sigma^y \rangle=-1$ at $\gamma = -2h = 1$. (c)-(d) as in (a)-(b) for a square lattice with $N=4\times4$ sites ($J_1=-1/4$ and $J_2=1/2$). Darker dashed lines are corresponding exact dynamics obtained for a reduced number of $N=10$ and $N=3\times 3$ sites, respectively. Lighter solid lines in (c-d) represent variational results for $N=3\times 3$ sites.}
    \label{fig: long range dynamics}
\end{figure}

We first consider a pair of competing long-ranged antiferromangetic dipolar ($J_1<0, \alpha_1 = 3$) and ferromagnetic van-der-Waals ($J_2>0, \alpha_2 = 6$) interactions. This particular choice for the decay exponents was motivated by recent work on Rydberg arrays \cite{PhysRevLett.131.203003}, where a competition between van-der-Waals and dipolar exchange interactions was found to lead to novel quantum phases in the ground state. 
In Fig. \ref{fig: long range dynamics} (a-b), we study the non-equilibrium dynamics of the magnetizations and nearest-neighbor correlation functions for a chain with up to $N=200$ sites.
Our implementation can also yield reliable results in the considerably more challenging two-dimensional case for modest system sizes; in Fig. \ref{fig: long range dynamics} (c) and (d) we repeat the previous results for a $N=4\times 4$ square lattice. We note good qualitative agreement in both cases
with exact results (darker dashed lines), obtained for smaller system sizes of $N=10$ and $N=3\times 3$, respectively, with larger discrepancies in the two-dimensional case expected due to larger finite-size effects.

As a more challenging example, we now consider a competition between a strongly long-ranged antiferromagnetic Coulomb and ferromagnetic dipolar interactions, with exponents $(\alpha_1,\alpha_2) = (1,3)$. Such interaction ranges are attainable in particular with trapped ions \cite{annurev:/content/journals/10.1146/annurev-conmatphys-032822-045619}.
Let us concentrate on the steady state of the model with the Hamiltonian \eqref{eq: competing interactions Hamiltonian} in one dimension and dissipation channels $\Gamma_k = \frac{1}{2}(\sigma^z_k-i\sigma^y_k)$, which destroy magnetic ordering in the steady state. 
Indeed, a mean-field analysis yields the paramagnetic product state $\langle \sigma^x \rangle=1$ as a fixed point in the dynamics, irrespective of the interactions.
However, signatures of magnetic ordering due to the long-ranged interactions can be observed by examining the spin-spin correlation functions and structure factor $S_{zz}(q) = \frac{1}{N^2}{ \sum_{n,m} e^{iq(n-m)} \langle\sigma^z_{n} \sigma^z_{m}\rangle }$; 
a finite value of (the square root of) $S_{zz}(q=0)$ and $S_{zz}(q=\pi)$ is indicative of ferromagnetic and antiferromagnetic ordering in the steady state, respectively \cite{Cui}. 
Our algorithm enables us to accurately quantify the correlations, even in the presence of strongly long-ranged interactions. 

\begin{figure}[t]
    \includegraphics[width=\columnwidth]{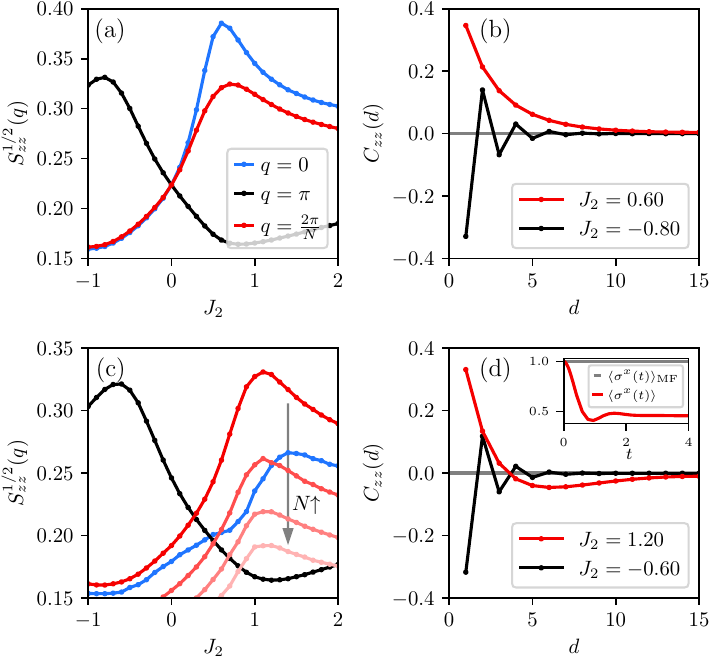}
    \caption{\textbf{Steady-state phase diagram analysis for spin chain with long-range competing Ising interactions.} (a) Steady-state structure factor phase diagrams for a dissipative spin chain with long-range dipolar $(\alpha_2 = 3)$ Ising interactions  as function of the interaction strength $J_2$ for $N=20$. (b) Steady-state spin-spin correlation functions as function of separation distance for $N=30$. (c)-(d) As in (a)-(b) with addition of competing antiferromagnetic Coulomb Ising interaction ($\alpha_1 = 1, J_1=-1.5$). In (c), the reduction of $S_{zz}(q=2\pi/N)$ with increasing system sizes from $N=20$ to $N=50$ is shown in lighter shades of red. To ensure extensivity with system size, the interaction strengths were renormalized by the Kac normalization factor. Inset in (d): reduction over time of the paramagnetic order parameter $\langle \sigma^x (t) \rangle$ for $J_1=-J_2=0.5$ and $N=50$ as compared to mean-field prediction $\langle \sigma^x (t) \rangle_\text{MF} = 1$.}
    \label{fig: NESS phase diagrams}
\end{figure}

In Fig. \ref{fig: NESS phase diagrams} we compare the steady-state correlation functions and $S_{zz}(q)$ for different values of $J_2$ and $q$, demonstrating signatures of ferromagnetic and antiferromagnetic behaviour in the steady state. In particular, we observe the emergence of a spatially-modulated magnetic ordering in the regime of competing interactions, as evidenced by the dominance of the order parameter $S_{zz}(q=2\pi/N)$ for $J_2\gtrsim 0.4$ in (c) and the unusual behavior of the correlation function with separation distance in (d). Nevertheless, with the magnetic ordering suppressed over longer distances by dissipation, the order parameters decrease with increasing system sizes (as illustrated for $q=2\pi/N$ in (c)), indicative of short-ranged order. The above is further corroborated by the observed reduction in the paramagnetic order parameter $\langle \sigma^x \rangle$ in the steady state to less than half (and decreasing with increasing system size), as compared to the mean-field prediction (inset).
We note that a large bond dimension of $\chi=30$ (corresponding to 3600 distinct variational parameters) was necessary to obtain reliable measurements in Fig. \ref{fig: NESS phase diagrams} (c-d) due to the strongly long-ranged interactions.

\begin{figure}[t]
    \includegraphics[width=\columnwidth]{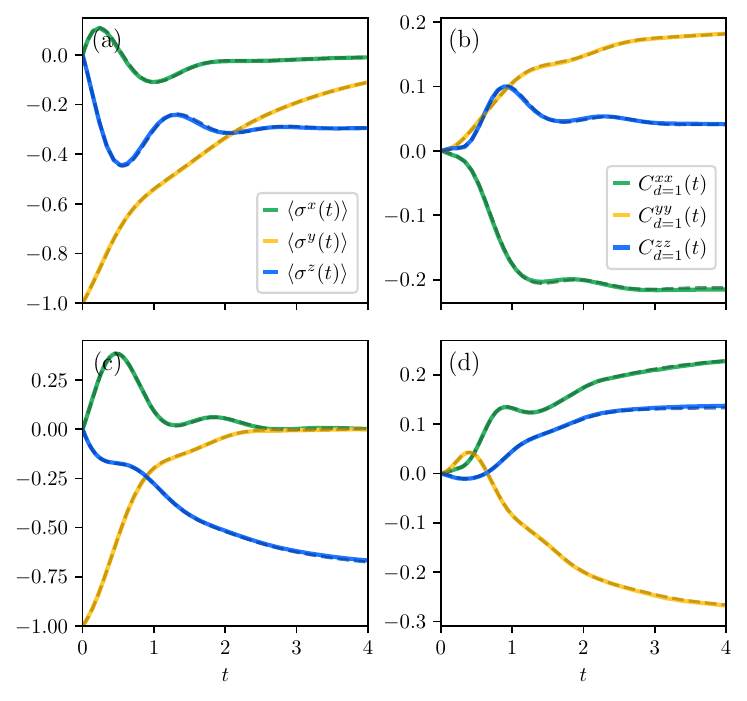}
    \caption{\textbf{Non-equilibrium relaxation dynamics of spin chain with long-range competing XYZ interactions.} Dynamics of (a) magnetizations and (b) nearest-neighbor correlation functions for a dissipative spin chain with $N=50$ sites and long-ranged dipolar anisotropic XYZ interactions ($\alpha_1=3$, $\bm{J}_1=(-0.6, 0.5, -0.4)$) as in Hamiltonian \eqref{eq: long-range XYZ}, starting from the product state $\langle \sigma^y \rangle=-1$ at $\gamma = -2h = 1$. (c)-(d) As in (a)-(b) with addition of competing long-range XYZ interactions ($\alpha_2 = 6$, $\bm{J}_2=(0.9, -1.0, 1.1)$). Darker dashed lines are corresponding exact dynamics obtained for a reduced number of $N=10$ sites in both cases.}
    \label{fig: non-diagonal long range dynamics}
\end{figure}

\begin{center}\noindent\textbf{{Non-diagonal long-range interactions}}\end{center}

The introduced approach can be applied also to models with non-diagonal spatially-decaying long-range (competing) interactions.
In contrast to diagonal terms in the Lindbladian, which we can treat exactly yet efficiently as showcased above, non-diagonal terms must be truncated at longer distances to reduce costly tensor contractions. As an example, we consider a variation on the quantum Heisenberg model with long-range competing interactions and local incoherent losses, described by the Hamiltonian
\begin{equation}
    H = - \sum_{n}  \sum_{\beta = \{x,y,z\}} J_n^{\beta} \sum_{i,j} d_{ij}^{-\alpha_n} \sigma^{\beta}_i \sigma^{\beta}_j - h\sum_i \sigma^z_i.
    \label{eq: long-range XYZ}
\end{equation}
Specifically, we consider the pair of anisotropic interactions $\bm{J}_1=(0.6, -0.5, 0.4)$ and $\bm{J}_2=(-0.9, 1.0, -1.1)$ with exponents $\alpha_1 = 3$ and $\alpha_2 = 6$. We model dissipation by the jump operators $\Gamma_k = \frac{\sqrt{\gamma}}{2}(\sigma^x_k - i\sigma^y_k)$ for $k=1,\dots,N$. To reduce computational cost, we truncate the interactions at a distance of $r = 4$, beyond which the combined interaction strength falls to less than 2\% of its value for nearest neighbors. In Fig. \ref{fig: non-diagonal long range dynamics} we plot the dynamics of the magnetizations and nearest-neighbor correlation functions for $N=50$ for this model, noting excellent agreement with exact results for $N=10$ with and without competition. For a comprehensive convergence analysis of this system see Supplementary Note 2.

\begin{center}\noindent\textbf{DISCUSSION}\end{center}

Understanding the non-equilibrium behaviour of many-body systems with complex non-local interactions is of significant fundamental and practical interest, necessitating the development of advanced simulation methods.
In this Letter, we introduced a variational method combining tensor network and Monte Carlo techniques for simulating the non-equilibrium dynamics and steady states of open many-body quantum systems. 
Our numerical implementation permits efficient simulation of large open quantum lattices while offering important technical advantages over comparable established tensor network and variational Monte Carlo approaches.
Crucially, by enabling the simulation of highly-non-local interactions in one and two dimensions, our method opens new avenues for exploring the properties of realistic open quantum systems with complex long-ranged interaction.

Our findings reveal a rich and hitherto unexplored domain of non-equilibrium physics at the intersection of long-ranged competing interactions, drive, and dissipation. An obvious question raised by our results regard the possible existence of modulated long-ranged magnetic order in the steady state, i.e. whether it can withstand, or be sustained by, dissipation. 
With long-ranged antiferromagnetic order shown to exist in the steady state of two-dimensional Rydberg lattices in the strong decoherence limit \cite{PhysRevA.90.021603}, one may be hopeful that similar mechanisms supporting long-ranged modulated phases can exist too.
However, while the present approach was found to be effective at characterizing the dynamical and steady-state properties of open quantum lattices with long-range competing Ising interactions, the frustration present in lattices with more complex interactions or geometries may prohibit accurate simulation with the MPO ansatz. In this respect, a reformulation of the present approach for a projected-entangled pair-operator or another suitable variational ansatz \cite{PhysRevLett.129.220401} may be worthwhile.  
Finally, we expect that our variational tensor network approach can be extended beyond solving the Lindblad master equation, and be applied to e.g. stochastic reaction-diffusion systems, for which the potential of tensor networks is only beginning to be realized \cite{PhysRevX.13.041006}.

\begin{center}\noindent\textbf{METHODS}\end{center}

\textbf{Tensor network results}. For an extended discussion of the design, implementation, and numerical cost analysis of the t-VMC+MPO method introduced in this work, see Supplementary Note 1. Below, we briefly summarize the main steps of a single iteration:
\begin{enumerate}
    \item Draw $N_{\text{MC}}$ samples from the distribution $p(\bm{x})=|\langle \bm{x} | \rho \rangle|^2/\sum_{\bm{x}} |\langle \bm{x} | \rho \rangle|^2$ via the sequential Metropolis sampling protocol, discussed in detail in Supplementary Note 1 A.
    \item Calculate the local estimator of the Lindbladian $\pazocal{L}_\text{loc}(\bm{x})$ and logarithmic derivatives $\Delta_i(\bm{x})$. For details on optimal tensor contractions see Supplementary Note 1 B-C.
    \item Calculate the variational forces $f_i = \mathbbm{E}_{\bm{x}\sim p(\bm{x})}[ \Delta^\dagger_i(\bm{x}) \pazocal{L}_\text{loc}(\bm{x})]$ and metric $S_{ij} = \mathbbm{E}_{\bm{x}\sim p(\bm{x})}[ \Delta^\dagger_i(\bm{x}) \Delta_i(\bm{x})]$. 
    \item Regularize the metric tensor $S_{ij}$ to ensure invertibility. For an explanation and convergence analysis of the regularization schemes considered in our implementation, see Supplementary Note 1 D and 2.
    \item Update the variational parameters as $a_i(t+\delta t) = a_i(t)+\delta t \sum_j S_{ij}^{-1}f_j$. Alternatively, a more advanced higher-order integration scheme may be used, in which case all previous steps must be repeated for all intermediate quantities involved. In our implementation, we utilize an adaptive Heun integrator as proposed in \cite{PhysRevLett.125.100503}.
\end{enumerate}
Steps 1-6 are iterated until a desired time is reached.

In Fig. \ref{fig: XYZ} we also obtained the dynamics using the t-MPS algorithm; for an explanation of t-MPS, see \cite{SCHOLLWOCK201196}.\newline\newline
\textbf{Exact results}.
Exact results in Fig. \ref{fig: long range dynamics} have been obtained with \textit{QuantumOptics.jl} \cite{KRAMER2018109}.\newline\newline
\textbf{Mean-field results}.
The mean-field results discussed in the manuscript and in Fig. \ref{fig: NESS phase diagrams} have been obtained from the mean-field Heisenberg equations of motion for the magnetizations of the model. For the Hamiltonian in Eq. (5) with Kac-renormalized interaction strengths ($\tilde{J}=J/\pazocal{N(\alpha)}$, where $\pazocal{N(\alpha)}$ is the Kac normalization factor) they are given by
\begin{align}
    \frac{\dd}{\dd t} \expval{\sigma^x} &= 2\left(\sum_n \tilde{J}_n\right)\expval{\sigma^y}\expval{\sigma^z} + \gamma(1-\expval{\sigma^x}) \\
    \frac{\dd}{\dd t} \expval{\sigma^y} &= -2\left(\sum_n \tilde{J}_n\right)\expval{\sigma^x}\expval{\sigma^z} + 2h\expval{\sigma^z} - \frac{\gamma}{2}\expval{\sigma^y} \\
    \frac{\dd}{\dd t} \expval{\sigma^z} &= -2h\expval{\sigma^y} - \frac{\gamma}{2}\expval{\sigma^z}.
\end{align}
\newline\newline

\begin{center}\noindent\textbf{CODE AVAILABILITY}\end{center}

The method introduced in this manuscript has been implemented in \textit{Julia} \cite{doi:10.1137/141000671} and is accessible on GitHub \footnote{Online GitHub repository, \url{https://github.com/dhryniuk/t-VMPOMC}}. The parameter and hyperparameter values used in our simulations are given in Table S1.\\\\

\begin{center}\noindent\textbf{DATA AVAILABILITY}\end{center}

The simulation data presented in this manuscript is available online at the UCL Research Data Repository \cite{Hryniuk2025data}.\\\\

\begin{center}\noindent\textbf{ACKNOWLEDGEMENTS}\end{center}

We would like to thank Filippo Vicentini for his valuable insights and discussions.
M. H. S. gratefully acknowledges financial support from Engineering and Physical Sciences Research Council (Grants No. EP/R04399X/1, No. EP/S019669/1, and No. EP/V026496/1).
This work was supported by the Engineering and Physical Sciences Research Council (Grant No. EP/R513143/1 and No. EP/S021582/1).
The authors acknowledge the use of the UCL HPC facilities and associated support services in the completion of this work.\\\\

\begin{center}\noindent\textbf{AUTHOR CONTRIBUTIONS}\end{center}

D.A.H. conceptualized the work, developed and implemented the method, performed the simulations, and wrote the manuscript. M.H.S. supervised the work and reviewed the manuscript.\\\\

\begin{center}\noindent\textbf{ADDITIONAL INFORMATION}\end{center}

\begin{center}\noindent\textbf{{Competing interests}}\end{center}

The authors declare no competing interests.\\

\begin{center}\noindent\textbf{Correspondence}\end{center}

Correspondence and requests for materials should be addressed to D.A.H.\\

\bibliography{references.bib}
\end{document}

\title{Supplementary Material: Variational approach to open quantum systems with long-range competing interactions}

\author{Dawid A. Hryniuk}
\email{d.hryniuk@ucl.ac.uk}
\affiliation{Department of Physics and Astronomy, University College London,
Gower Street, London, WC1E 6BT, United Kingdom}
\affiliation{London Centre for Nanotechnology, University College London, Gordon Street, London WC1H 0AH, United Kingdom}
\author{Marzena H. Szymańska}
\affiliation{Department of Physics and Astronomy, University College London,
Gower Street, London, WC1E 6BT, United Kingdom}

\date{\today}

\begin{abstract}
\end{abstract}

\maketitle

\tableofcontents
\bigskip

\section[SN\thesection]{t-VMC+MPO method -- additional details}

\subsection{Sequential Metropolis sweep}

The MPO ansatz for the density matrix after vectorization into a MPS is explicitly given by
\begin{equation}
    \ket{\rho} = \sum_{\{x_i\}} \sum_{\{\alpha_i\}} A_{\alpha_1\alpha_2}^{x_1}\dots A_{\alpha_N\alpha_1}^{x_N} \ket{x_1\dots x_N},
\end{equation}
where $\{x_i\}$ and $\{\alpha_i\}$ are called the "physical" and "virtual" indices of the MPO, respectively.
Our approach relies on sampling representative many-body configurations $\{\bm{x}\}$ from the distribution $p(\bm{x}) = |\langle \bm{x}|\rho (t) \rangle|^2/\sum_{\bm{y}} |\langle \bm{y}|\rho (t) \rangle|^2$, where the probability amplitudes become
\begin{equation}
    \bra{\bm{x}}\ket{\rho} = \sum_{\{\alpha_i\}} A^{x_1}_{\alpha_1\alpha_2}\dots A^{x_N}_{\alpha_N\alpha_1}.%
\end{equation}
An efficient and cost-effective approach is to apply a sequence of Metropolis updates on each site of the lattice \cite{Hryniuk2024tensornetworkbased}. Let us define the partial matrix products
\begin{equation}
    L_j = \sum_{\{\alpha_i\}} A^{x_1}_{\alpha_1\alpha_2}\dots A^{x_j}_{\alpha_j\alpha_{j+1}}, \qquad R_j = \sum_{\{\alpha_i\}} A^{x_j}_{\alpha_j\alpha_j+1}\dots A^{x_N}_{\alpha_N\alpha_{1}},
\end{equation}
which satisfy the recurrence relations $L_{j+1} = L_j A^{x_{j+1}}_{\alpha_{j+1}\alpha_{j+2}}$ and $R_{j-1} = A^{x_{j-1}}_{\alpha_{j-1}\alpha_{j}} R_j$ with $L_0 = R_{N+1} = \mathbb{1}$. We also note that $\bra{\bm{x}}\ket{\rho} = L_N = R_1$.
To generate a new sample $\bm{x}'$ given an existing sample $\bm{x}$ with $q=\bra{\bm{x}}\ket{\rho}$ and a set $\{R_j\}$ we proceed for $j=1,\dots,N$ as follows:
\begin{enumerate}
    \item Propose a new single-body configuration $x_j'$ from a uniform proposal distribution.
    \item Calculate the Metropolis acceptance probability $p(x_j\leftarrow x_j') = q'/q$ where $q' = \tr L_{j-1} A(x_j') R_{N-j}$.
    \item With probability $p(x_j\leftarrow x_j')$, set $x_j\leftarrow x_j'$ and $q\leftarrow q'$.
    \item Compute and store $L_j = L_{j-1}A(x_j)$.
\end{enumerate}
The above algorithm produces a new Monte Carlo sample $\bm{x}'$ and a set of partial matrix products $\{L_j\}$, to be reused at later points of the method, at a computational cost of $2N\chi^3$. One can also proceed in reverse, generating $\{R_j\}$.

\subsection{Contraction of the local estimator}

Here we discuss the efficient computation of the local estimator of the Lindbladian, $\pazocal{L}_\text{loc}(\bm{x}) = \frac{\bra{\bm{x}}\pazocal{L}\ket{\rho}}{\bra{\bm{x}}\ket{\rho}}$, for Lindbladian and density operators in MPO form. 
An arbitrary Liouville-space operator $l$ can be put into MPO form:
\begin{equation}
    l = \sum_{\{u_i\}, \{v_i\}} \sum_{\{\beta_i\}} B_{\beta_1\beta_2}^{u_1 v_1} \dots B_{\beta_N\beta_1}^{u_N v_N} \ket{u_1\dots u_N} \bra{v_1\dots v_N}
\end{equation}
One therefore obtains for the numerator:
\begin{equation}
    \bra{\bm{x}}l\ket{\rho} = \sum_{\{\alpha_i\}, \{\beta_i\}} M_{(\beta_1 \alpha_1)(\beta_2 \alpha_2)}^{x_1} \dots M_{(\beta_N \alpha_N)(\beta_1 \alpha_1)}^{x_N},
\end{equation}
where $M^{x_i}_{(\beta_i \alpha_i )(\beta_{i+1} \alpha_{i+1})} := \sum_{v_i} B^{x_i v_i}_{\beta_i \beta_{i+1}} A^{v_i}_{\alpha_i \alpha_{i+1}}$. 
An important simplification occurs for quasi-local operators $l_j^{[n]}$, with explicit MPO form
\begin{equation}
    l_j^{[n]} = \sum_{\{u_i\}, \{v_i\}} \sum_{\{\beta_i\}} \delta_{\beta_1\beta_2}^{u_1 v_1} \dots \delta_{\beta_{j-1}\beta_{j}}^{u_{j-1} v_{j-1}}  B_{\beta_{j}\beta_{j+1}}^{u_j v_j} \dots B_{\beta_{j+n-1}\beta_{j+n}}^{u_{j+n-1} v_{j+n-1}} \delta_{\beta_{j+n}\beta_{j+n+1}}^{u_{j+n} v_{j+n}} \dots \delta_{\beta_{N}\beta_{1}}^{u_N v_N} \ket{u_1\dots u_N} \bra{v_1\dots v_N},
\end{equation}
for which the contracted numerator in the local estimator can be written as (suppressing virtual indices)
\begin{align}
    \bra{\bm{x}} l_j^{[n]} \ket{\rho} &= A^{x_1} \dots A^{x_{j-1}} M^{x_j} \dots M^{x_{j+n-1}} A^{x_{j+n}} \dots A^{x_N} \\
    &= L_{j-1}  M^{x_j} \dots M^{x_{j+n-1}} R_{j+n}.
\end{align}
Since $\bm{x}$ is a sample drawn during a sequential Metropolis sweep, either of the partial matrix products $\{L_j\}$ and $\{R_j\}$ have already been computed and stored to memory. As such, they can be recalled to reduce the total number of contractions necessary in the computation of the numerator. For the $n$-local Lindbladian $\pazocal{L} = \sum_{j=1}^N l_j^{[n]}$ we therefore obtain
\begin{equation}
    \pazocal{L}_\text{loc}(\bm{x}) = \bra{\bm{x}}\ket{\rho}^{-1} \sum_{j=1}^N L_{j-1}  M^{x_j} \dots M^{x_{j+n-1}} R_{j+n}.
    \label{local estimator}
\end{equation}

Another critical simplification occurs for the diagonal part of the Lindbladian, for which all tensor contractions vanish. To see this, we rewrite the local estimator as
\begin{equation}
    \pazocal{L}_\text{loc}(\bm{x}) = \frac{\bra{\bm{x}}\pazocal{L}\ket{\rho}}{\bra{\bm{x}}\ket{\rho}} = \sum_{\{\bm{y}\}} \bra{\bm{x}}\pazocal{L}\ket{\bm{y}} \frac{\bra{\bm{y}}\ket{\rho}}{\bra{\bm{x}}\ket{\rho}},
    \label{local estimator expanded}
\end{equation}
where the ratio of probability amplitudes cancels out for the diagonal part of $\pazocal{L}$.
The contraction procedures described above significantly reduce the computational cost of computing the local estimator and log-derivative gradient for most Lindbladians of interest, and especially for periodic lattices with a period $D<N$, as compared to introducing an explicit summation over auxiliary configuration $\{\bm{y}\}$ in Eq. \eqref{local estimator expanded}, a necessity in other VMC algorithms.
The computational cost of computing the numerator in Eq. \eqref{local estimator} for $N$ (translationally-symmetric) operators $\{l_j^{[n]}\}_{j=1}^N$ is $2N\chi^3$ (assuming $\{M\}$, $L_{j-1}$, and $R_{j+1}$ are known, and excluding the cost of contracting the $\{M\}$ matrices with each other, which would depend on the specific bond dimensions of $\{M\}$ and the period $D$). 
This should be contrasted with the cost of computing 
\begin{equation}
    \sum_{j=1}^N \bra{\bm{x}} l_j^{[n]} \ket{\rho} = \sum_{j=1}^N \sum_{\{\bm{y}\}} \bra{\bm{x}} l_j^{[n]} \ket{\bm{y}} \bra{\bm{y}}\ket{\rho} = \sum_{j=1}^N  \sum_{\{\bm{y}\}} \bra{\bm{x}} l_j^{[n]} \ket{\bm{y}} L_{j-1}  A^{y_j} \dots A^{y_{j+n-1}} R_{j+n},
\end{equation}
which is $d^{2n}N(n+1)\chi^3$, explicitly scaled by an additional factor of $d^{2n}$. Admittedly, the former approach additionally requires the computation and contraction of the $\{M\}$ tensors, which we have omitted from the cost-analysis above, but which is nevertheless far cheaper for sufficiently-local Lindbladian operators with small period $D$ (for $n=1$, this cost is only $Dd^4\chi^2$, while for $n=2$, it is $D(d^4\chi^3 + d^8\chi^2)$).

\subsection{Computation of log-derivative operator}

The other essential quantity in the computation of the metric tensor and variational forces is the log-derivative operator. Fortunately, an efficient computation of the operator is yet again afforded by the $\{L\}$ and $\{R\}$ matrix products. As the most general case, we consider a periodic MPO with period $D\leq N$ (possibly aperiodic for $D=N$) and assign an index $r$ with $1\leq r \leq D$ to the variational parameter $a^{rs}_{uv}$. Defining $\bar{j} :=\text{mod}{(j-1,D)+1}$, one obtains, 
\begin{align}
    \frac{\partial}{\partial a^{rs}_{uv}} \bra{\bm{x}}\ket{\rho} &= \frac{\partial}{\partial a^{rs}_{uv}} \sum_{\{\alpha_i\}} A_{\alpha_1\alpha_2}^{x_1} A_{\alpha_2\alpha_3}^{x_2}\dots A_{\alpha_N\alpha_1}^{x_N} \\
    &= \sum_{\{a_i\}} \delta_{r\bar{1}} \delta_{sx_1} \delta_{u\alpha_1} \delta_{v\alpha_2} A_{\alpha_2\alpha_3}^{x_2}\dots A_{\alpha_N\alpha_1}^{x_N} + A_{\alpha_1\alpha_2}^{x_1} \delta_{r\bar{2}} \delta_{sx_2} \delta_{u\alpha_2} \delta_{v\alpha_3} \dots A_{\alpha_N\alpha_1}^{x_N} + \dots\\
    &= \sum_{j=1}^N \delta_{r\bar{j}}  \delta_{sx_j} \left[ R_{j+1} L_{j-1} \right]_{vu}.
\end{align}
Hence,
\begin{equation}
    \Delta^{rs}_{uv}(\bm{x}) =  \bra{\bm{x}}\ket{\rho}^{-1} \sum_{j=1}^N \delta_{r\bar{j}} \delta_{sx_j} \left[ R_{j+1} L_{j-1} \right]_{vu}.
\end{equation}
The above has a worst-case computational cost of $N\chi^3$ (assuming $\bra{\bm{x}}\ket{\rho}$, $\{L\}$, and $\{R\}$ are known).

\subsection{Algorithm summary}

\begin{figure}[h]
    \begin{tikzpicture}
        % MPI worker 1 box
        \node[draw, rectangle, rounded corners, fill=lightgray!30, 
              text width=4.5cm, inner sep=8pt] at (0,-1) (box1)
            {
                \texttt{TensorComputeGradient!}:\\
                \textbf{for} $k=1:N_\text{MC}$ \textbf{do}\\
                \quad \texttt{MetropolisSweepLeft!}:\\
                \quad \quad draw sample $\bm{x}$\\
                \quad \quad compute $\{R_j\}$\\
                \quad \texttt{TensorUpdate!}:\\
                \quad \quad compute $\pazocal{L}_\text{loc}(\bm{x})$\\
                \quad \quad compute $\{L_j\}$\\
                \quad \quad compute $\{\Delta_i(\bm{x})\}$\\
                \quad \quad compute $\{\Delta^\dagger_i(\bm{x})\Delta_j(\bm{x})\}$\\
                \quad \quad update $\mathbbm{E}\left[\Delta^\dagger_i(\bm{x})\pazocal{L}_\text{loc}(\bm{x})\right]$\\
                \quad \quad update $\mathbbm{E}\left[\Delta^\dagger_i(\bm{x})\Delta_j(\bm{x})\right]$\\
                \textbf{end do}
            };

        % MPI worker 2 box
        \node[draw, rectangle, rounded corners, fill=lightgray!30, 
              text width=4.5cm, inner sep=8pt] at (5.5,-1) (box2)
            {
                \texttt{TensorComputeGradient!}:\\
                \textbf{for} $k=1:N_\text{MC}$ \textbf{do}\\
                \quad \texttt{MetropolisSweepLeft!}:\\
                \quad \quad draw sample $\bm{x}$\\
                \quad \quad compute $\{R_j\}$\\
                \quad \texttt{TensorUpdate!}:\\
                \quad \quad compute $\pazocal{L}_\text{loc}(\bm{x})$\\
                \quad \quad compute $\{L_j\}$\\
                \quad \quad compute $\{\Delta_i(\bm{x})\}$\\
                \quad \quad compute $\{\Delta^\dagger_i(\bm{x})\Delta_j(\bm{x})\}$\\
                \quad \quad update $\mathbbm{E}\left[\Delta^\dagger_i(\bm{x})\pazocal{L}_\text{loc}(\bm{x})\right]$\\
                \quad \quad update $\mathbbm{E}\left[\Delta^\dagger_i(\bm{x})\Delta_j(\bm{x})\right]$\\
                \textbf{end do}
            };
    
        % Optimize box
        \node[draw, rectangle, rounded corners, fill=lightgray!30, 
              text width=3cm, inner sep=8pt] at (0,-8) (optimize)
            {
                \texttt{Optimize!}:\\
                \quad compute $\{f_i\}$\\
                \quad compute $\{S_{ij}\}$\\
                \quad regularize $\{S_{ij}\}$\\
                \quad update $\bm{a}$\\
                \quad renormalize $\bm{a}$\\
                \texttt{MPI.Bcast}:\\
                \quad broadcast $\bm{a}$
            };
    
            \node at (9.5, -1)  {\scalebox{2}{\Huge ...}};

                % Optimize box
                \node[draw, rectangle, rounded corners, fill=lightgray!30, 
                text width=5.5cm, inner sep=8pt] at (0,6) (initialize)
              {
                \texttt{params = Parameters(\dots)}\\
                \texttt{initial$\textunderscore\,$mpo = MPO(\dots)}\\
                \texttt{sampler = MetropolisSampler(\dots)}\\
                \texttt{optimizer = TDVP(\dots)}
              };
    
        % Arrows
        \draw[thick, postaction={decorate}, 
        decoration={markings, mark=at position 0.5 with {\arrow[line width=1pt, scale=1.5]{>}}}] 
        (box1.south) -- (optimize.north);
        \draw[thick, postaction={decorate}, 
        decoration={markings, mark=at position 0.5 with {\arrow[line width=1pt, scale=1.5]{>}}}] 
        (box2.south) -- (optimize.north);
        \draw[thick, postaction={decorate}, 
        decoration={markings, mark=at position 0.5 with {\arrow[line width=1pt, scale=1.5]{>}}}] 
        ($(box2.south) + (4,0)$) -- (optimize.north);

        \draw[z=0, thick, postaction={decorate}, rounded corners=10,
        decoration={markings, mark=at position 0.5 with {\arrow[line width=1pt, scale=1.5]{>}}}] 
        (initialize.south) -- ($(box1.north) + (0,1)$) -- (box1.north);
        \draw[z=0, thick, postaction={decorate}, rounded corners=10,
        decoration={markings, mark=at position 0.55 with {\arrow[line width=1pt, scale=1.5]{>}}}] 
        (initialize.south) -- ($(box1.north) + (0,1)$) -- ($(box2.north) + (0,1)$) -- (box2.north);
        \draw[z=0, thick, postaction={decorate}, rounded corners=10,
        decoration={markings, mark=at position 0.825 with {\arrow[line width=1pt, scale=1.5]{>}}}] 
        (initialize.south) -- ($(box1.north) + (0,1)$) -- ($(box2.north) + (0,1)$) -- ($(box2.north) + (4,1)$) -- ($(box2.north) + (4,0)$);

        \draw[thick, postaction={decorate}, rounded corners=10,
        decoration={markings, mark=at position 0.5 with {\arrow[line width=1pt, scale=1.5]{>}}}] 
        (optimize.south) -- ($(optimize.south) + (0,-0.5)$) -- ($(optimize.south) + (-3,-0.5)$) -- ($(optimize.south) + (-3,14)$) -- ($(optimize.south) + (0,14)$) -- ($(optimize.south) + (0,13)$);

        \node[draw] at (-1.3, -5.2) {$\texttt{MPI\textunderscore \,mean!}$};

        \node[draw] at (-4.0, 6) {(a)};
        \node[draw] at (-4.0, -1) {(b)};
        \node[draw] at (-4.0, -8) {(c)};

        \node[draw, z=1, fill=white] at (0, 2.2) (worker1) {MPI worker 1};
        \node[draw, z=1, fill=white] at (5.5, 2.2) (worker2) {MPI worker 2};
        
    \end{tikzpicture}
    \caption{\textbf{Flowchart of the essential elements of a single iteration of the t-VMC+MPO algorithm with first-order forward-Euler integration, referencing the names of functions called in our Julia implementation.} After (a) specifying the parameter values, the Metropolis sampler, the initial MPO, and the optimizer, (b) the core part of the overall computation can be split into separate MPI processes to be reduced before (c) updating the tensor elements. The (b)-(c) loop is then repeated for a chosen number of time steps.}
    \label{fig: flowchart}
\end{figure}
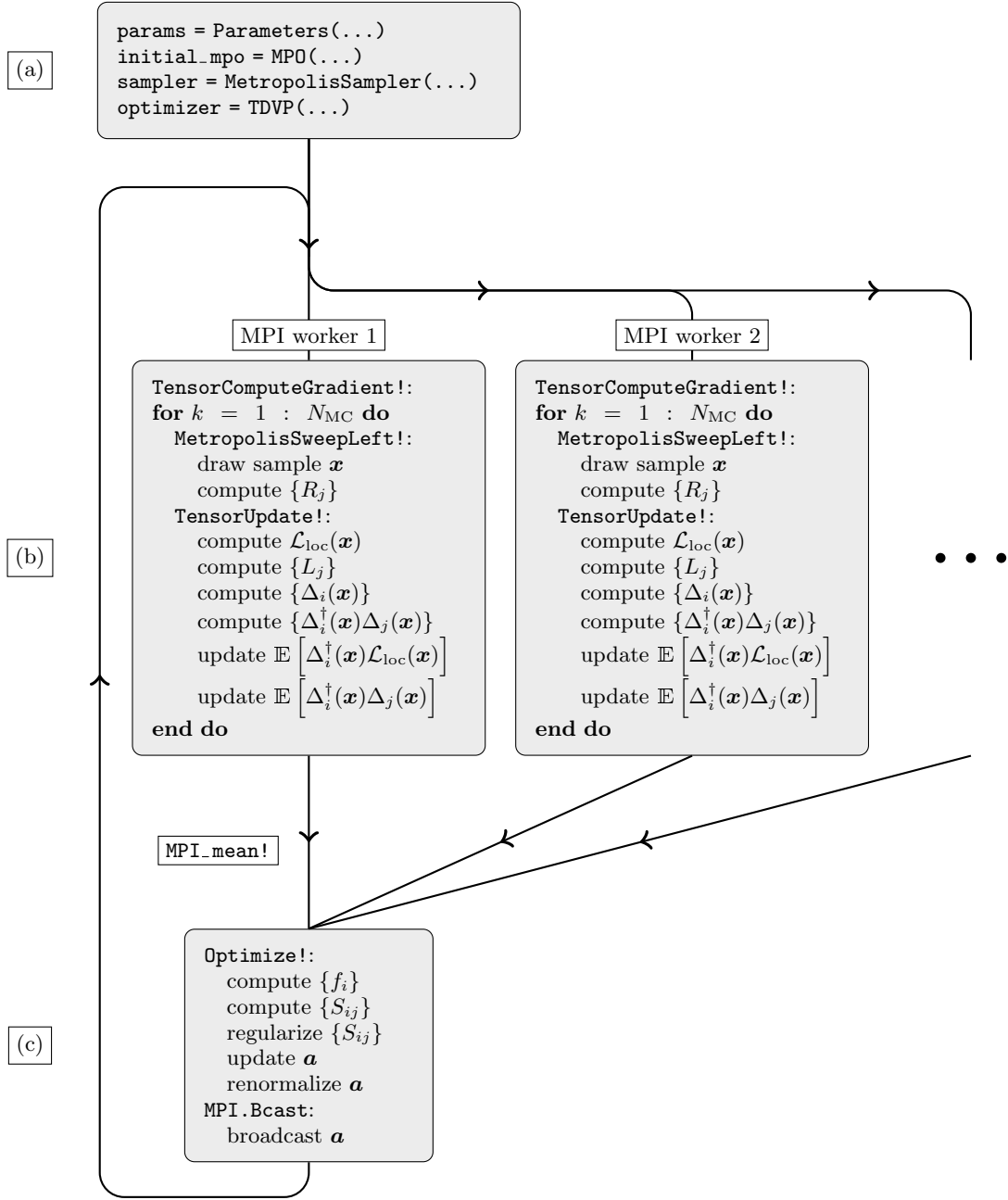

We summarize the key steps of our algorithm in Fig. \ref{fig: flowchart}, where we also sketch how to efficiently parallelize the computations. We regularize the metric tensor by a combination of a diagonal shift $\epsilon_\text{shift}$ and the $S$-regularization scheme of \cite{PhysRevLett.125.100503} with soft cutoff $\epsilon_\text{SNR}$. When integrating the equations of motion, we either use a fixed step-sized forward-Euler integrator or a second order adaptive Heun integrator of \cite{PhysRevLett.125.100503} with tolerance $\epsilon_\text{tol}$. After the tensor elements have been updated, they are renormalized to ensure $\tr \rho = 1$. 

For two-dimensional translationally-invariant latices, we wrap the MPO around all sites of the lattices row-by-row (with periodic boundary conditions imposed in both dimensions), defining the row as the unit-cell of the lattice, and mapping the two-dimensional problem onto a one-dimensional one, with the nearest-neighbor interactions between rows becoming effective long-ranged interactions. This is worth contrasting with other variational MPO approaches for two-dimensional lattices, where cylindrical boundary conditions are used \cite{doi:10.1146/annurev-conmatphys-020911-125018}.

\subsection{Computational cost}

Below we summarize the computational cost of a single iteration of the t-VMC+MPO algorithm for $N_\text{MC}$ Monte Carlo samples with a fixed step-size first-order Euler integration without regularization:
\begin{itemize}
    \item $2(N-1)\chi^3 N_\text{MC}$ due to the computation of the $\{L\}$ and $\{R\}$ matrix products.
    \item $2N\chi^3 N_\text{MC}$ due to single sequential Metropolis sweep.
    \item $2N\chi^3 N_\text{MC}$ due to contraction of the local estimator of a Lindbladian.
    \item $2D^2d^4\chi^4 N_\text{MC} + D^3d^6\chi^6$ due to computation and inversion of the metric tensor.
\end{itemize}
Therefore, the total computational cost for a Lindbladian with $N$ $n$-local terms has an asymptotic scaling of $O(N\chi^3N_\text{MC} + D^2d^4\chi^4 N_\text{MC} + D^3d^6\chi^6)$ (omitting an $n$- and $D$-dependent term for the computation and contraction of the $\{M\}$ tensors in the contraction of the local estimator)

\section[SN\thesection]{Additional convergence tests}

\subsection{Overview}

In this section we provide extensive convergence studies of our approach for six distinct models of dissipative spin chains:
\begin{enumerate}
    \item Transverse-field Ising model with short-ranged interactions.
    \item Transverse-field Ising model with weak long-ranged competing interactions.
    \item Transverse-field Ising model with strong long-ranged competing interactions.
    \item Transverse-field Ising model with super-long-ranged competing interactions.
    \item XYZ model with short-ranged interactions.
    \item XYZ model with long-ranged competing interactions.
\end{enumerate}
In each model, the incoherent dissipation is induced by a one-local spin-loss Lindblad jump operators $\Gamma_k = \frac{\sqrt{\gamma}}{2}(\sigma_k^x - i\sigma_k^y)$ for $k=1,\dots,N$ with $\gamma=1$ throughout, and all simulations start from the same initial condition $\langle \sigma^y \rangle=-1$. In addition, each model is considered for short ($N=10$) and long ($N=50$ or $N=200$) chains, with the former being also treatable via exact integration of the Lindblad master equation, permitting direct comparison with exact dynamics. The exact dynamics for short chains are obtained with the help of the \textit{QuantumOptics.jl} package \cite{KRAMER2018109}. For each variant of each model, we test the convergence of our algorithm with respect to five key hyperparameters:
\begin{enumerate}
    \item MPO density operator bond dimension $\chi$.
    \item Number of Monte Carlo samples per iteration $N_\text{MC}$.
    \item Second-order adaptive step-sizing tolerance parameter $\epsilon_{\text{tol}}$.
    \item Diagonal shift metric regularization parameter $\epsilon_{\text{shift}}$.
    \item Signal-to-noise metric regularization parameter $\epsilon_{\text{SNR}}$.
\end{enumerate}
For every considered hyperparameter values, we plot the non-equilibrium dynamics of the polarizations and connected 2-point correlation functions at 1, 2, and 3-site separation in each spatial direction, for a total of 12 distinct plots. In addition, for short chains, we also provide:
\begin{enumerate}
    \item Complementary plots of the absolute errors over exact results, for another 12 distinct plots.
    \item The evolution of the Rényi-2 entropy for the full density operator.
    \item The smallest eigenvalue of the variational density matrix.
    \item Schmidt singular values for a single, representative simulation.
    \item A time-analysis for each simulation, where we plot the total number of iterations for each simulation and the average time per iteration (with error bars representing the standard deviation in iteration time over all iterations in each simulation).
\end{enumerate}
In grand total, we provide 60 distinct figures below.

Each simulation was run for a maximum of 48 hours on 10 MPI processes each. As such, some of the most demanding calculations, particularly with very small adaptive step-size tolerances, did not complete within this timeframe.

\subsection{Hamiltonians and parameters}

The two Hamiltonians considered in this convergence analysis are:
\begin{enumerate}
    \item Transverse-field Ising model: \\
    \begin{equation}
        H = \sum_{n}  J_n \sum_{i,j} d_{ij}^{-\alpha_n} \sigma^z_i \sigma^z_j + h\sum_i \sigma^x_i,
        \label{eq: competing interactions Hamiltonian}
    \end{equation}\\
    For Figs. 2-11 (short-range), $J=0.5$, $\alpha=\infty$, $h=1$.\\ For Figs. 12-21 (weak long-range), $J_1=0.5$, $\alpha_1=3$, $J_2=-1.0$, $\alpha_2=6$, $h=1$.\\ For Figs. 22-31 (strong long-range), $J_1=1.0$, $\alpha_1=3$, $J_2=-2.0$, $\alpha_2=6$, $h=1$.\\ For Figs. 12-21 (super-long-range), $J_1=0.5$, $\alpha_1=2$, $J_2=-1.0$, $\alpha_2=4$, $h=1$.
    \item XYZ model: \\
    \begin{equation}
        H = \sum_{n}  \sum_{\beta = \{x,y,z\}} J_n^{\beta} \sum_{i,j} d_{ij}^{-\alpha_n} \sigma^{\beta}_i \sigma^{\beta}_j + h\sum_i \sigma^z_i.
        \label{eq: long-range XYZ}
    \end{equation}\\
    For Figs. 42-51 (short-range), $\bm{J}=(0.6, -0.5, 0.4)$, $\alpha=\infty$, $h=0.5$.\\
    For Figs. 52-61 (long-range), $\bm{J}_1=(0.6, -0.5, 0.4)$, $\alpha_1=3$, $\bm{J}_2=(-0.9, 1.0, -1.1)$, $\alpha_2=6$, $h=0.5$.
\end{enumerate}
The interactions strengths are not renormalized.

\subsection{Auxiliary hyperparameters}

Besides the five key hyperparameters tested in this section, our results rely of a number of other minor hyperparameters. Those include:
\begin{enumerate}
    \item Number of Monte Carlo sweeps in between samples: set to 5. A single sweep was found sufficient to give rise to virtually uncorrelated samples \cite{Hryniuk2024tensornetworkbased}, but additional sweeps can be performed at very little added computational cost.
    \item Number of burn-in Monte Carlo sweeps: set to 5.
    \item Initial step size during adaptive step-sizing evolution: set to $10^{-8}$. Must be initially set to a small number; it is increased to an appropriate size automatically.
    \item Maximum allowed step size: set to $0.1$. Improves resolution of the trajectories at larger tolerances.
\end{enumerate}

\subsection{Insights and recommendations}

\begin{enumerate}
    \item Convergence is improved for larger bond dimensions, larger number of samples, and smaller tolerance, although all come at a cost of more expensive computations.
    \item Convergence is optimal for regularization parameters at around $\epsilon=10^{-8}$ to $\epsilon=10^{-10}$.
    \item A larger interaction strength requires a larger bond dimension, number of samples, or step-size tolerance for convergence.
    \item Longer-ranged interactions require a larger bond dimension, number of samples, or step-size tolerance for convergence.
    \item Convergence does not seem to be adversely affected when the size of the system is increased, but becomes more expensive. 
    \item Smallest eigenvalue of the reduced density matrix for $N=10$ is strictly non-negative in all simulations for $N=10$, demonstrating that positive semi-definiteness of the ansatz is maintained throughout every simulation.
    \item Step size decreases with decreasing $\epsilon_{\text{shift}}$, but doesn't change with $\epsilon_{\text{SNR}}$.
\end{enumerate}

On the basis of the above observation, we recommend the following baseline hyperparameter values for the types of models studied in this work, which are to be refined for the specific problem at hand and desired accuracy:
\begin{enumerate}
    \item $N_\text{MC} = 5000$
    \item $\epsilon_\text{tol} = 0.01$
    \item $\epsilon_\text{shift} = 10^{-8}$ (may be less if $N_\text{MC}$ is sufficiently large and vice versa)
    \item $\epsilon_\text{SNR} = 10^{-8}$ (may be less if $N_\text{MC}$ is sufficiently large and vice versa)
\end{enumerate}
The appropriate bond dimension must be determined empirically, depending on the desired accuracy.

\subsection{Panel descriptions, $N=10$}
Panels (1a)-(6d): variational dynamics of magnetization and pair correlation functions obtained with t-VMC+MPO and compared with exact results; Columns (a) -- magnetizations $\{\langle \sigma^\beta \rangle(t)\}_{\beta=\{x,y,z\}}$, (b) - nearest-neighbour correlation functions $\{[\langle \sigma_i^\beta \sigma_{i+1}^\beta \rangle - \langle \sigma_i^\beta \rangle\langle \sigma_{i+1}^\beta \rangle](t)\}_{\beta=\{x,y,z\}}$, (c) -- next-nearest-neighbour correlation functions $\{[\langle \sigma_i^\beta \sigma_{i+2}^\beta \rangle - \langle \sigma_i^\beta \rangle\langle \sigma_{i+2}^\beta \rangle](t)\}_{\beta=\{x,y,z\}}$, (d) -- next-next-nearest-neighbour correlation functions $\{[\langle \sigma_i^\beta \sigma_{i+3}^\beta \rangle - \langle \sigma_i^\beta \rangle\langle \sigma_{i+3}^\beta \rangle](t)\}_{\beta=\{x,y,z\}}$; Rows (1) -- above referenced magnetizations and correlation functions for $\beta=x$, (3) -- $\beta=y$, (5) -- $\beta=z$; Rows (2) -- log plots of the absolute values of the difference of the variational and exact dynamics from row (1) above, (4) -- as in row (2) but for results in row (3), (6) -- as in row (2) but for results in row (5).

Panels (7a)-(7d): (7a) -- variational dynamics for the Rényi-2 entropy $S_2(t)$, (7b) -- value over time of the smallest eigenvalue of the variational density matrix, (7c) -- Schmidt singular values over time for a single simulation, (7d) -- total number of t-VMC+MPO iterations recorded during respective simulations (red) and average time in seconds for a single iteration with standard deviation shown as error bars (blue).

For parameter and hyperparameter values used, see Tables S2 or S3 at the end of this document.

\subsection{Panel descriptions, $N=50$ and $N=200$}
Variational dynamics of magnetization and pair correlation functions obtained with t-VMC+MPO; Columns (a) -- magnetizations $\{\langle \sigma^\beta \rangle(t)\}_{\beta=\{x,y,z\}}$, (b) - nearest-neighbour correlation functions $\{[\langle \sigma_i^\beta \sigma_{i+1}^\beta \rangle - \langle \sigma_i^\beta \rangle\langle \sigma_{i+1}^\beta \rangle](t)\}_{\beta=\{x,y,z\}}$, (c) -- next-nearest-neighbour correlation functions $\{[\langle \sigma_i^\beta \sigma_{i+2}^\beta \rangle - \langle \sigma_i^\beta \rangle\langle \sigma_{i+2}^\beta \rangle](t)\}_{\beta=\{x,y,z\}}$, (d) -- next-next-nearest-neighbour correlation functions $\{[\langle \sigma_i^\beta \sigma_{i+3}^\beta \rangle - \langle \sigma_i^\beta \rangle\langle \sigma_{i+3}^\beta \rangle](t)\}_{\beta=\{x,y,z\}}$; Rows (1) -- above referenced magnetizations and correlation functions for $\beta=x$, (2) -- $\beta=y$, (3) -- $\beta=z$.

For parameter and hyperparameter values used, see Tables S2 or S3 at the end of this document.

\newpage
\subsection{Transverse-field Ising model with short-range interactions, $N=10$}

\begin{figure}[H]
    \includegraphics[width=1.0\textwidth]{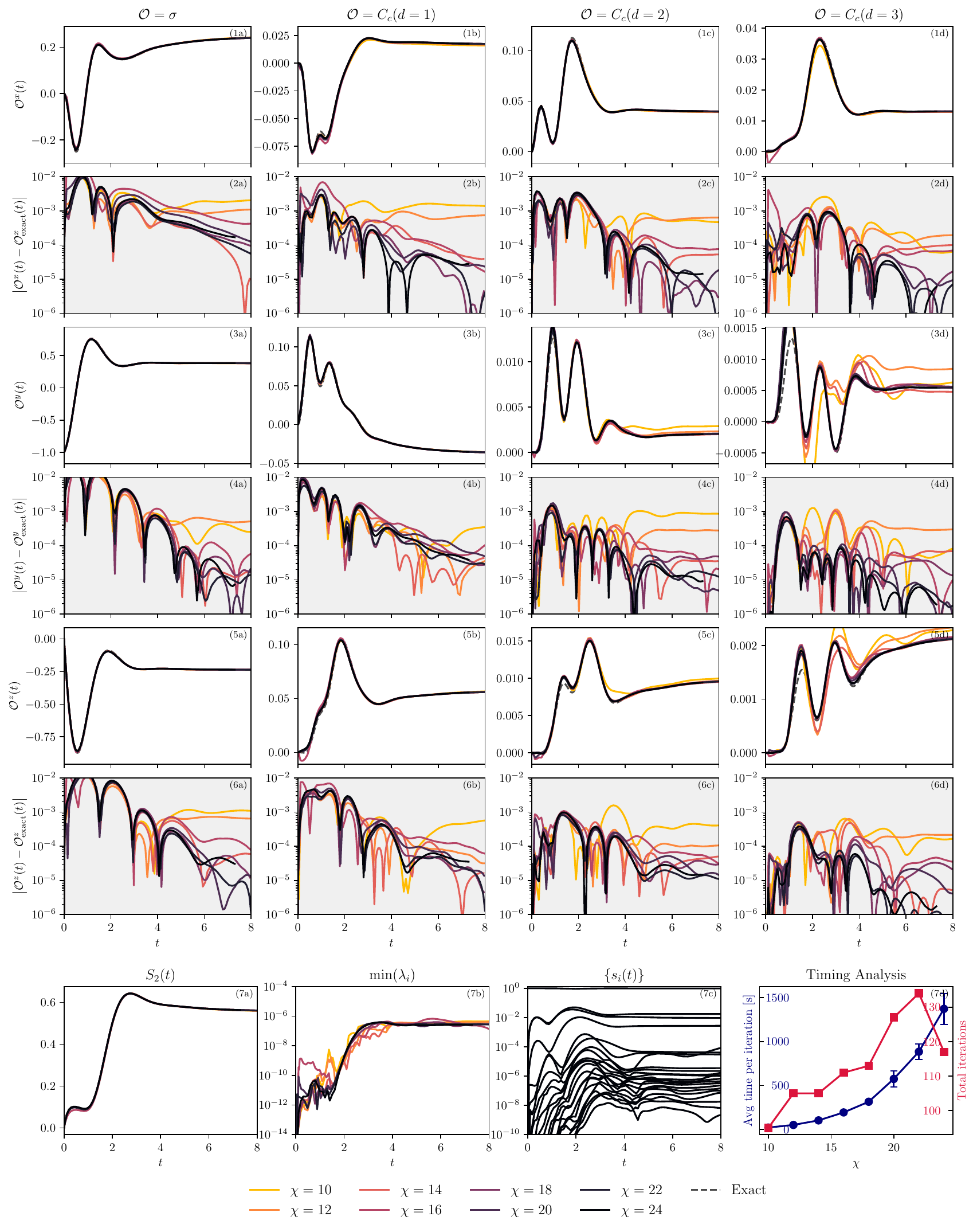}
    \caption{\textbf{Convergence with $\chi$. Transverse-field Ising model with short-range interactions interactions, $N=10$.} For panel descriptions, see Supplementary Note 2 E.}
    \label{fig: I_N10_chi}
\end{figure}

\begin{figure}[H]
    \includegraphics[width=1.0\textwidth]{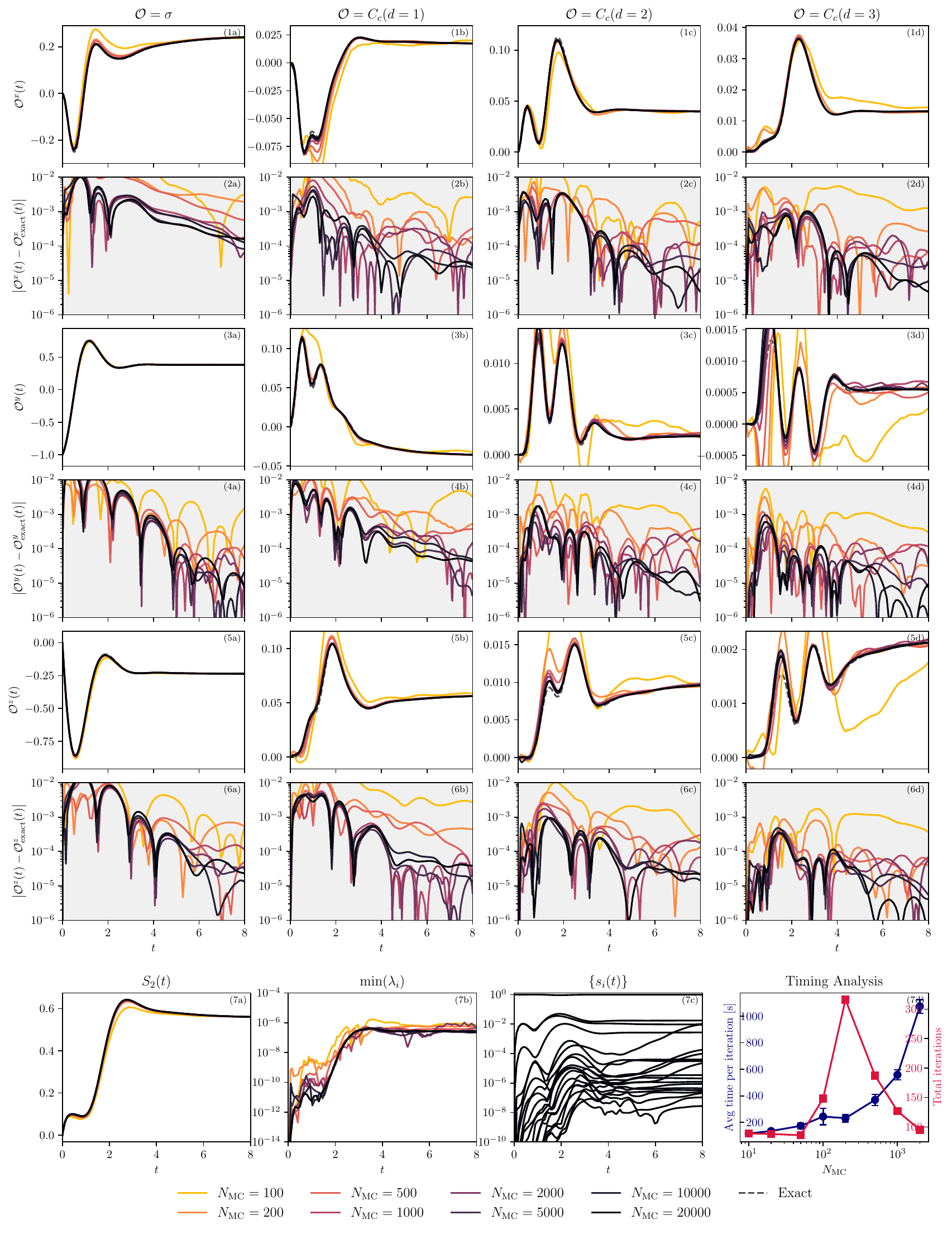}
    \caption{\textbf{Convergence with $N_\text{MC}$. Transverse-field Ising model with short-range interactions, $N=10$.} For panel descriptions, see Supplementary Note 2 E.}
    \label{fig: I_N10_samples}
\end{figure}

\begin{figure}[H]
    \includegraphics[width=1.0\textwidth]{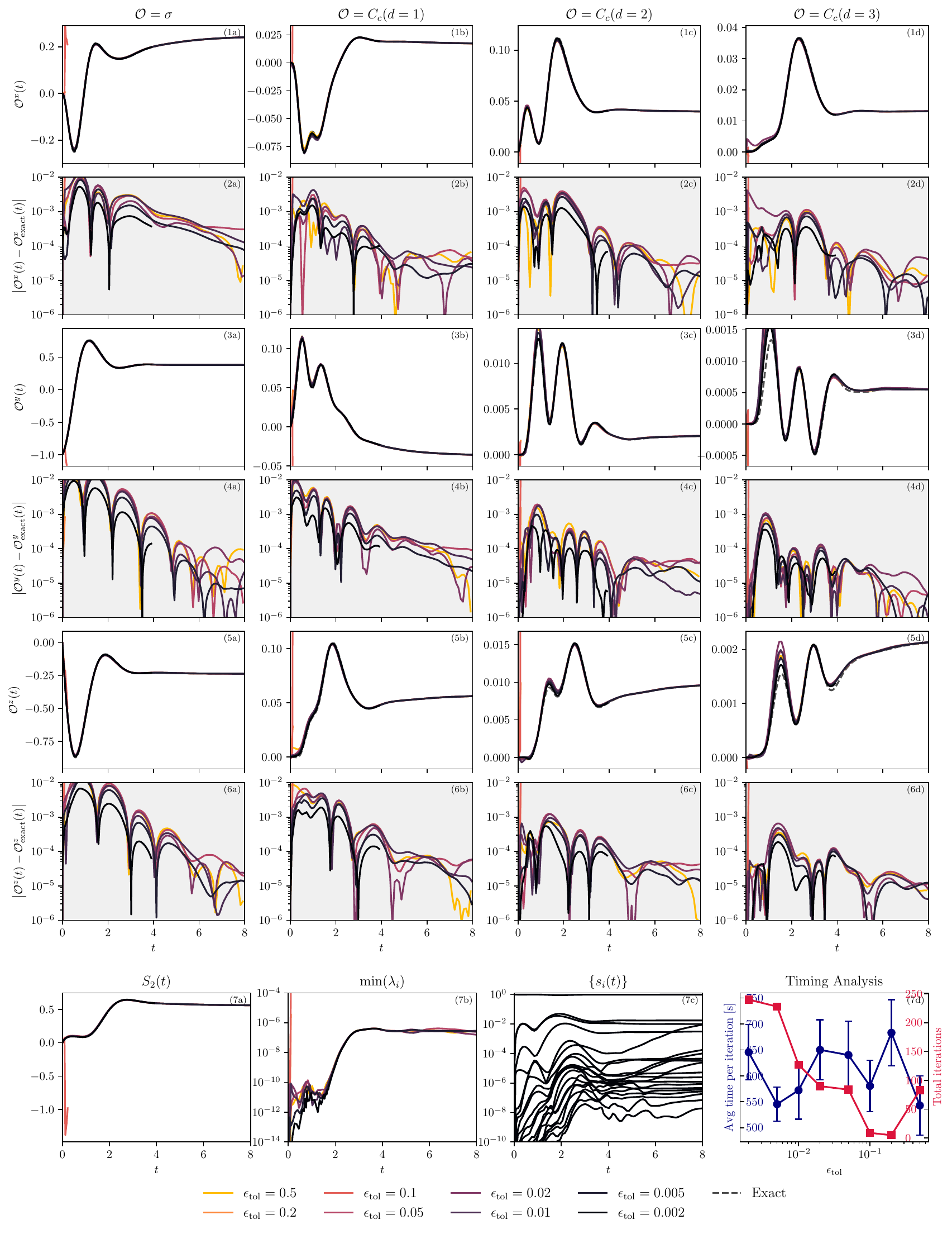}
    \caption{\textbf{Convergence with $\epsilon_\text{tol}$. Transverse-field Ising model with short-range interactions, $N=10$.} For panel descriptions, see Supplementary Note 2 E.}
    \label{fig: I_N10_eps_tol}
\end{figure}

\begin{figure}[H]
    \includegraphics[width=1.0\textwidth]{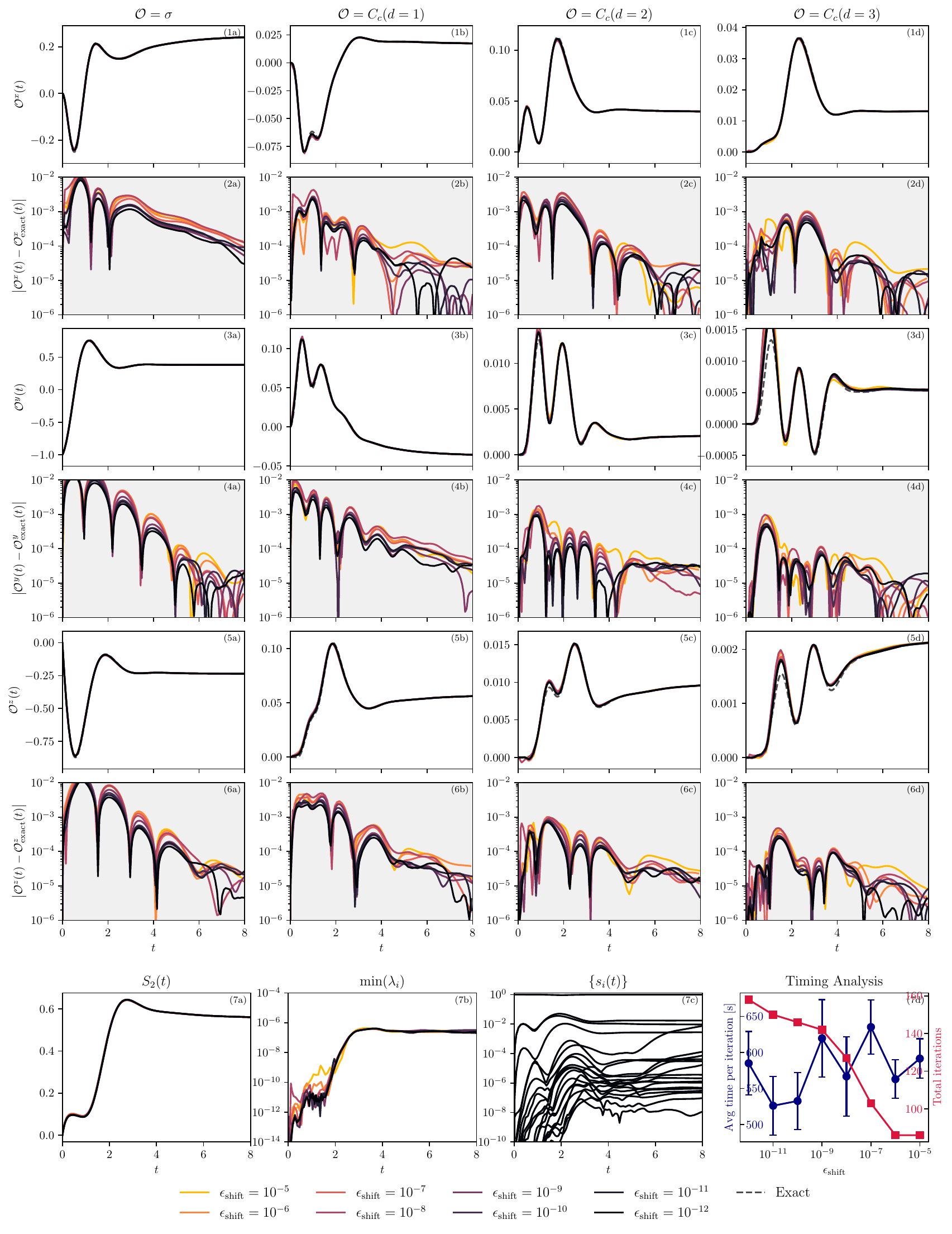}
    \caption{\textbf{Convergence with $\epsilon_\text{shift}$. Transverse-field Ising model with short-range interactions, $N=10$.} For panel descriptions, see Supplementary Note 2 E.}
    \label{fig: I_N10_eps_shift}
\end{figure}

\begin{figure}[H]
    \includegraphics[width=1.0\textwidth]{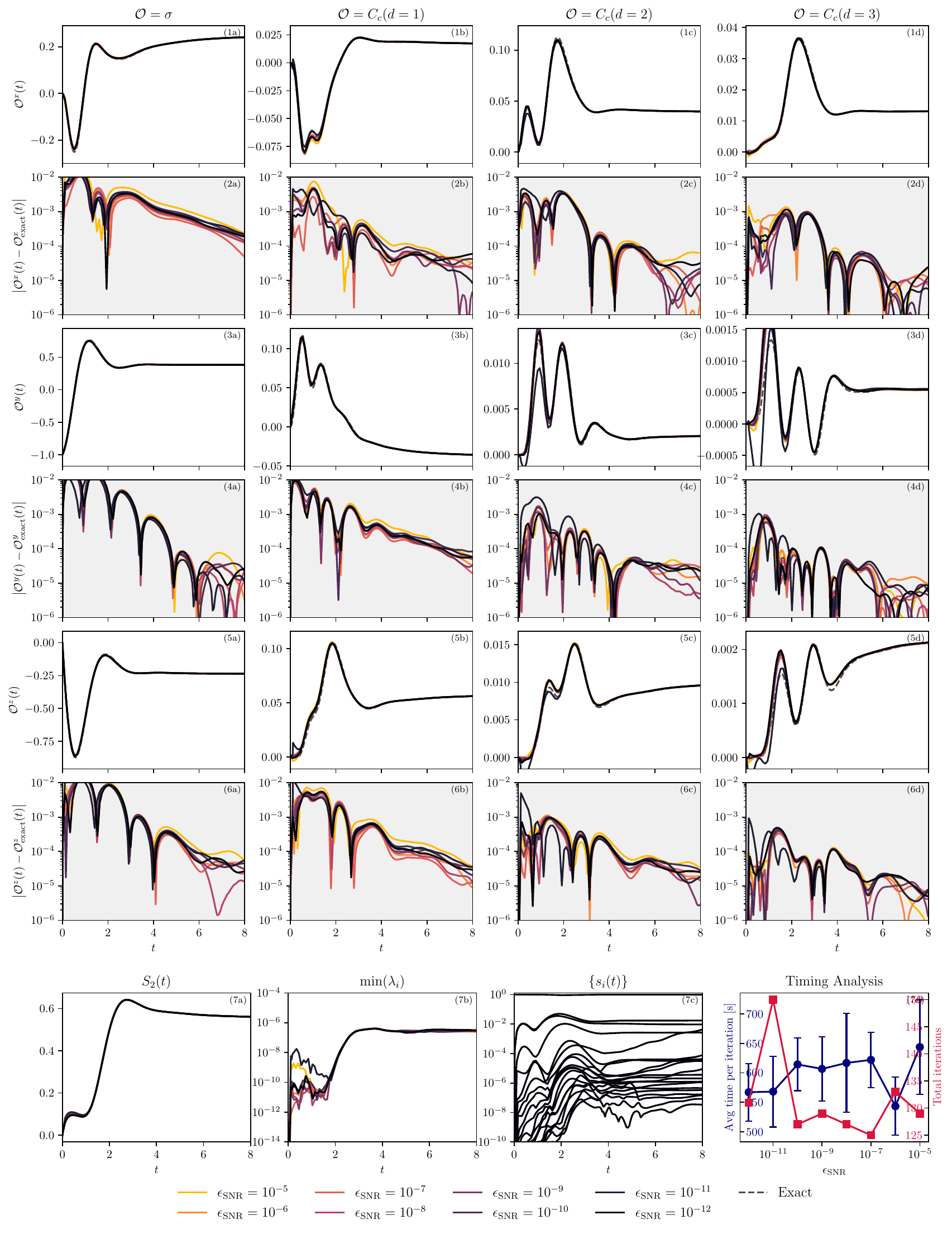}
    \caption{\textbf{Convergence with $\epsilon_\text{SNR}$. Transverse-field Ising model with short-range interactions, $N=10$.} For panel descriptions, see Supplementary Note 2 E.}
    \label{fig: I_N10_eps_snr}
\end{figure}

\newpage
\subsection{Transverse-field Ising model with short-range interactions, $N=200$}

\begin{figure}[H]
    \includegraphics[width=1.0\textwidth]{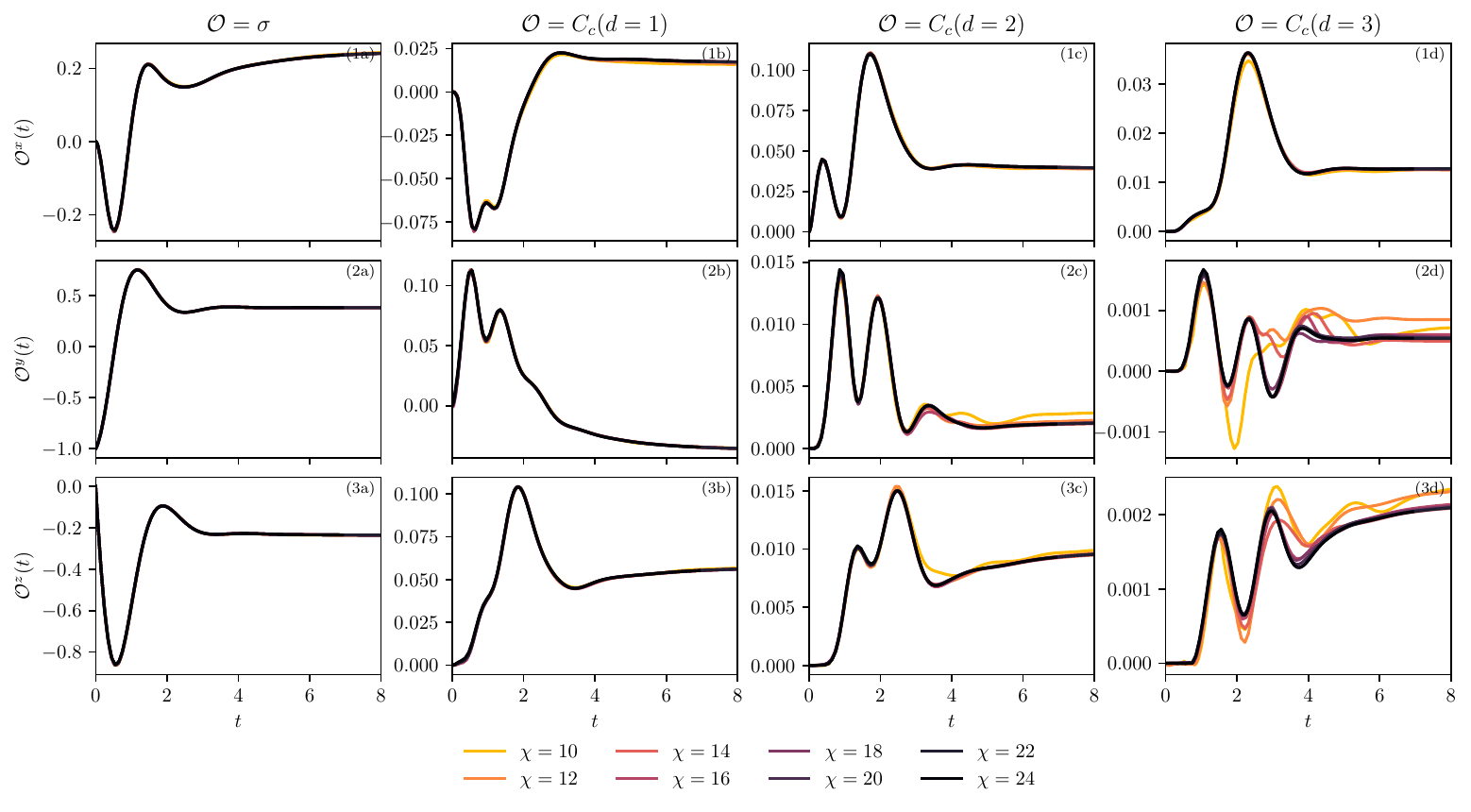}
    \caption{\textbf{Convergence with $\chi$. Transverse-field Ising model with short-range interactions interactions, $N=200$.} For panel descriptions, see Supplementary Note 2 F.}
    \label{fig: I_N200_chi}
\end{figure}

\begin{figure}[H]
    \includegraphics[width=1.0\textwidth]{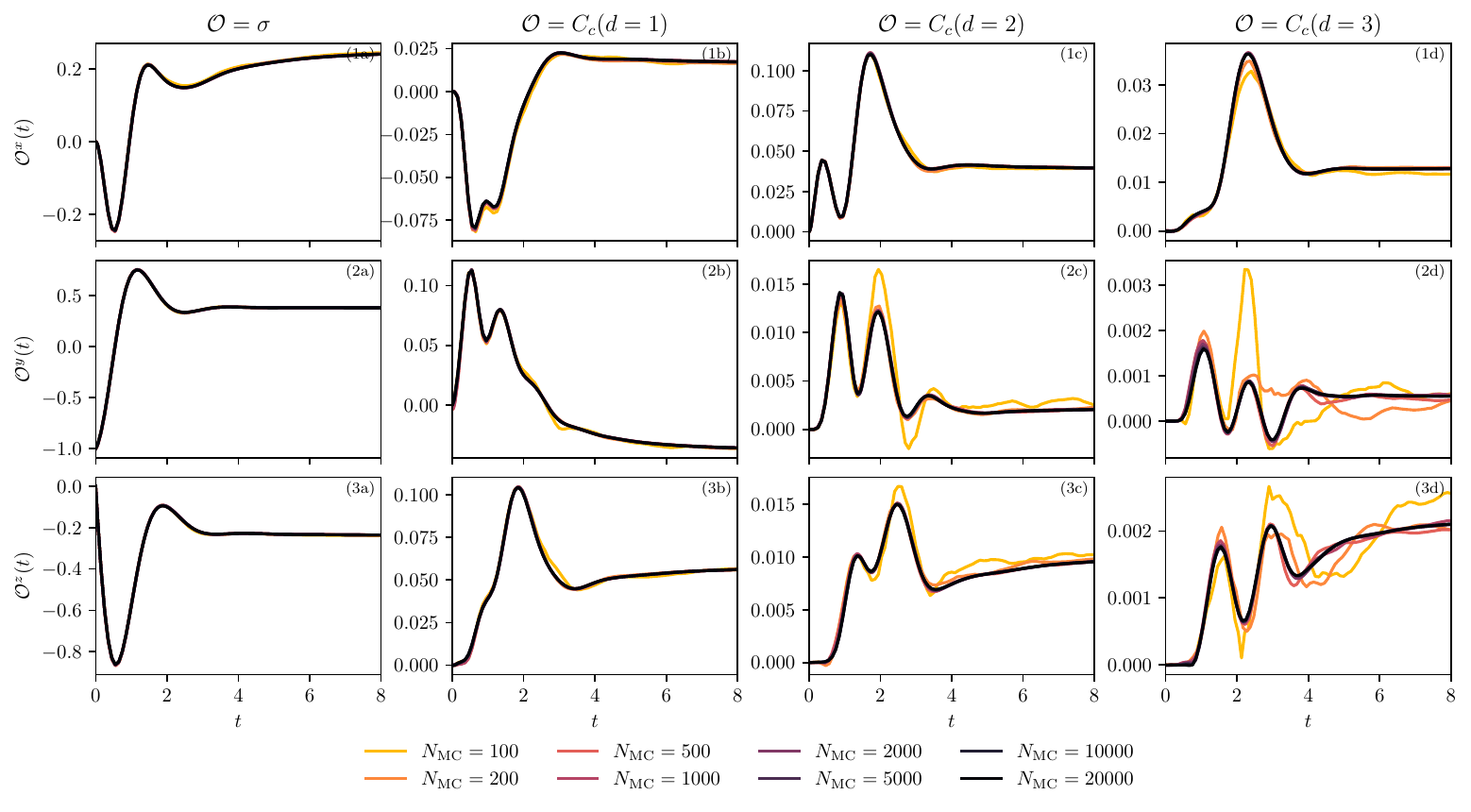}
    \caption{\textbf{Convergence with $N_\text{MC}$. Transverse-field Ising model with short-range interactions, $N=200$.} For panel descriptions, see Supplementary Note 2 F.}
    \label{fig: I_N200_samples}
\end{figure}

\begin{figure}[H]
    \includegraphics[width=1.0\textwidth]{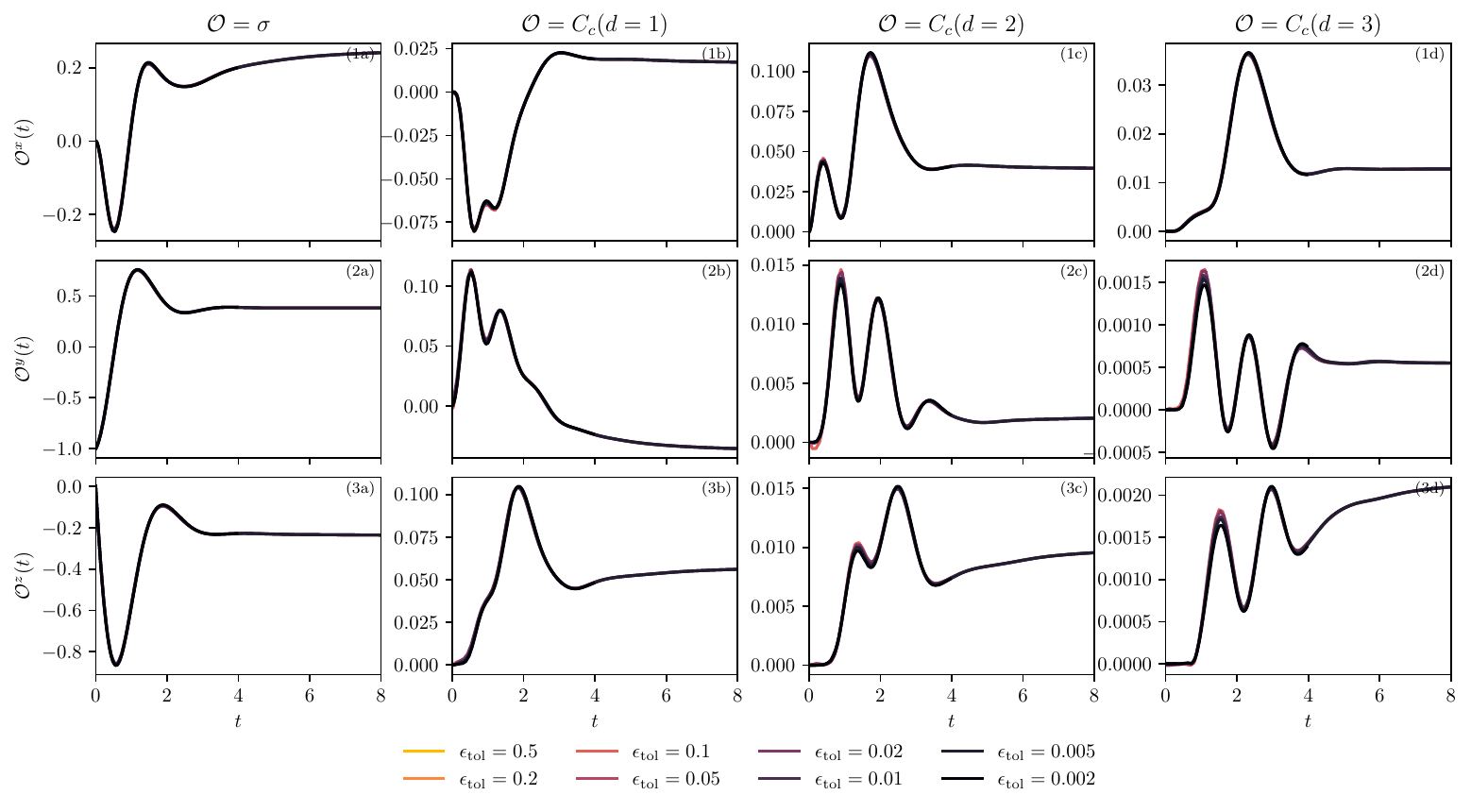}
    \caption{\textbf{Convergence with $\epsilon_\text{tol}$. Transverse-field Ising model with short-range interactions, $N=200$.} For panel descriptions, see Supplementary Note 2 F.}
    \label{fig: I_N200_eps_tol}
\end{figure}

\begin{figure}[H]
    \includegraphics[width=1.0\textwidth]{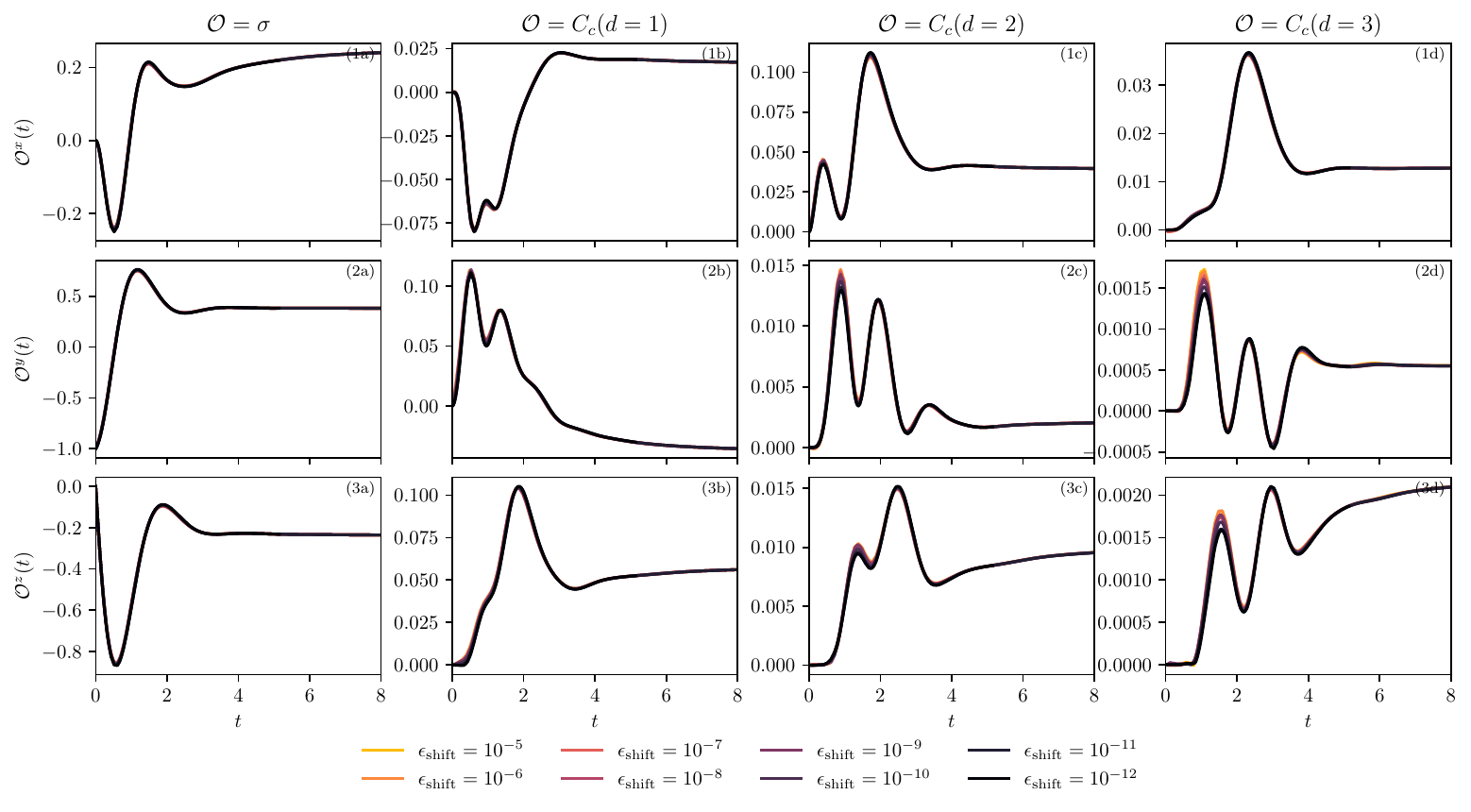}
    \caption{\textbf{Convergence with $\epsilon_\text{shift}$. Transverse-field Ising model with short-range interactions, $N=200$.} For panel descriptions, see Supplementary Note 2 F.}
    \label{fig: I_N200_eps_shift}
\end{figure}

\begin{figure}[H]
    \includegraphics[width=1.0\textwidth]{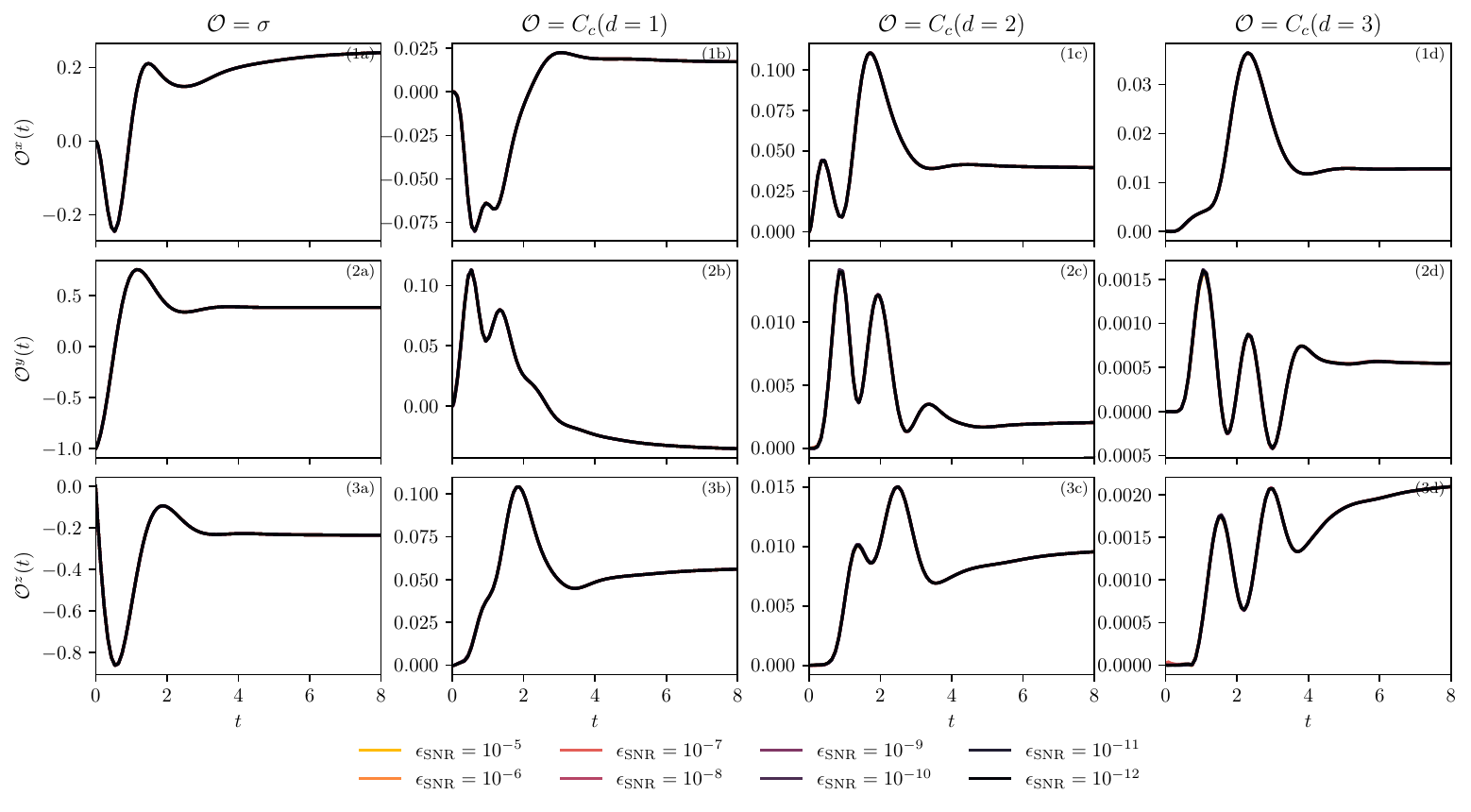}
    \caption{\textbf{Convergence with $\epsilon_\text{SNR}$. Transverse-field Ising model with short-range interactions, $N=200$.} For panel descriptions, see Supplementary Note 2 F.}
    \label{fig: I_N200_eps_snr}
\end{figure}

\newpage
\subsection{Transverse-field Ising model with weak long-range competing interactions, $N=10$}

\begin{figure}[H]
    \includegraphics[width=1.0\textwidth]{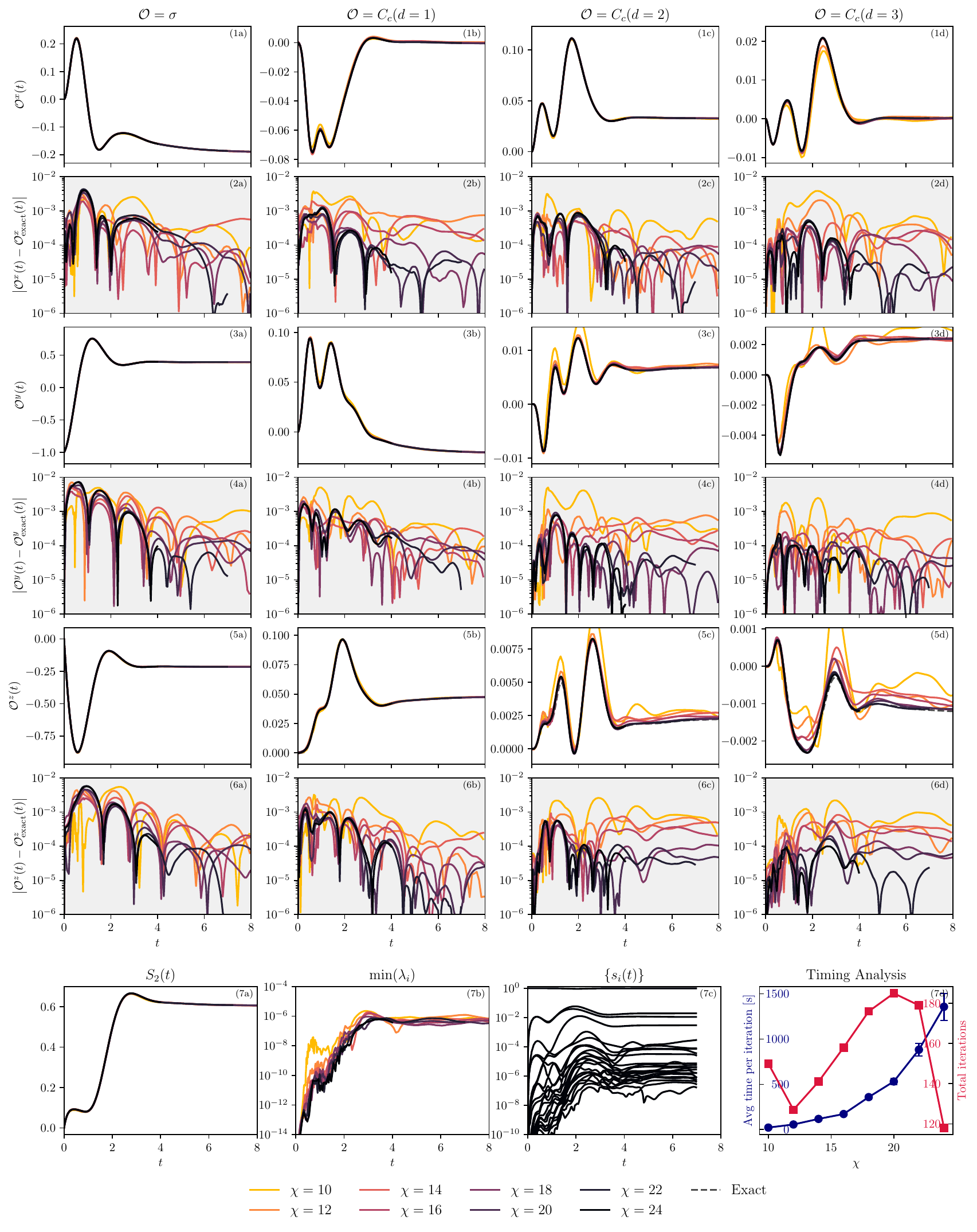}
    \caption{\textbf{Convergence with $\chi$. Transverse-field Ising model with weak long-range competing interactions, $N=10$.} For panel descriptions, see Supplementary Note 2 E.}
    \label{fig: CI_N10_chi}
\end{figure}

\begin{figure}[H]
    \includegraphics[width=1.0\textwidth]{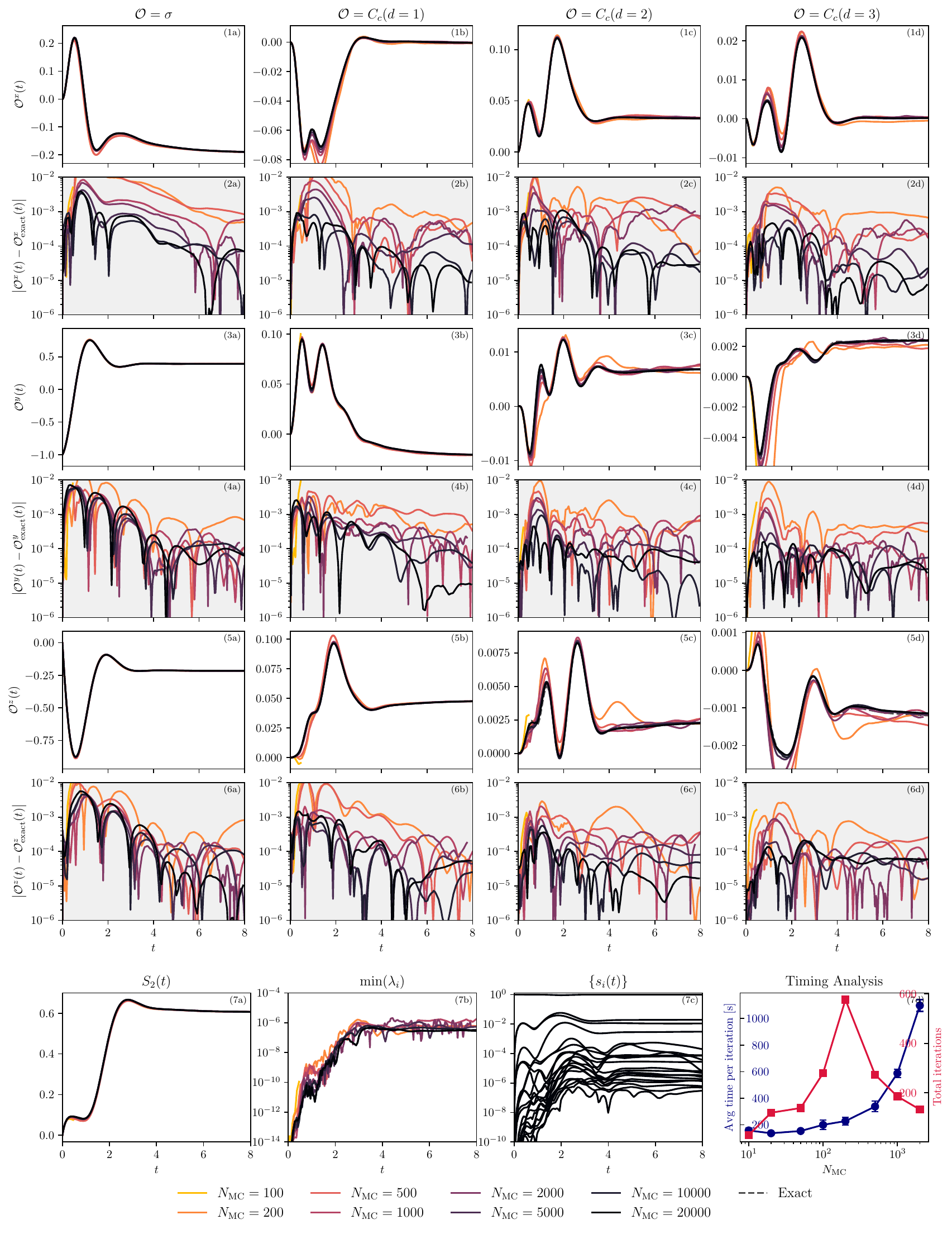}
    \caption{\textbf{Convergence with $N_\text{MC}$. Transverse-field Ising model with weak long-range competing interactions, $N=10$.} For panel descriptions, see Supplementary Note 2 E.}
    \label{fig: CI_N10_samples}
\end{figure}

\begin{figure}[H]
    \includegraphics[width=1.0\textwidth]{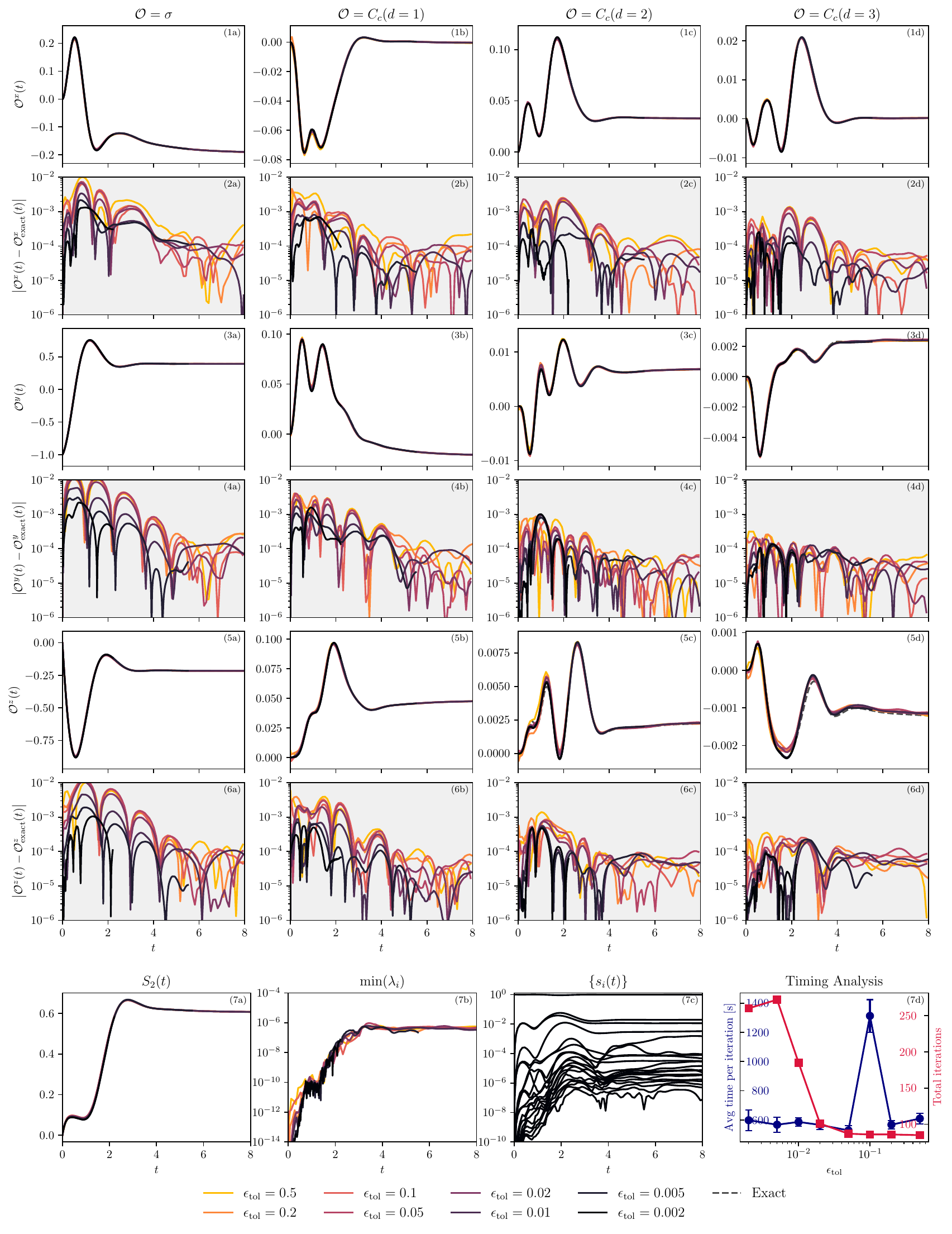}
    \caption{\textbf{Convergence with $\epsilon_\text{tol}$. Transverse-field Ising model with weak long-range competing interactions, $N=10$.} For panel descriptions, see Supplementary Note 2 E.}
    \label{fig: CI_N10_eps_tol}
\end{figure}

\begin{figure}[H]
    \includegraphics[width=1.0\textwidth]{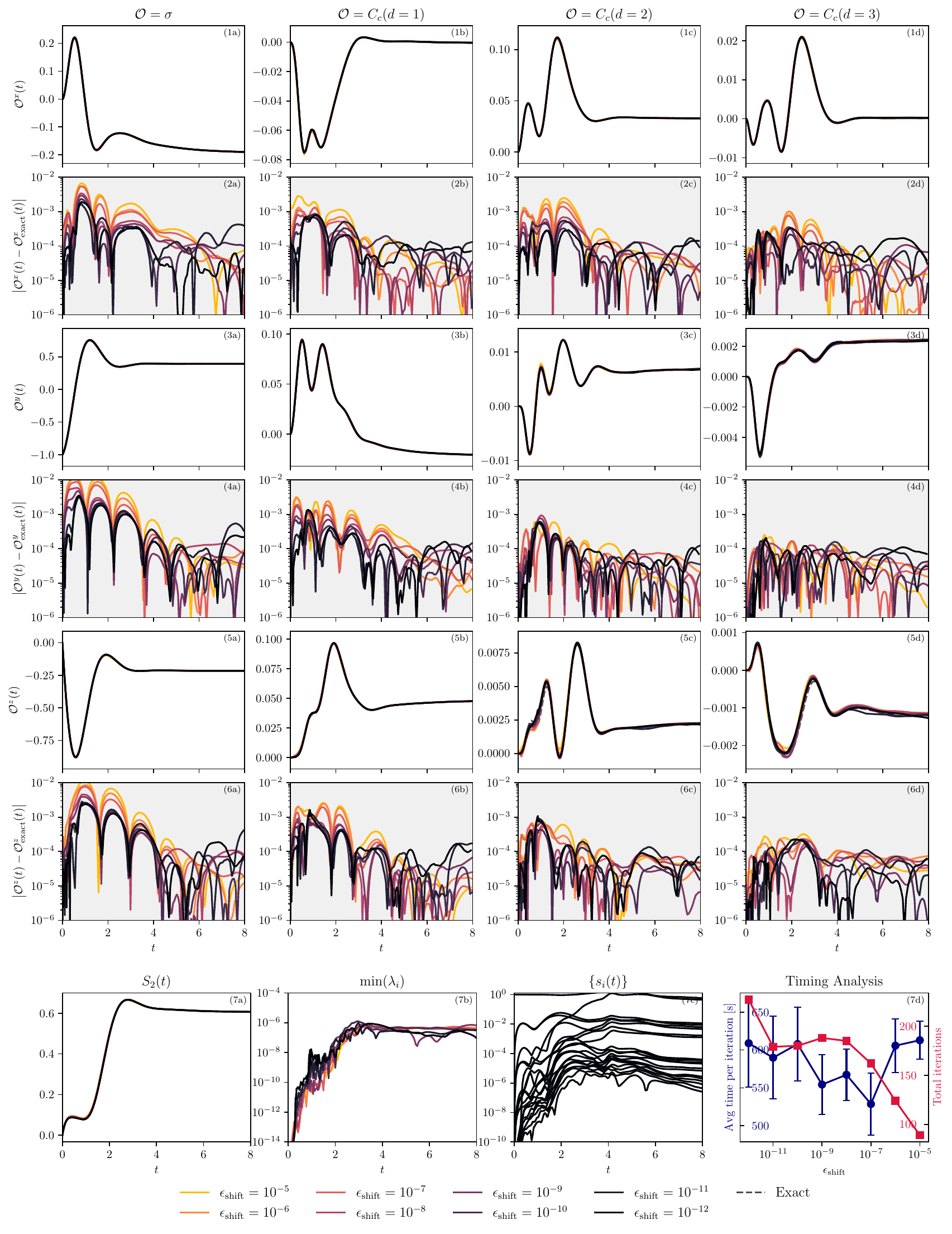}
    \caption{\textbf{Convergence with $\epsilon_\text{shift}$. Transverse-field Ising model with weak long-range competing interactions, $N=10$.} For panel descriptions, see Supplementary Note 2 E.}
    \label{fig: CI_N10_eps_shift}
\end{figure}

\begin{figure}[H]
    \includegraphics[width=1.0\textwidth]{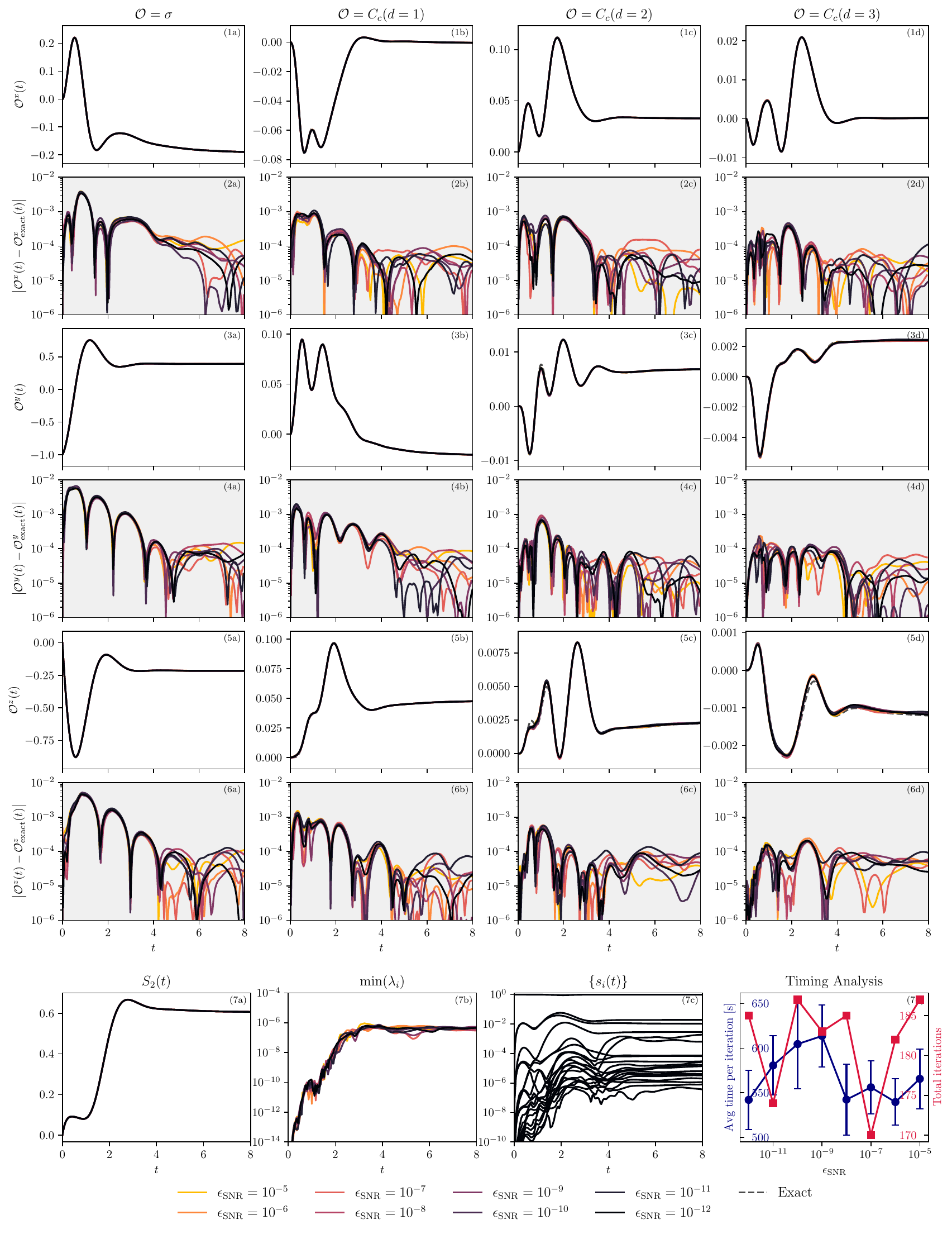}
    \caption{\textbf{Convergence with $\epsilon_\text{SNR}$. Transverse-field Ising model with weak long-range competing interactions, $N=10$.} For panel descriptions, see Supplementary Note 2 E.}
    \label{fig: CI_N10_eps_snr}
\end{figure}

\newpage
\subsection{Transverse-field Ising model with weak long-range competing interactions, $N=200$}

\begin{figure}[H]
    \includegraphics[width=1.0\textwidth]{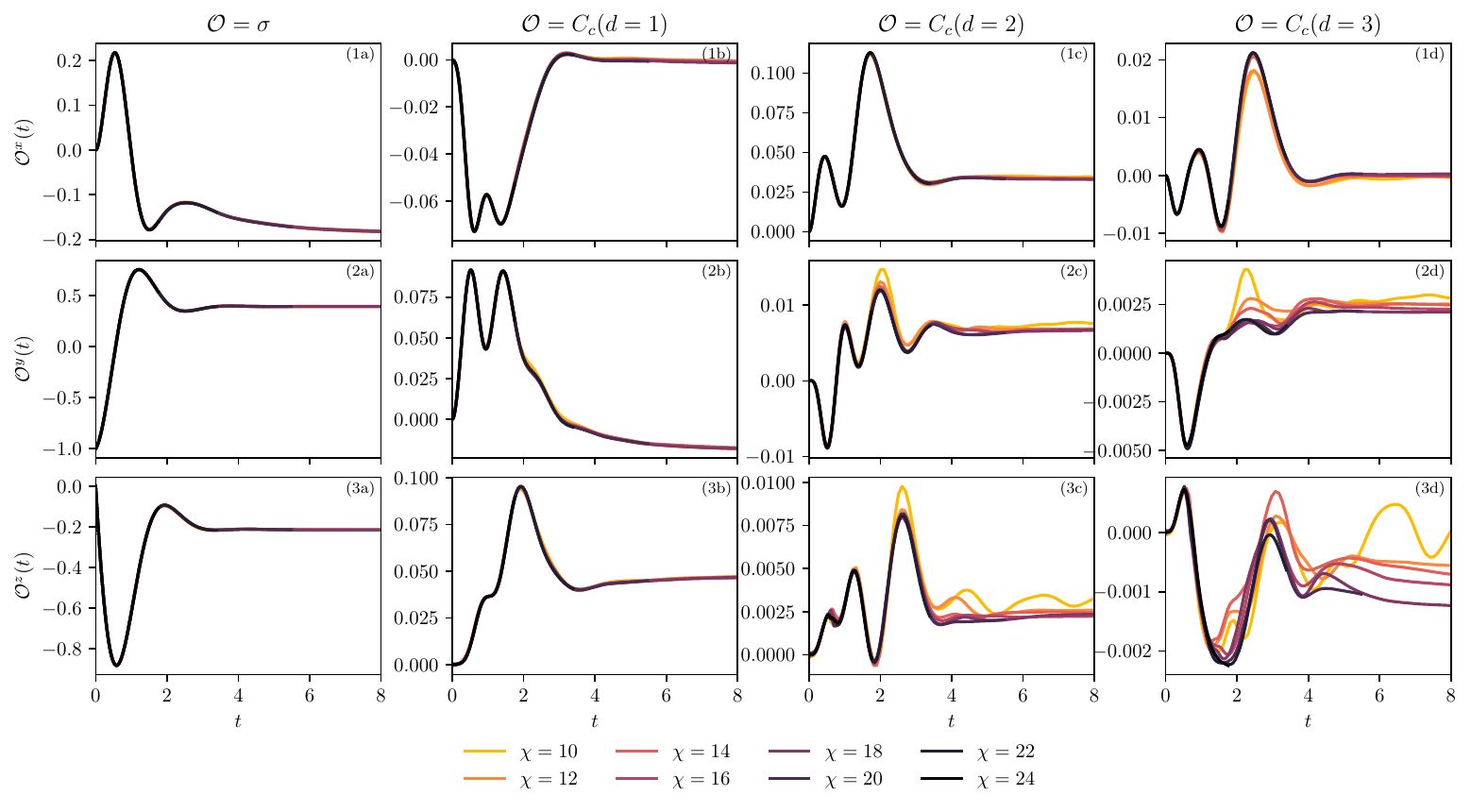}
    \caption{\textbf{Convergence with $\chi$. Transverse-field Ising model with weak long-range competing interactions, $N=200$.} For panel descriptions, see Supplementary Note 2 F.}
    \label{fig: CI_N200_chi}
\end{figure}

\begin{figure}[H]
    \includegraphics[width=1.0\textwidth]{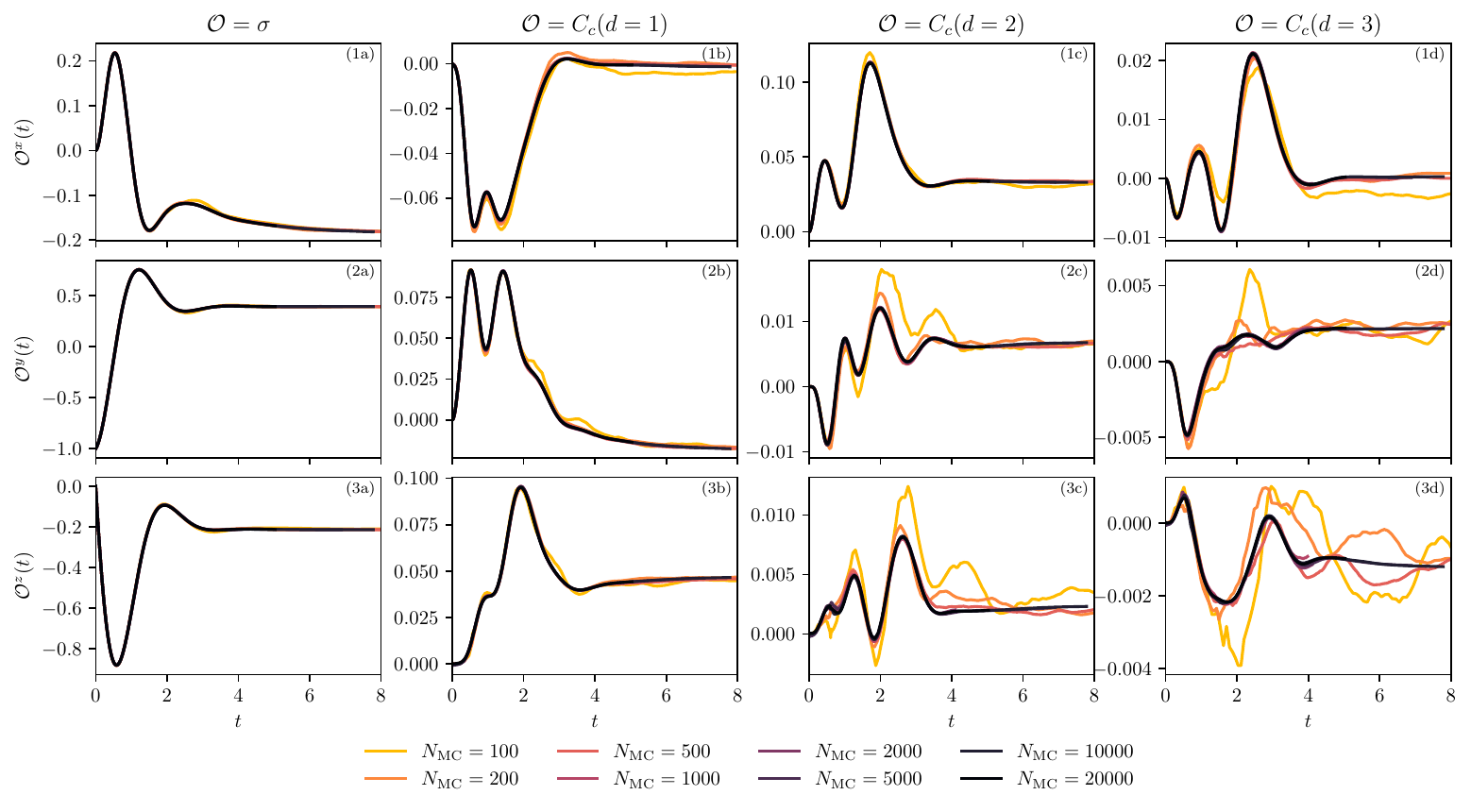}
    \caption{\textbf{Convergence with $N_\text{MC}$. Transverse-field Ising model with weak long-range competing interactions, $N=200$.} For panel descriptions, see Supplementary Note 2 F.}
    \label{fig: CI_N200_samples}
\end{figure}

\begin{figure}[H]
    \includegraphics[width=1.0\textwidth]{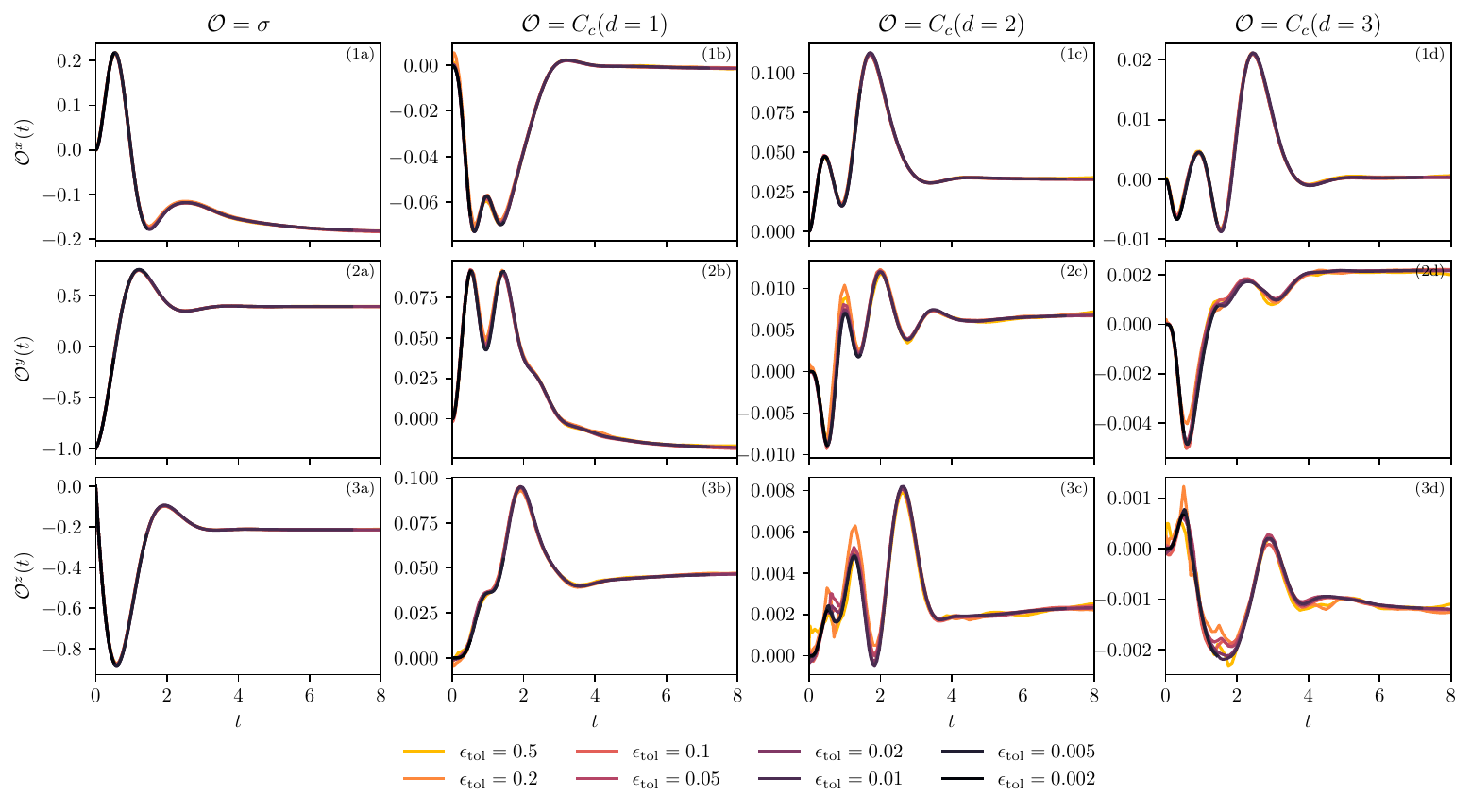}
    \caption{\textbf{Convergence with $\epsilon_\text{tol}$. Transverse-field Ising model with weak long-range competing interactions, $N=200$.} For panel descriptions, see Supplementary Note 2 F.}
    \label{fig: CI_N200_eps_tol}
\end{figure}

\begin{figure}[H]
    \includegraphics[width=1.0\textwidth]{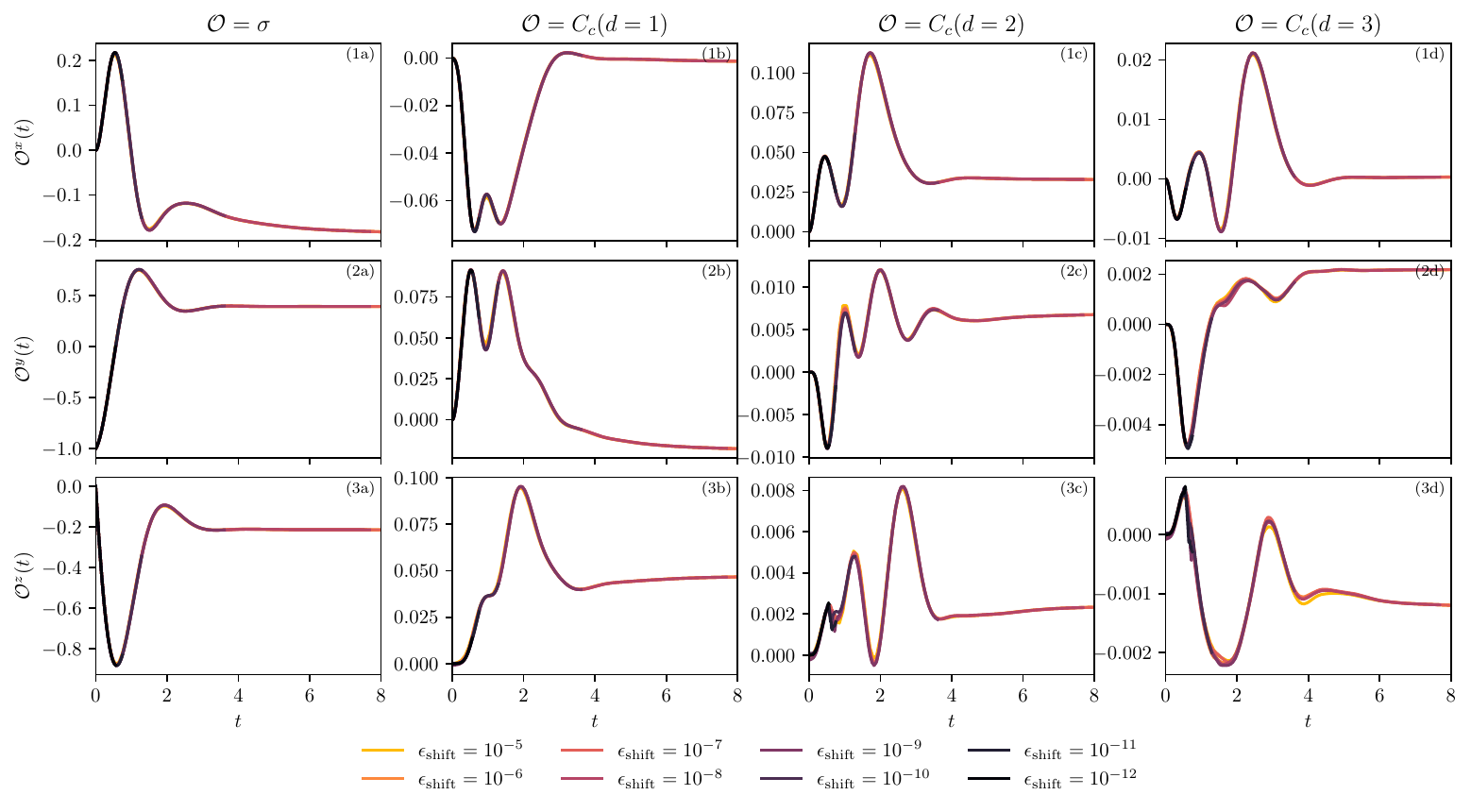}
    \caption{\textbf{Convergence with $\epsilon_\text{shift}$. Transverse-field Ising model with weak long-range competing interactions, $N=200$.} For panel descriptions, see Supplementary Note 2 F.}
    \label{fig: CI_N200_eps_shift}
\end{figure}

\begin{figure}[H]
    \includegraphics[width=1.0\textwidth]{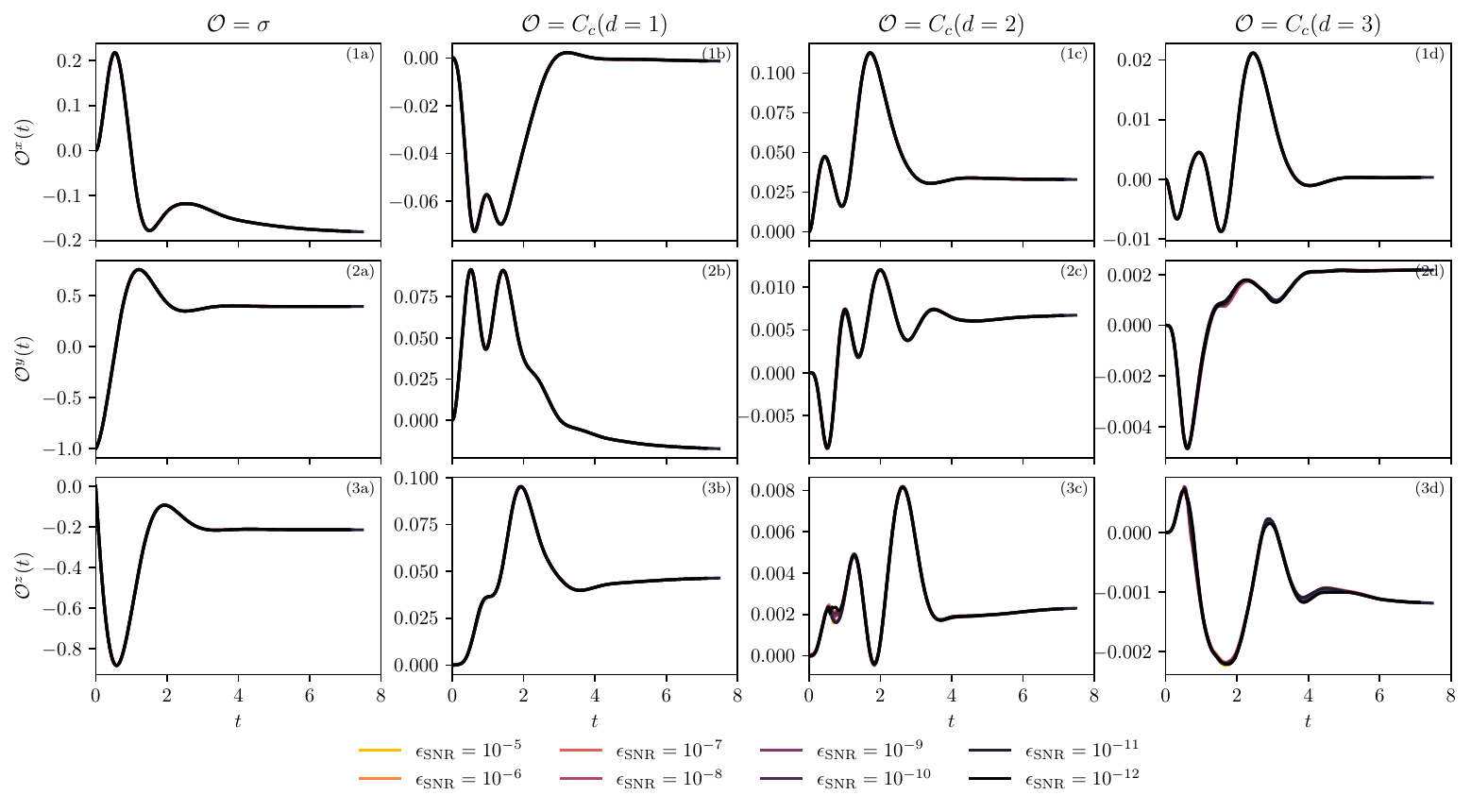}
    \caption{\textbf{Convergence with $\epsilon_\text{SNR}$. Transverse-field Ising model with weak long-range competing interactions, $N=200$.} For panel descriptions, see Supplementary Note 2 F.}
    \label{fig: CI_N200_eps_snr}
\end{figure}

\newpage
\subsection{Transverse-field Ising model with strong long-range competing interactions, $N=10$}

\begin{figure}[H]
    \includegraphics[width=1.0\textwidth]{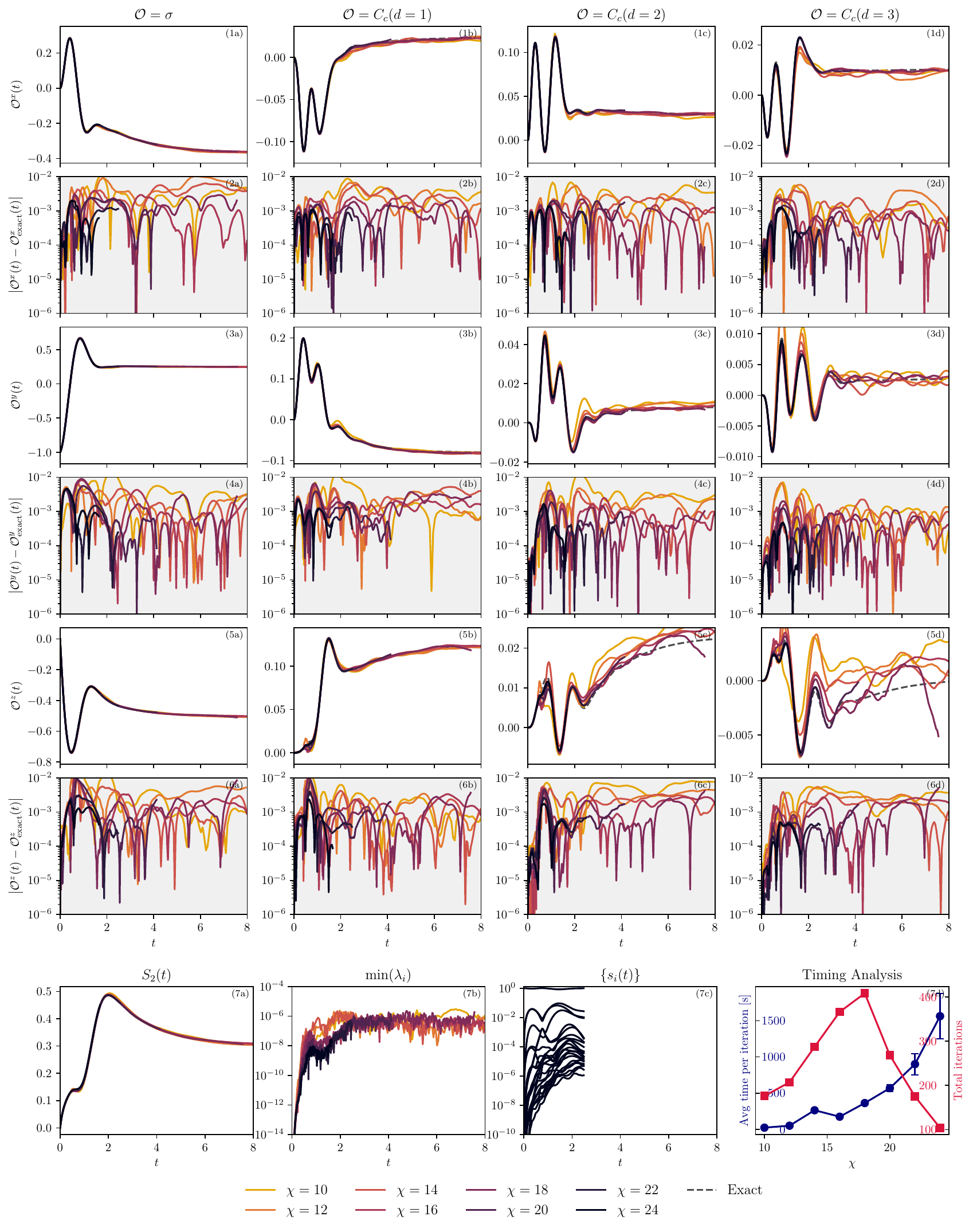}
    \caption{\textbf{Convergence with $\chi$. Transverse-field Ising model with strong long-range competing interactions, $N=10$.} For panel descriptions, see Supplementary Note 2 E.}
    \label{fig: CI_N10_strong_chi}
\end{figure}

\begin{figure}[H]
    \includegraphics[width=1.0\textwidth]{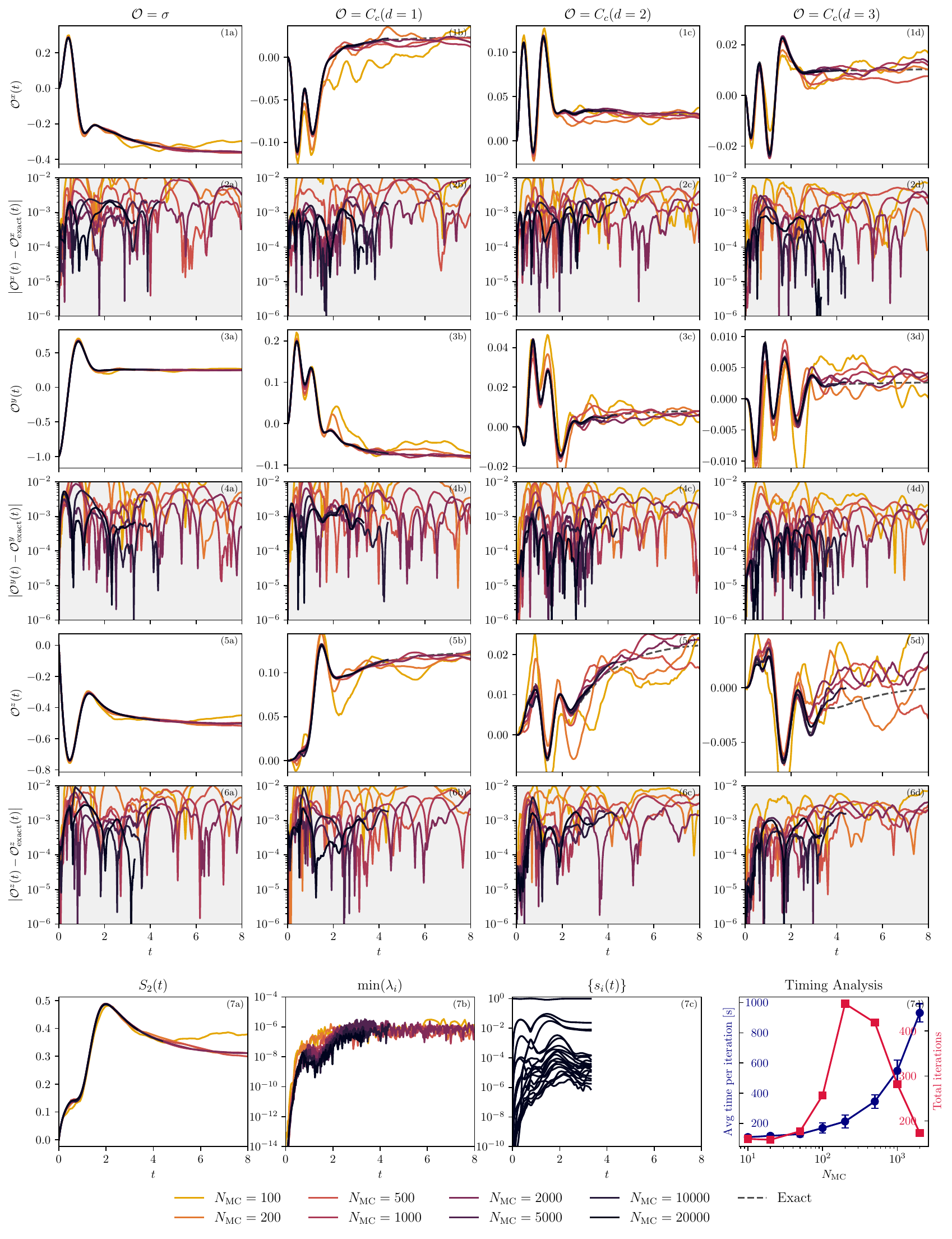}
    \caption{\textbf{Convergence with $N_\text{MC}$. Transverse-field Ising model with strong long-range competing interactions, $N=10$.} For panel descriptions, see Supplementary Note 2 E.}
    \label{fig: CI_N10_strong_samples}
\end{figure}

\begin{figure}[H]
    \includegraphics[width=1.0\textwidth]{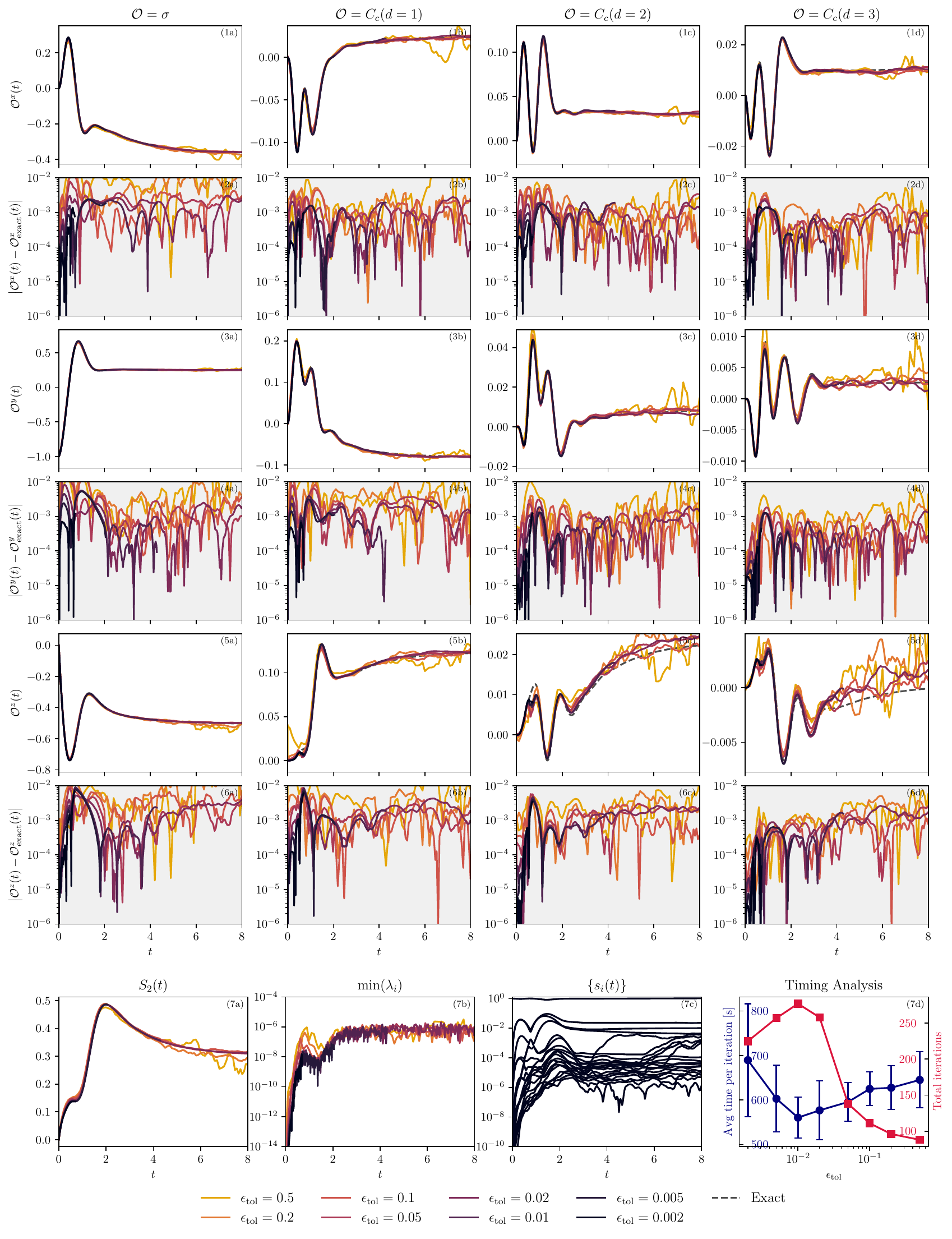}
    \caption{\textbf{Convergence with $\epsilon_\text{tol}$. Transverse-field Ising model with strong long-range competing interactions, $N=10$.} For panel descriptions, see Supplementary Note 2 E.}
    \label{fig: CI_N10_strong_eps_tol}
\end{figure}

\begin{figure}[H]
    \includegraphics[width=1.0\textwidth]{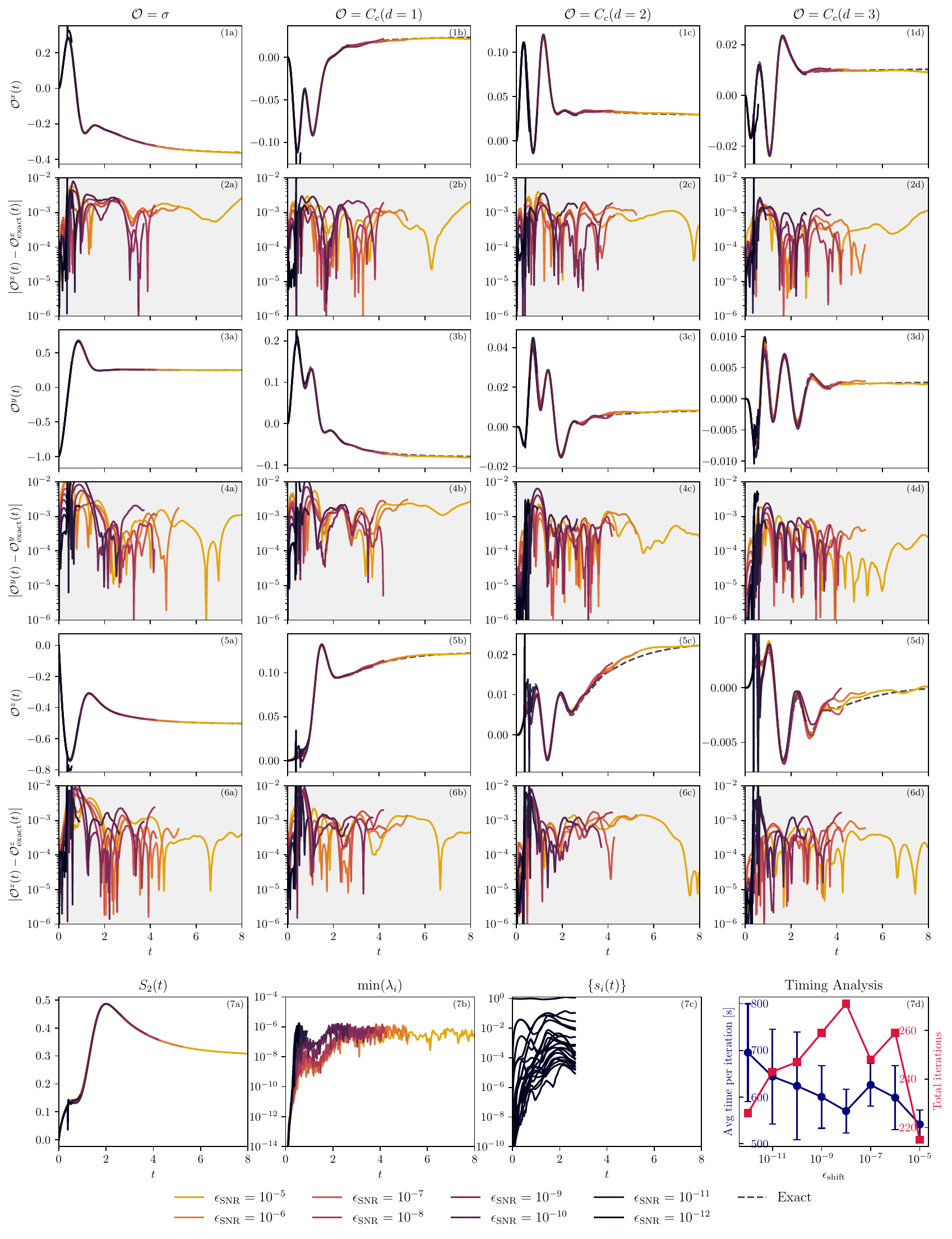}
    \caption{\textbf{Convergence with $\epsilon_\text{shift}$. Transverse-field Ising model with strong long-range competing interactions, $N=10$.} For panel descriptions, see Supplementary Note 2 E.}
    \label{fig: CI_N10_strong_eps_shift}
\end{figure}

\begin{figure}[H]
    \includegraphics[width=1.0\textwidth]{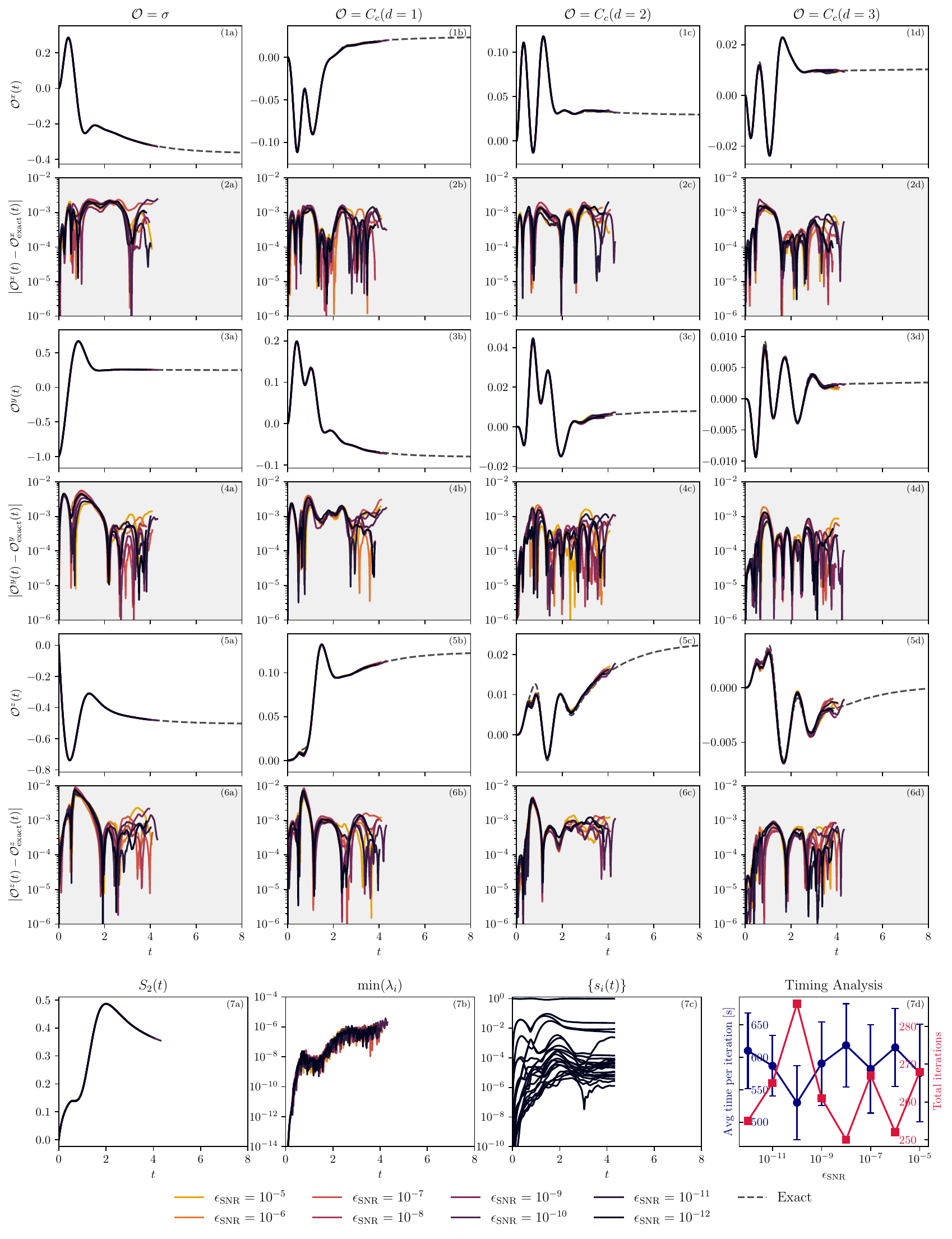}
    \caption{\textbf{Convergence with $\epsilon_\text{SNR}$. Transverse-field Ising model with strong long-range competing interactions, $N=10$.} For panel descriptions, see Supplementary Note 2 E.}
    \label{fig: CI_N10_strong_eps_snr}
\end{figure}

\newpage
\subsection{Transverse-field Ising model with strong long-range competing interactions, $N=200$}

\begin{figure}[H]
    \includegraphics[width=1.0\textwidth]{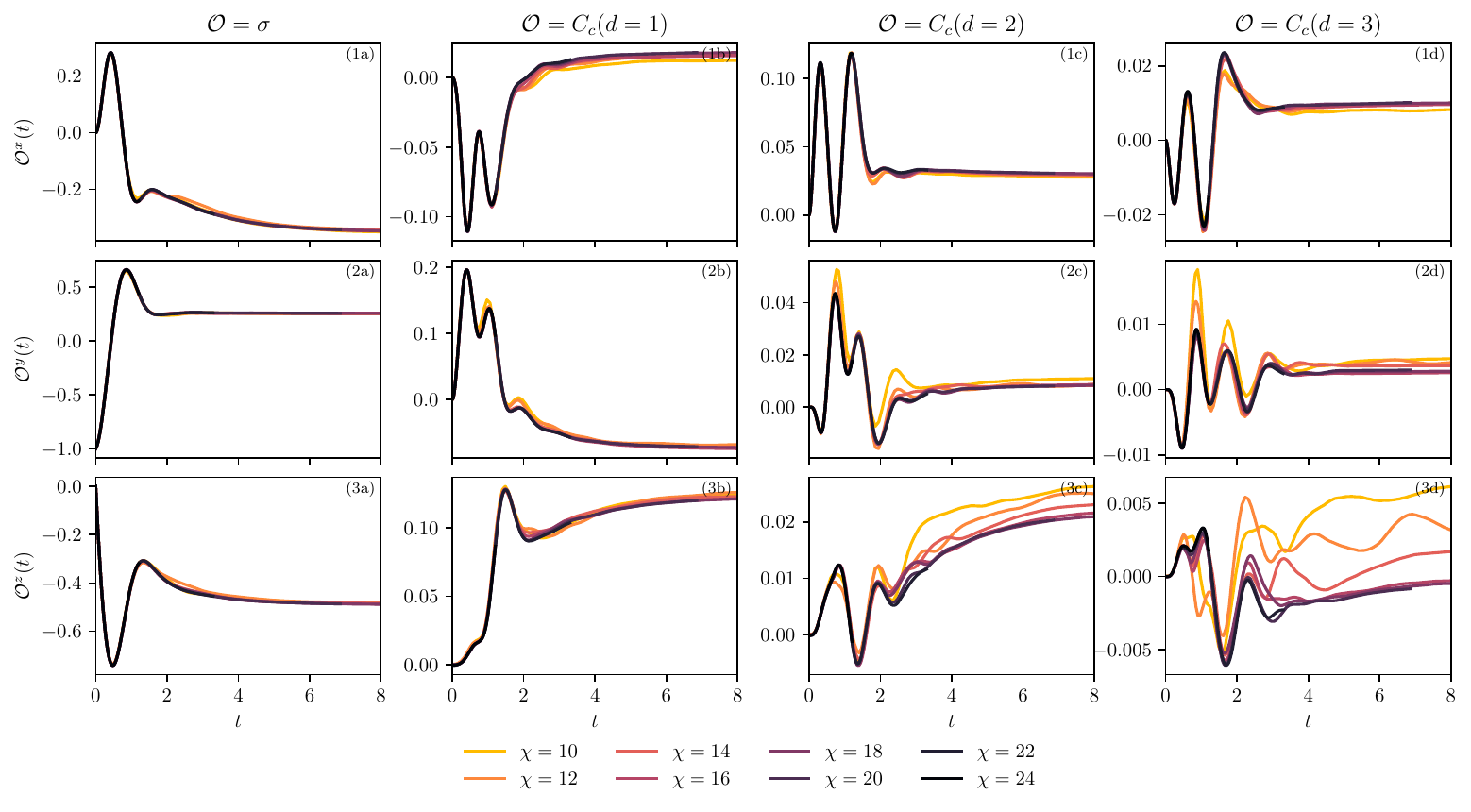}
    \caption{\textbf{Convergence with $\chi$. Transverse-field Ising model with strong long-range competing interactions, $N=200$.} For panel descriptions, see Supplementary Note 2 F.}
    \label{fig: CI_N200_strong_chi}
\end{figure}

\begin{figure}[H]
    \includegraphics[width=1.0\textwidth]{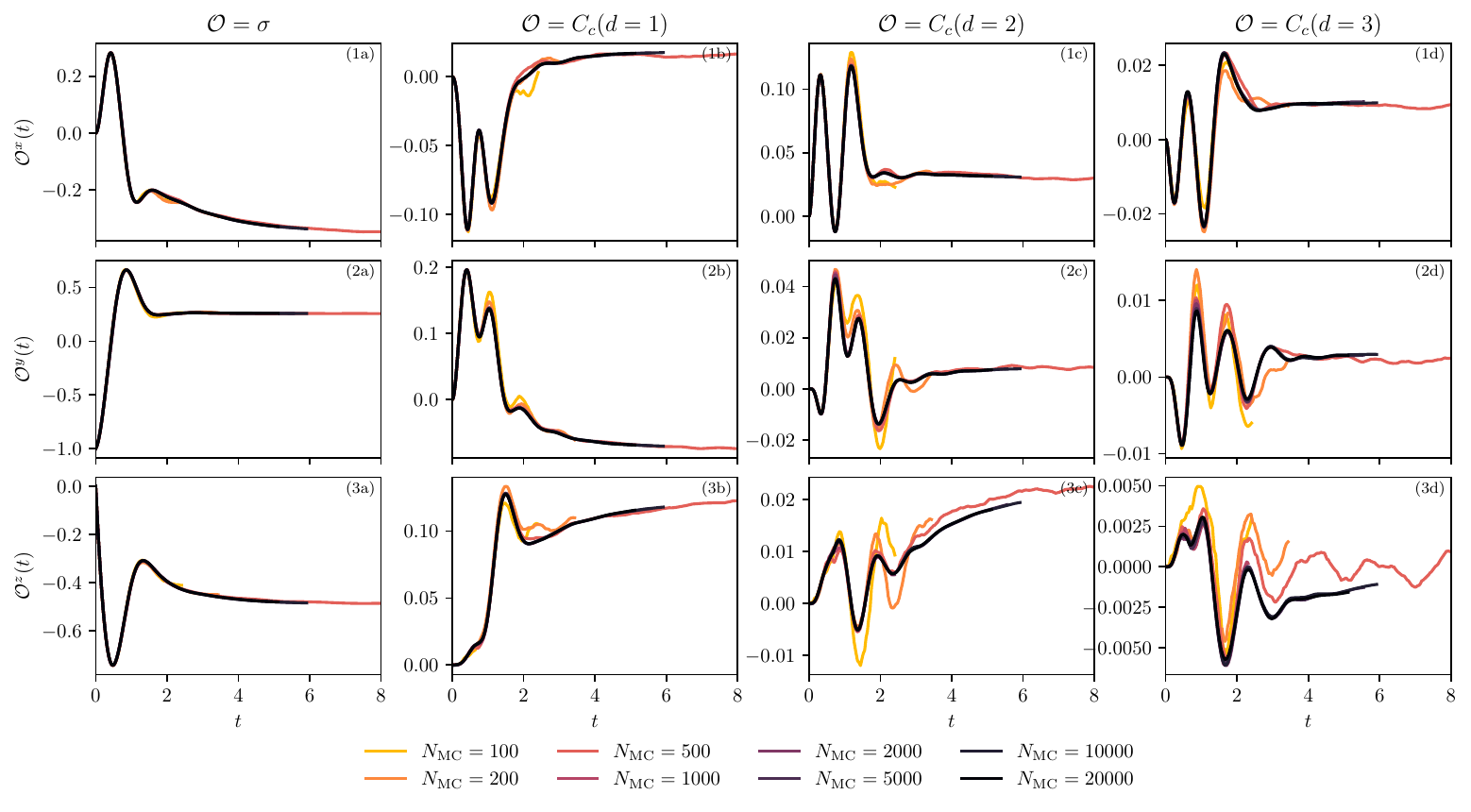}
    \caption{\textbf{Convergence with $N_\text{MC}$. Transverse-field Ising model with strong long-range competing interactions, $N=200$.} For panel descriptions, see Supplementary Note 2 F.}
    \label{fig: CI_N200_strong_samples}
\end{figure}

\begin{figure}[H]
    \includegraphics[width=1.0\textwidth]{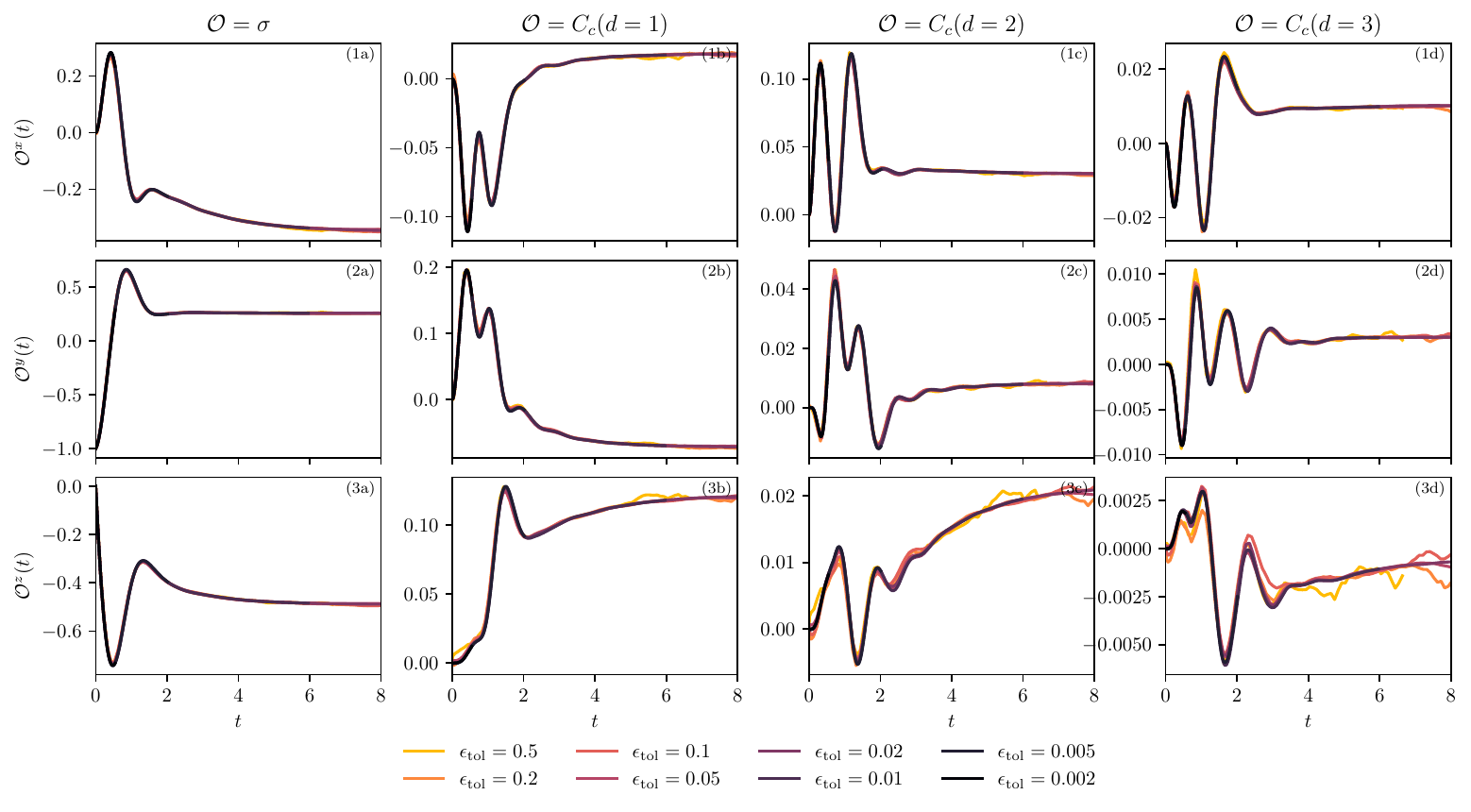}
    \caption{\textbf{Convergence with $\epsilon_\text{tol}$. Transverse-field Ising model with strong long-range competing interactions, $N=200$.} For panel descriptions, see Supplementary Note 2 F.}
    \label{fig: CI_N200_strong_eps_tol}
\end{figure}

\begin{figure}[H]
    \includegraphics[width=1.0\textwidth]{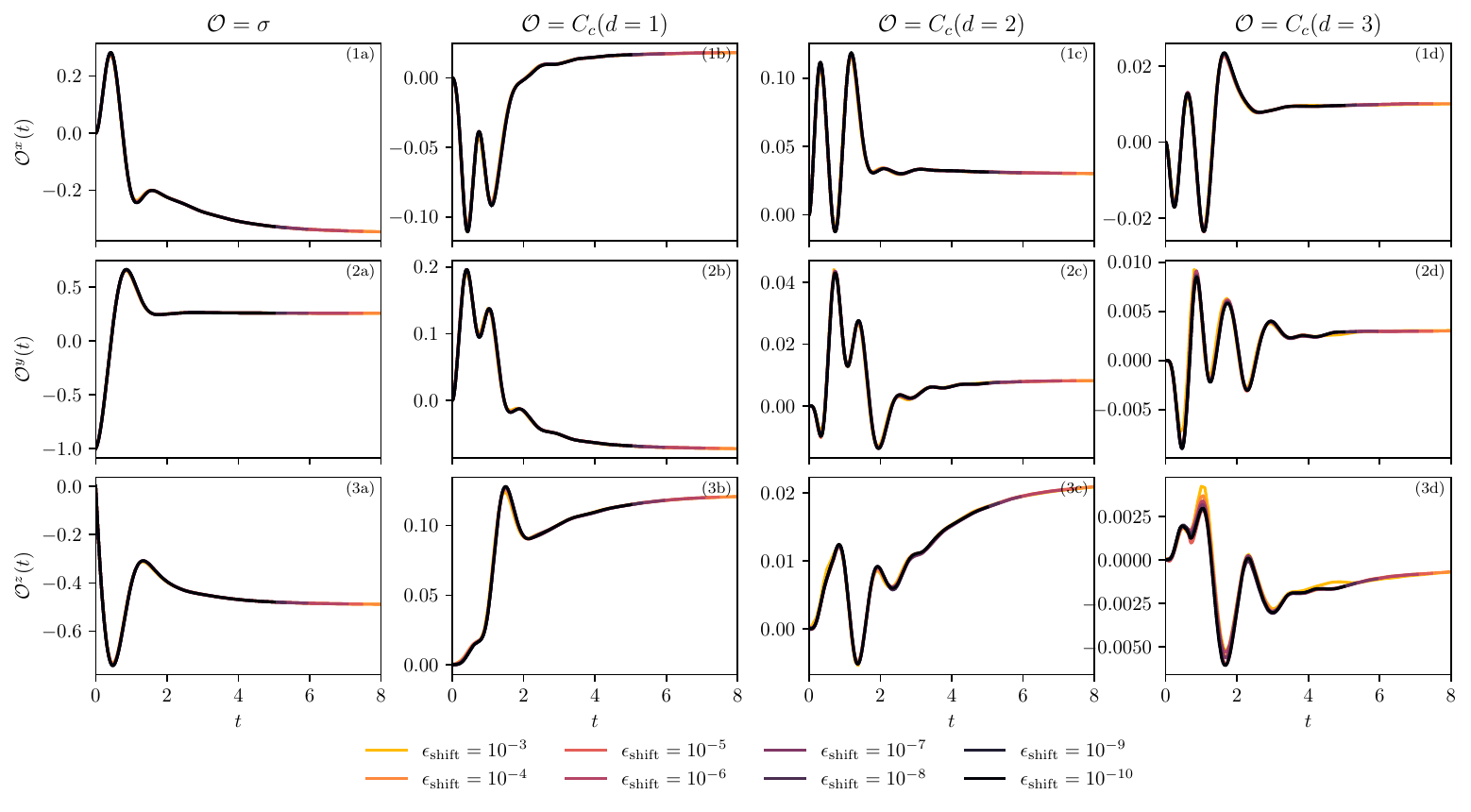}
    \caption{\textbf{Convergence with $\epsilon_\text{shift}$. Transverse-field Ising model with strong long-range competing interactions, $N=200$.} For panel descriptions, see Supplementary Note 2 F.}
    \label{fig: CI_N200_strong_eps_shift}
\end{figure}

\begin{figure}[H]
    \includegraphics[width=1.0\textwidth]{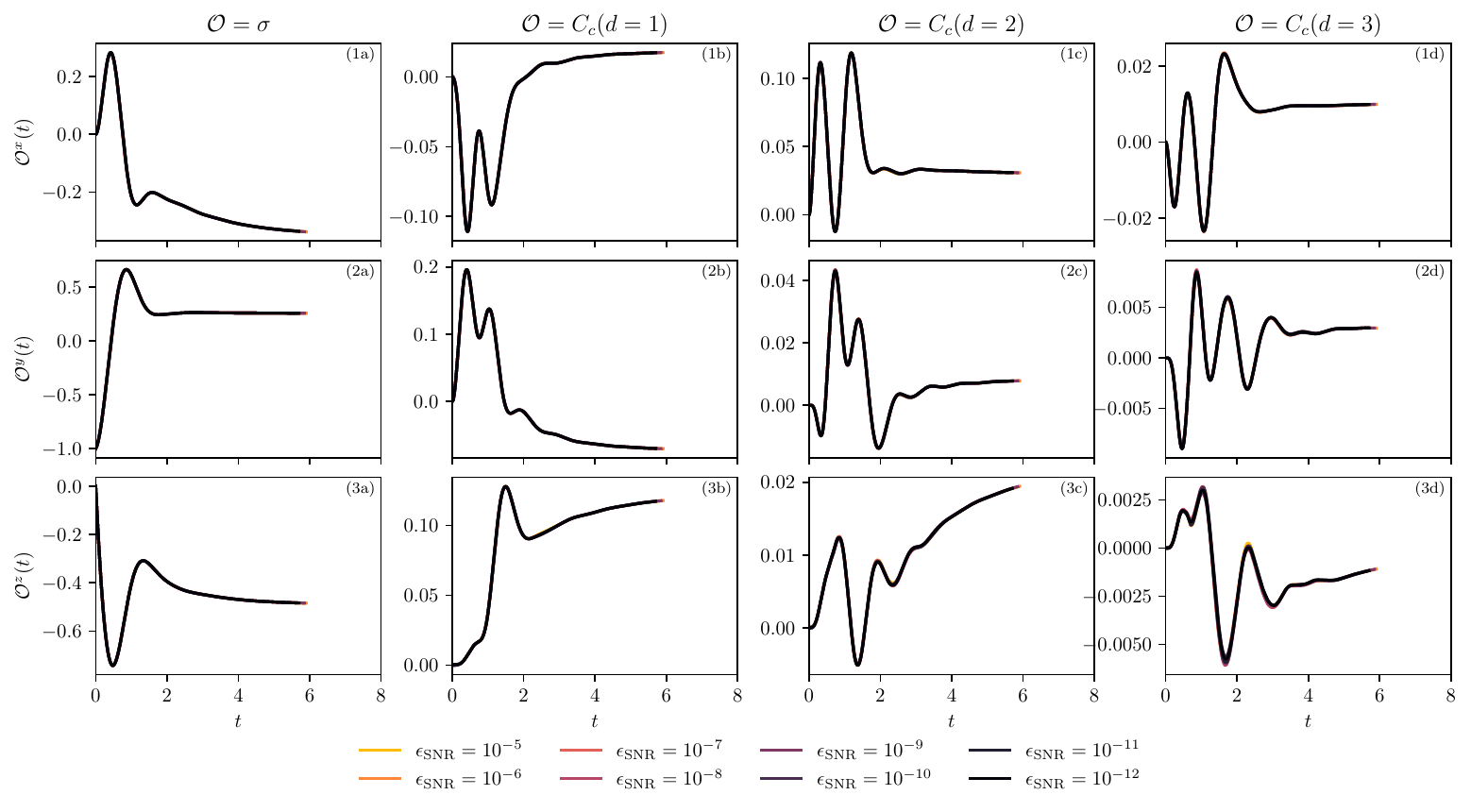}
    \caption{\textbf{Convergence with $\epsilon_\text{SNR}$. Transverse-field Ising model with strong long-range competing interactions, $N=200$.} For panel descriptions, see Supplementary Note 2 F.}
    \label{fig: CI_N200_strong_eps_snr}
\end{figure}

\newpage
\subsection{Transverse-field Ising model with super-long-range competing interactions, $N=10$}

\begin{figure}[H]
    \includegraphics[width=1.0\textwidth]{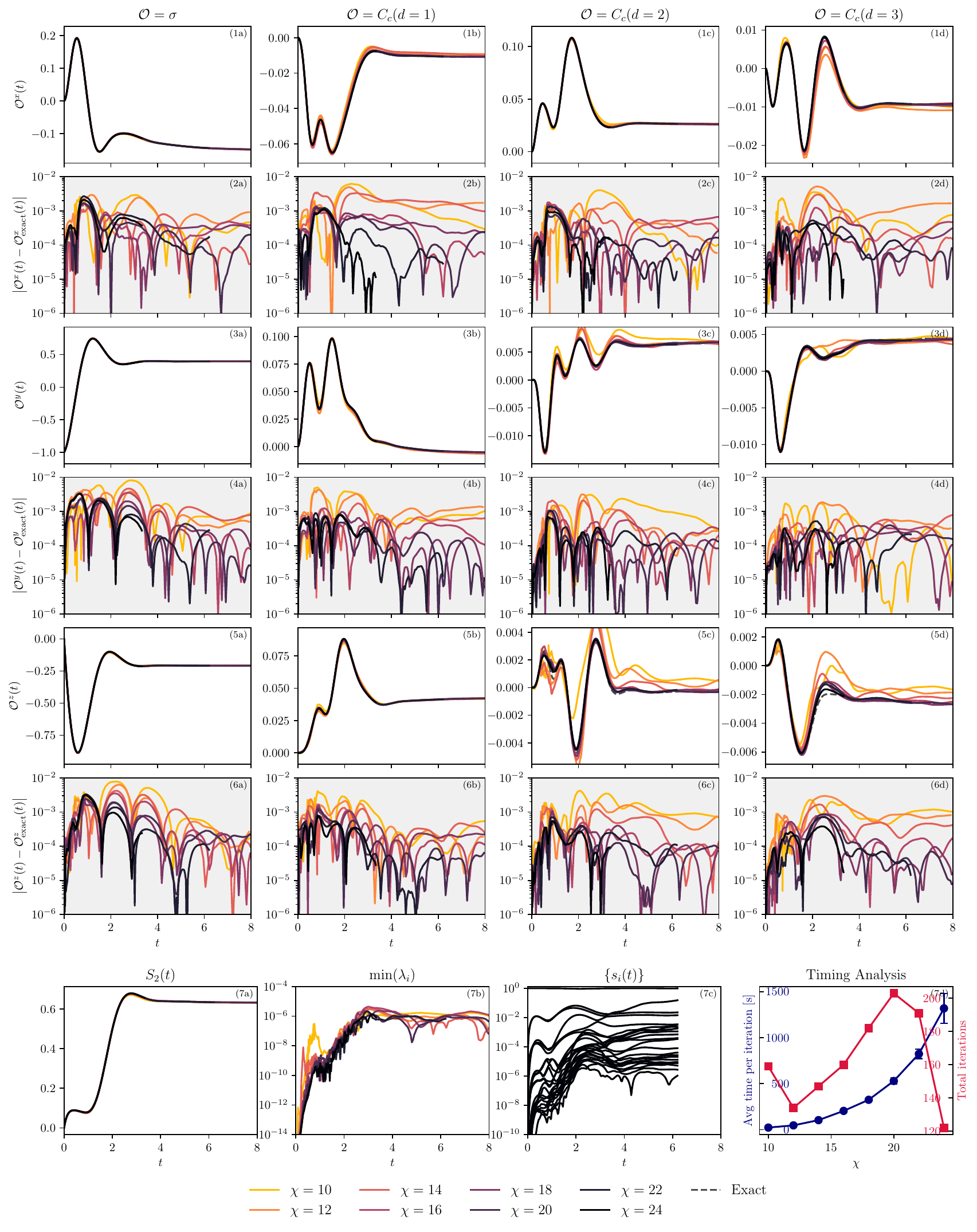}
    \caption{\textbf{Convergence with $\chi$. Transverse-field Ising model with super-long-range competing interactions, $N=10$.} For panel descriptions, see Supplementary Note 2 E.}
    \label{fig: CI_N10_far_chi}
\end{figure}

\begin{figure}[H]
    \includegraphics[width=1.0\textwidth]{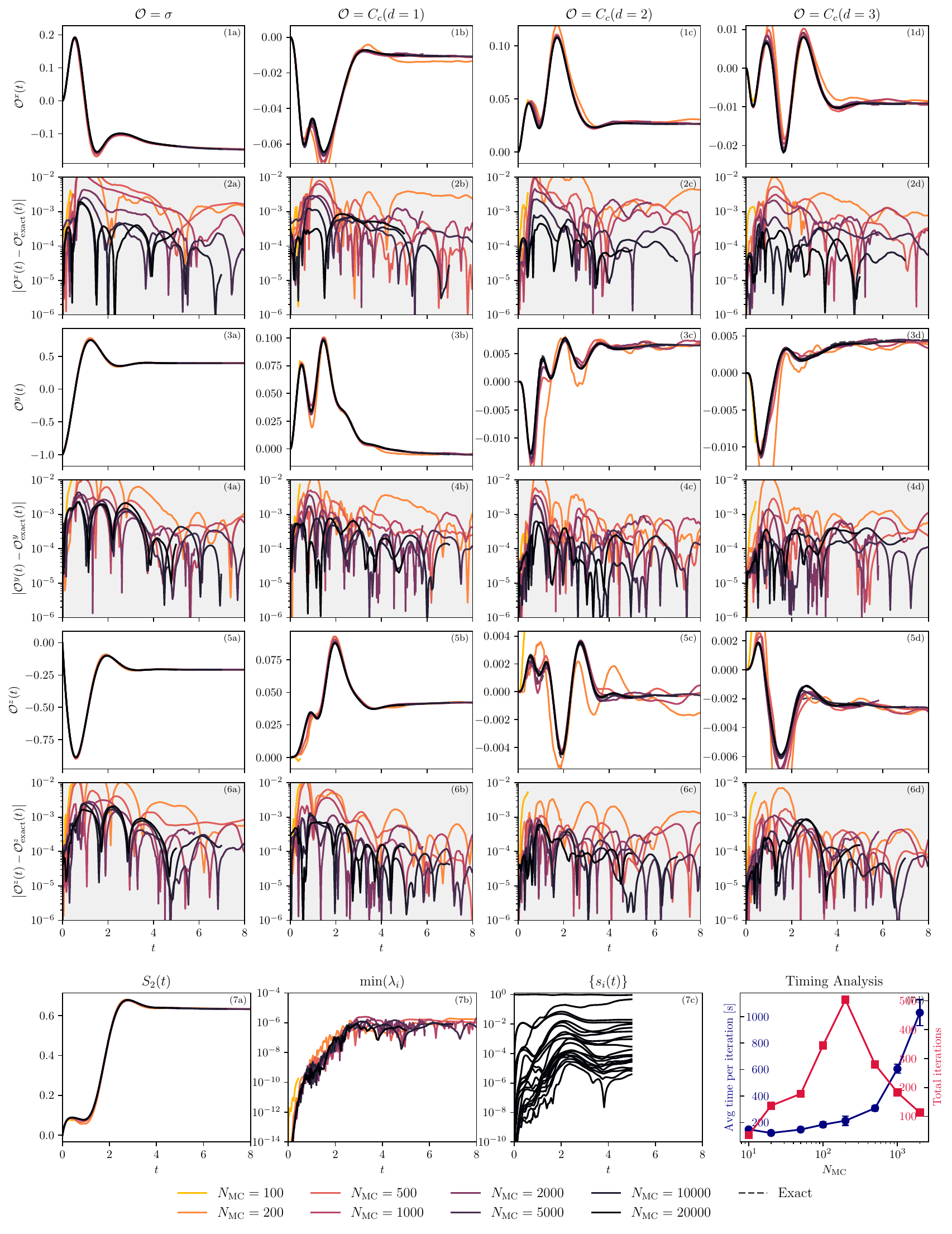}
    \caption{\textbf{Convergence with $N_\text{MC}$. Transverse-field Ising model with super-long-range competing interactions, $N=10$.} For panel descriptions, see Supplementary Note 2 E.}
    \label{fig: CI_N10_far_samples}
\end{figure}

\begin{figure}[H]
    \includegraphics[width=1.0\textwidth]{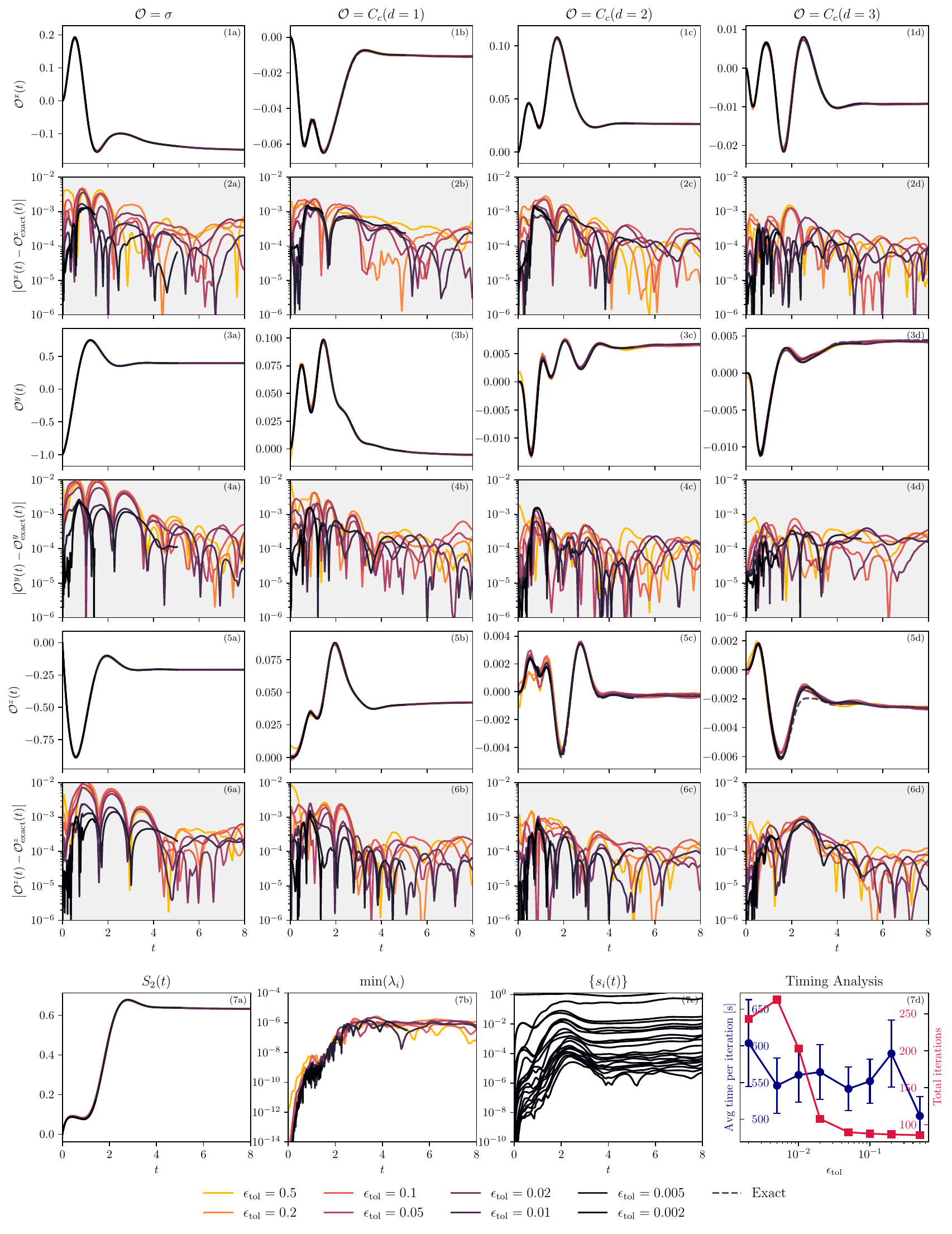}
    \caption{\textbf{Convergence with $\epsilon_\text{tol}$. Transverse-field Ising model with super-long-range competing interactions, $N=10$.} For panel descriptions, see Supplementary Note 2 E.}
    \label{fig: CI_N10_far_eps_tol}
\end{figure}

\begin{figure}[H]
    \includegraphics[width=1.0\textwidth]{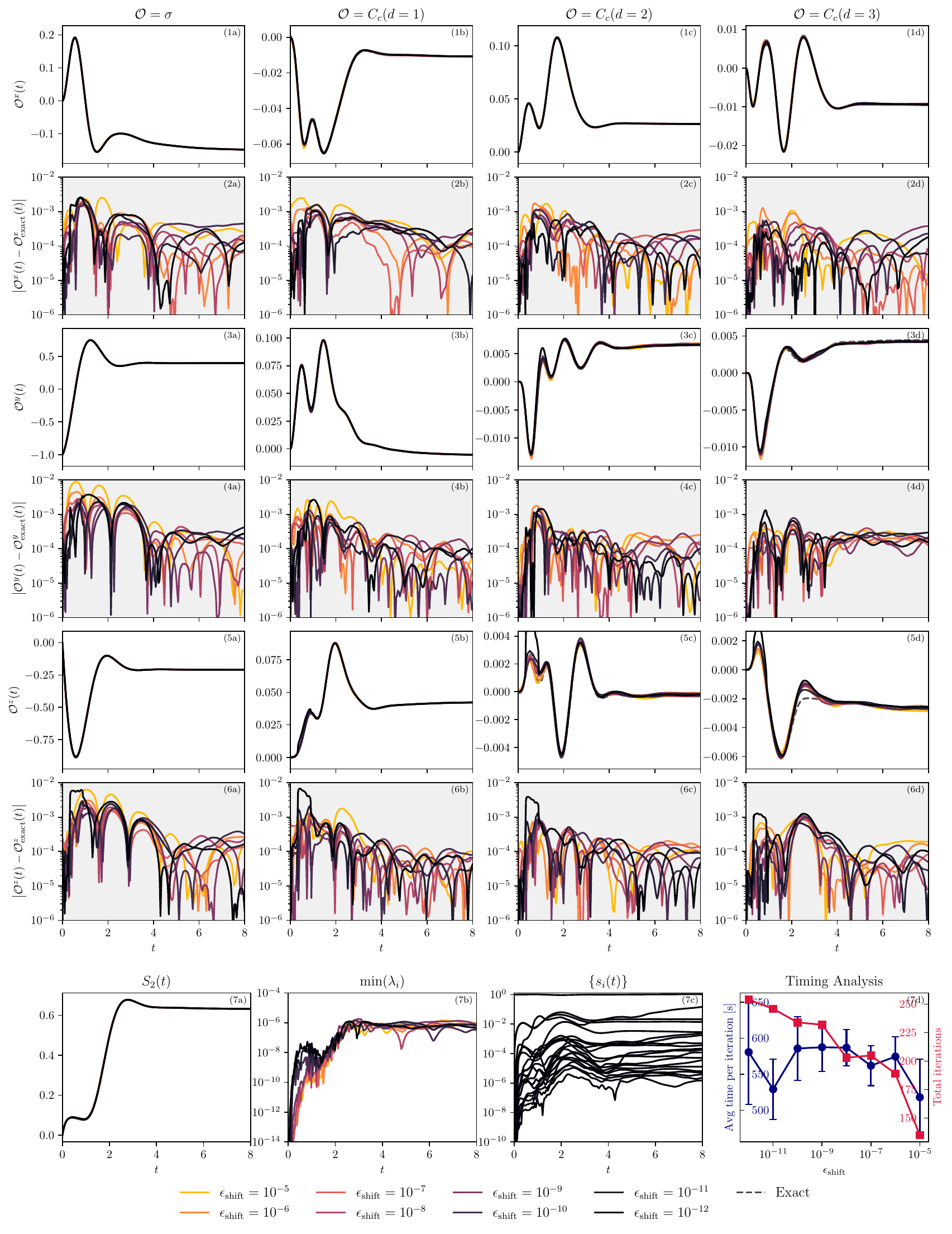}
    \caption{\textbf{Convergence with $\epsilon_\text{shift}$. Transverse-field Ising model with super-long-range competing interactions, $N=10$.} For panel descriptions, see Supplementary Note 2 E.}
    \label{fig: CI_N10_far_eps_shift}
\end{figure}

\begin{figure}[H]
    \includegraphics[width=1.0\textwidth]{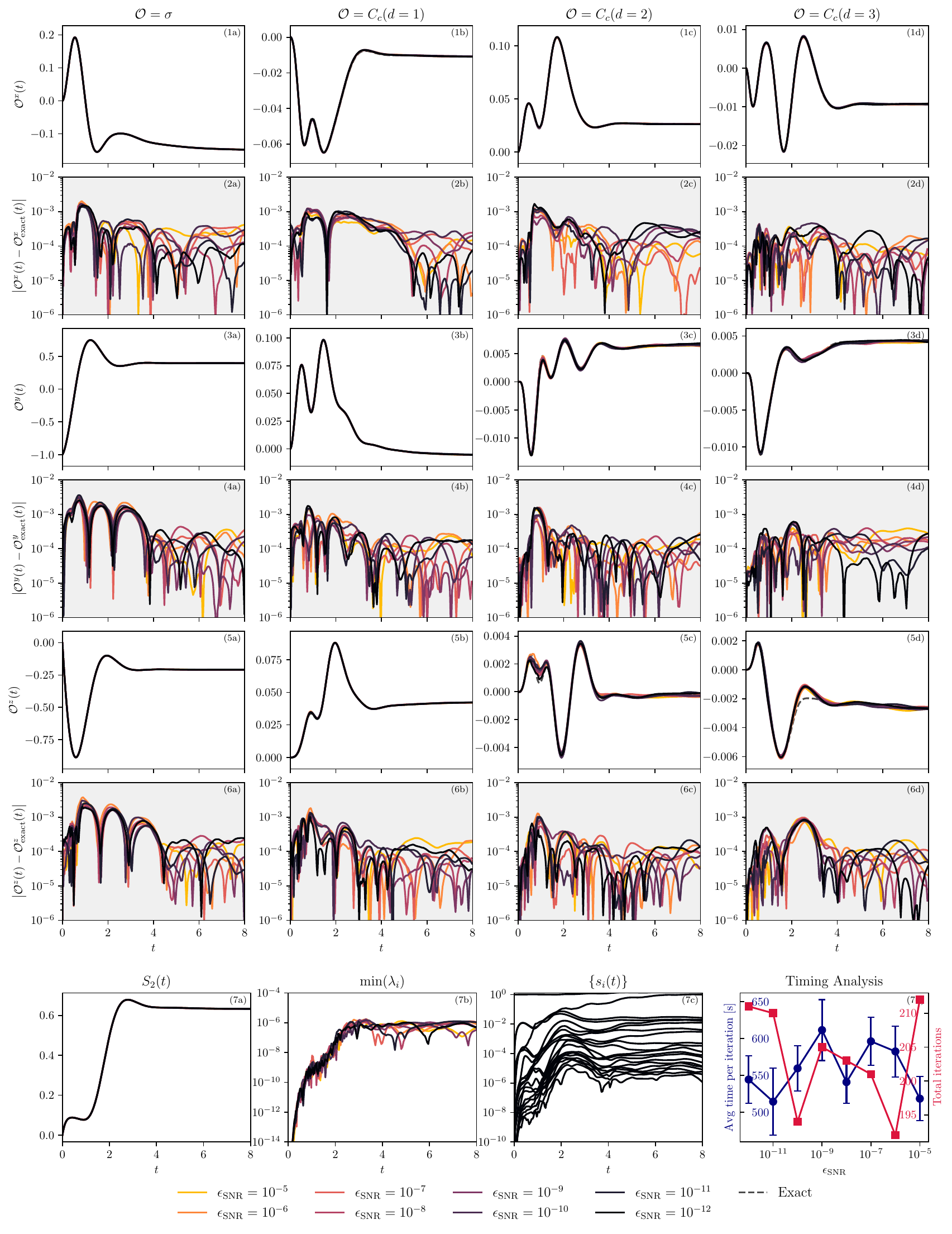}
    \caption{\textbf{Convergence with $\epsilon_\text{SNR}$. Transverse-field Ising model with super-long-range competing interactions, $N=10$.} For panel descriptions, see Supplementary Note 2 E.}
    \label{fig: CI_N10_far_eps_snr}
\end{figure}

\newpage
\subsection{Transverse-field Ising model with super-long-range competing interactions, $N=200$}

\begin{figure}[H]
    \includegraphics[width=1.0\textwidth]{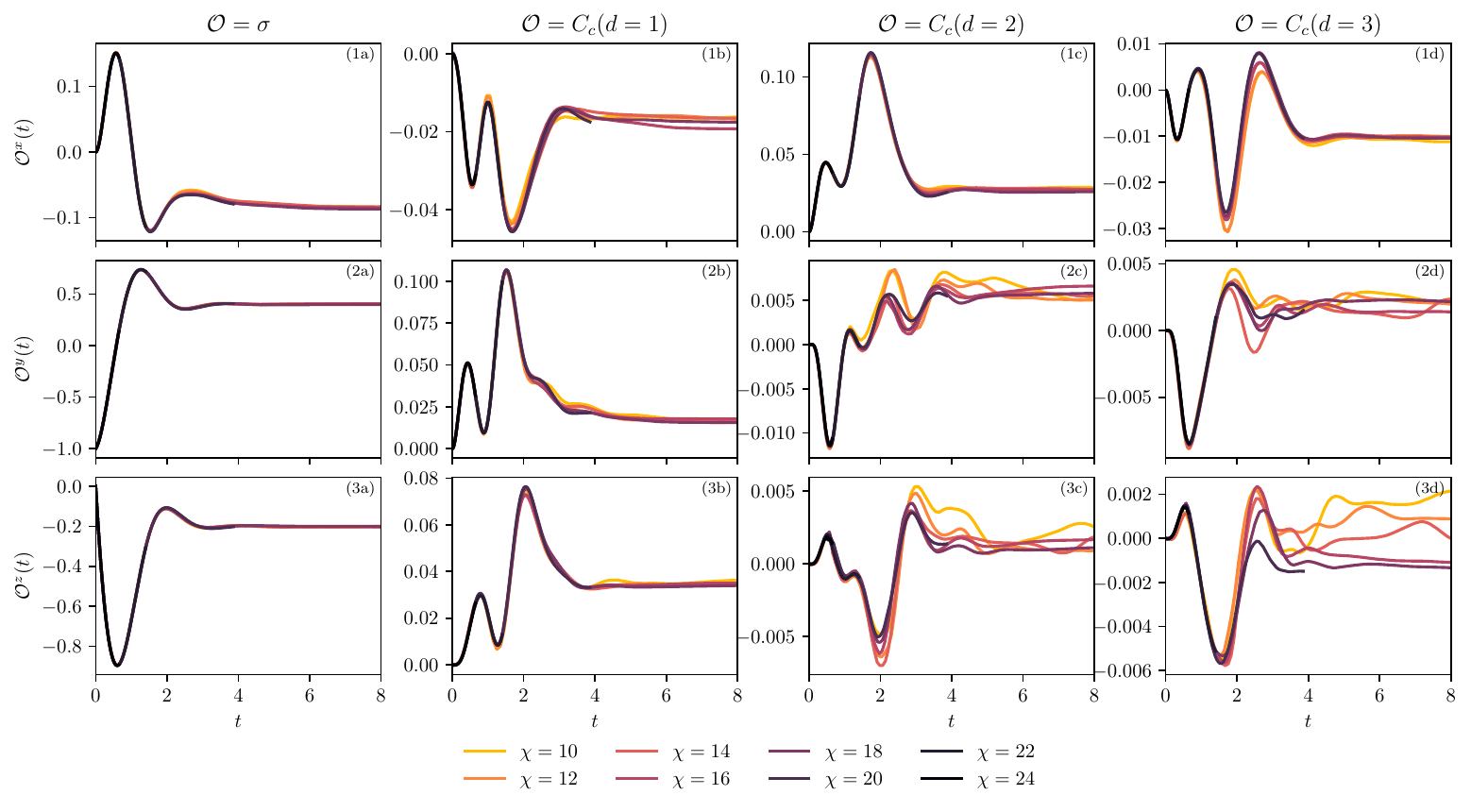}
    \caption{\textbf{Convergence with $\chi$. Transverse-field Ising model with super-long-range competing interactions, $N=200$.} For panel descriptions, see Supplementary Note 2 F.}
    \label{fig: CI_N200_far_chi}
\end{figure}

\begin{figure}[H]
    \includegraphics[width=1.0\textwidth]{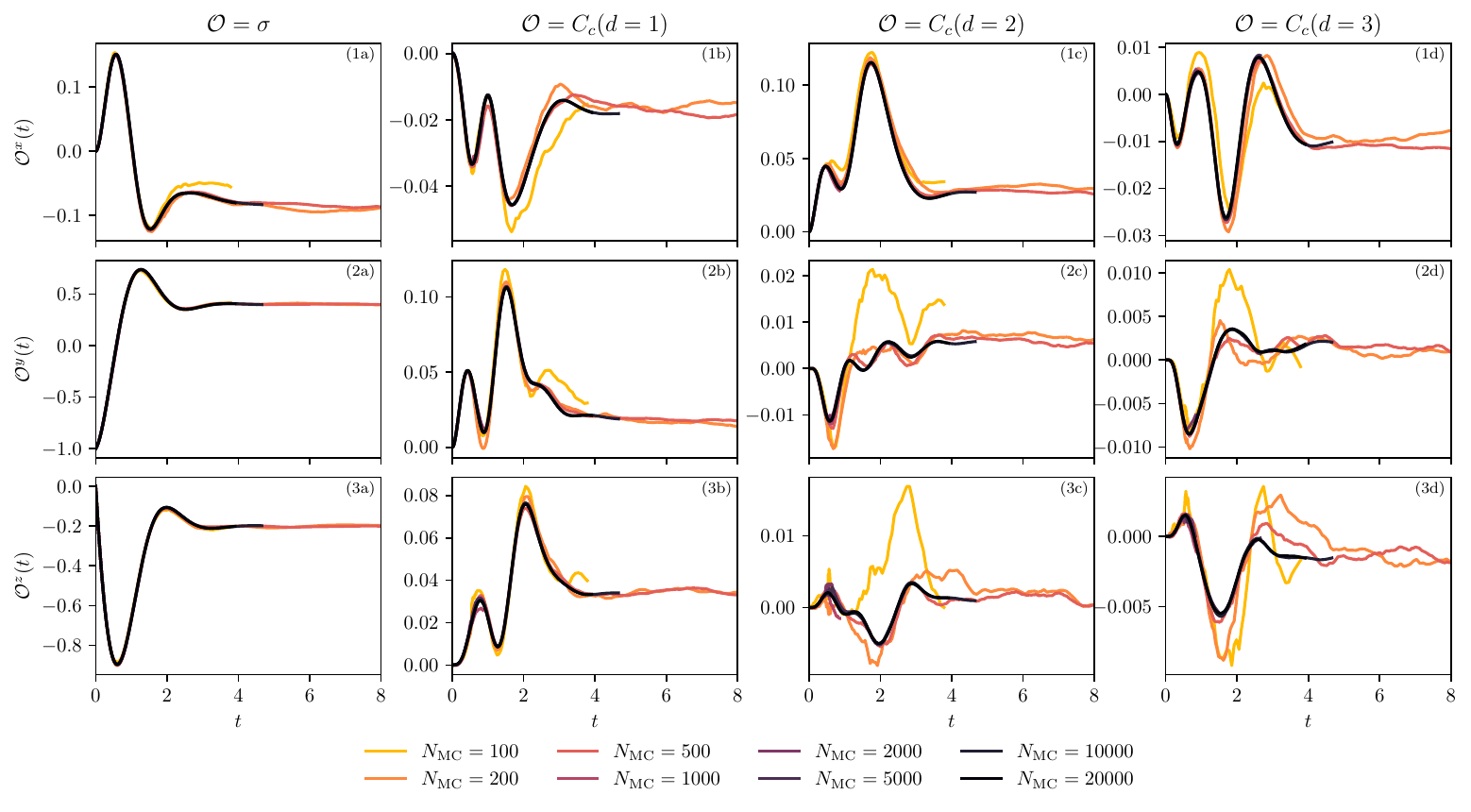}
    \caption{\textbf{Convergence with $N_\text{MC}$. Transverse-field Ising model with super-long-range competing interactions, $N=200$.} For panel descriptions, see Supplementary Note 2 F.}
    \label{fig: CI_N200_far_samples}
\end{figure}

\begin{figure}[H]
    \includegraphics[width=1.0\textwidth]{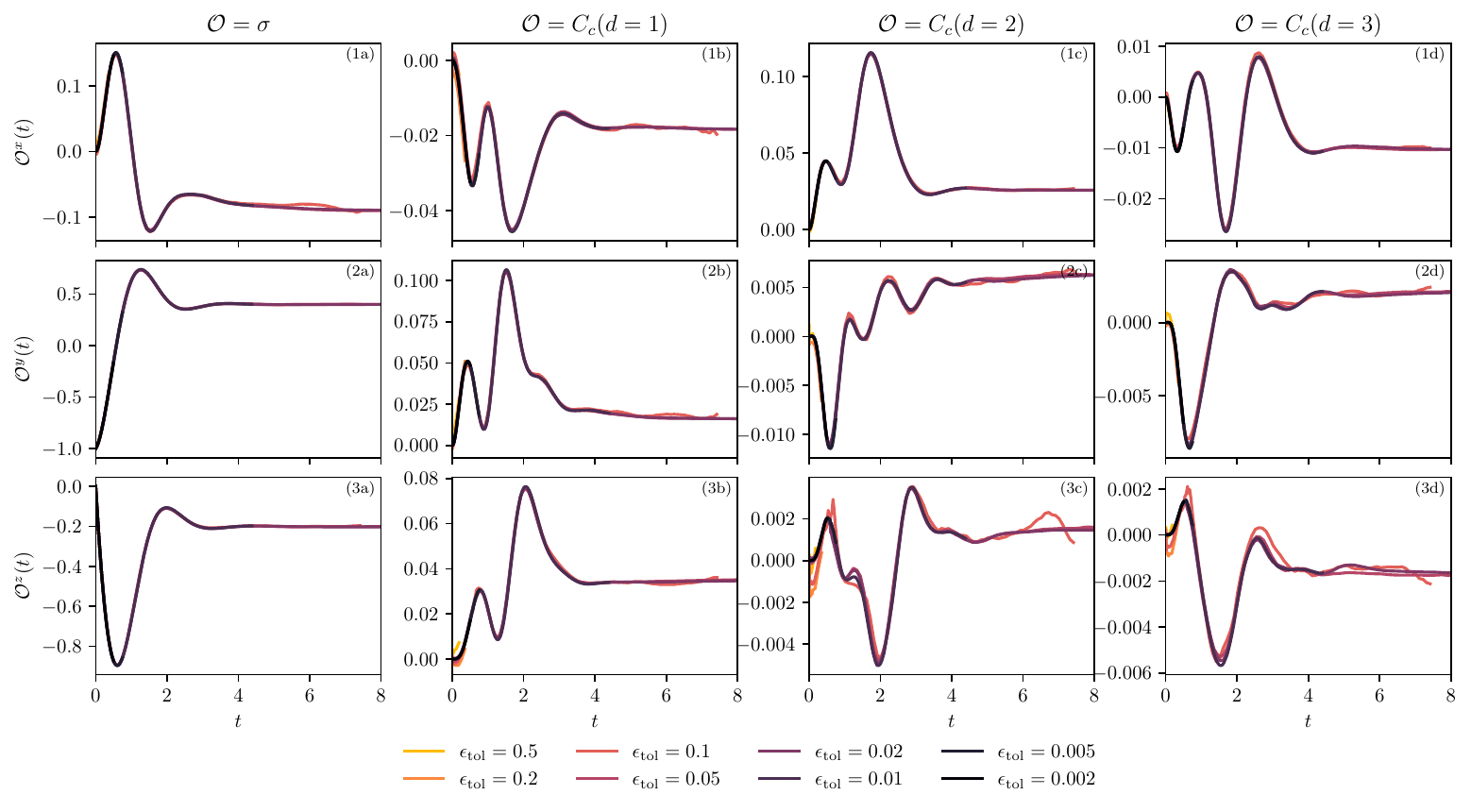}
    \caption{\textbf{Convergence with $\epsilon_\text{tol}$. Transverse-field Ising model with super-long-range competing interactions, $N=200$.} For panel descriptions, see Supplementary Note 2 F.}
    \label{fig: CI_N200_far_eps_tol}
\end{figure}

\begin{figure}[H]
    \includegraphics[width=1.0\textwidth]{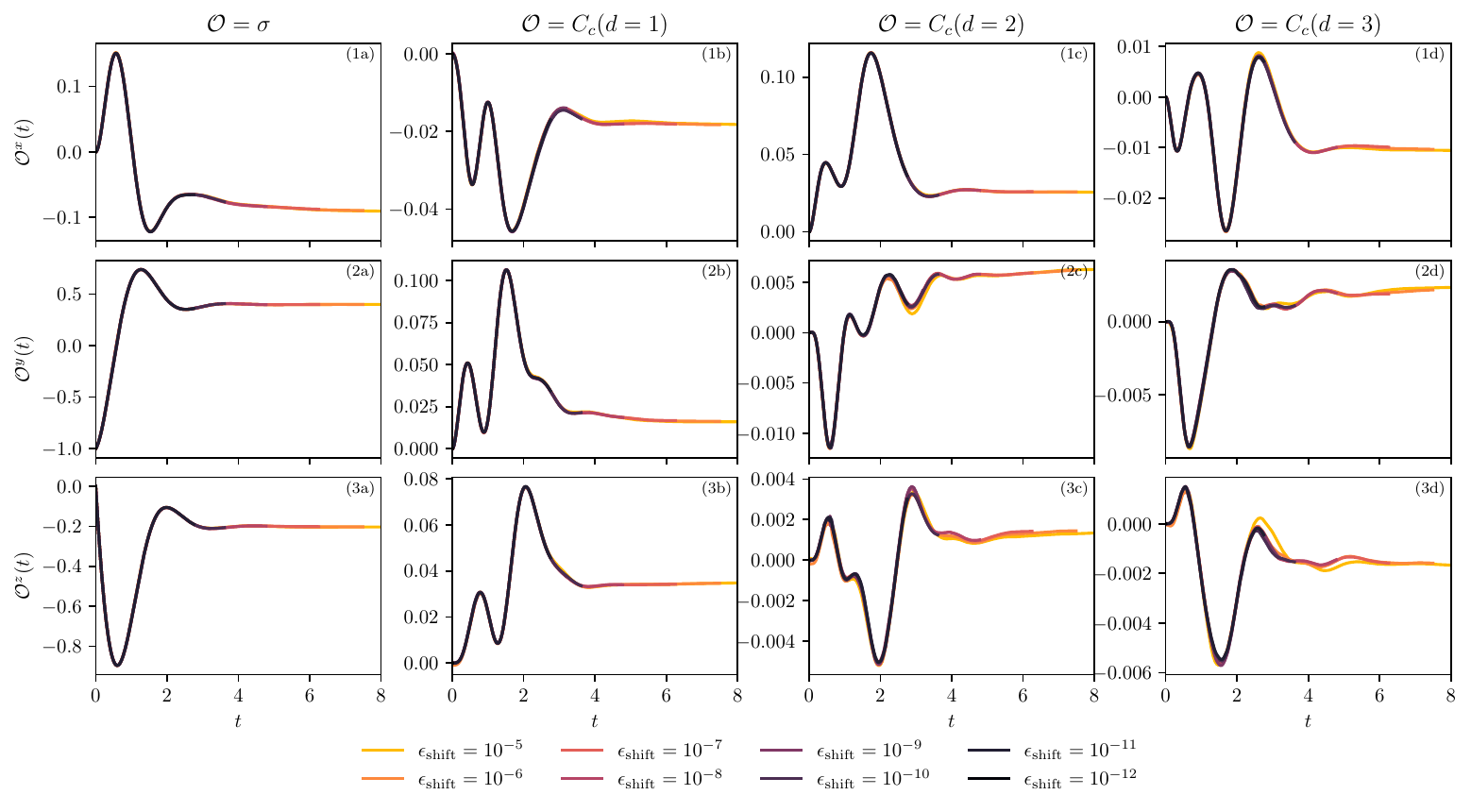}
    \caption{\textbf{Convergence with $\epsilon_\text{shift}$. Transverse-field Ising model with super-long-range competing interactions, $N=200$.} For panel descriptions, see Supplementary Note 2 F.}
    \label{fig: CI_N200_far_eps_shift}
\end{figure}

\begin{figure}[H]
    \includegraphics[width=1.0\textwidth]{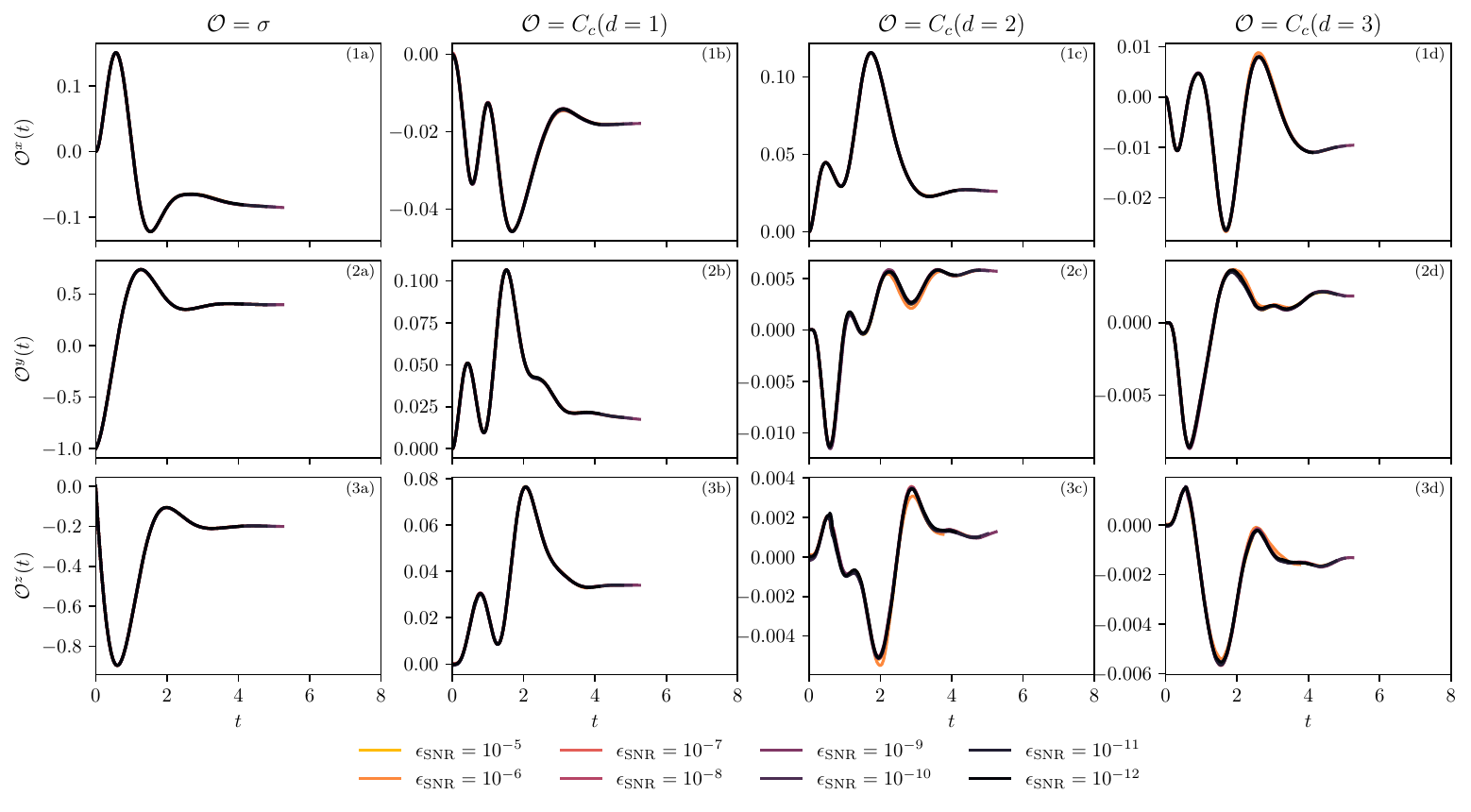}
    \caption{\textbf{Convergence with $\epsilon_\text{SNR}$. Transverse-field Ising model with super-long-range competing interactions, $N=200$.} For panel descriptions, see Supplementary Note 2 F.}
    \label{fig: CI_N200_far_eps_snr}
\end{figure}

\newpage
\subsection{XYZ model with short-range interactions, $N=10$}

\begin{figure}[H]
    \includegraphics[width=1.0\textwidth]{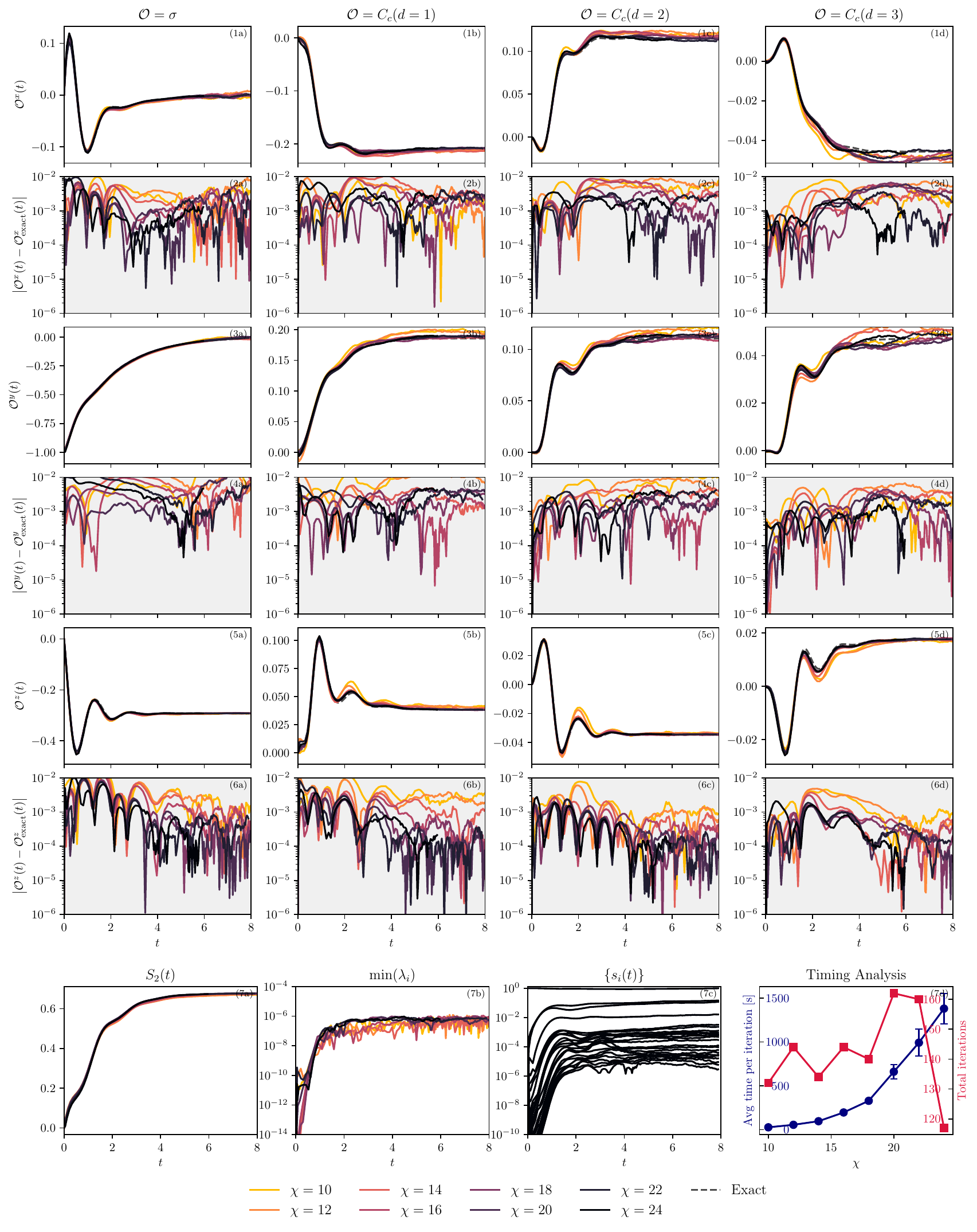}
    \caption{\textbf{Convergence with $\chi$. XYZ model with short-range interactions, $N=10$.} For panel descriptions, see Supplementary Note 2 E.}
    \label{fig: XYZ_N10_chi}
\end{figure}

\begin{figure}[H]
    \includegraphics[width=1.0\textwidth]{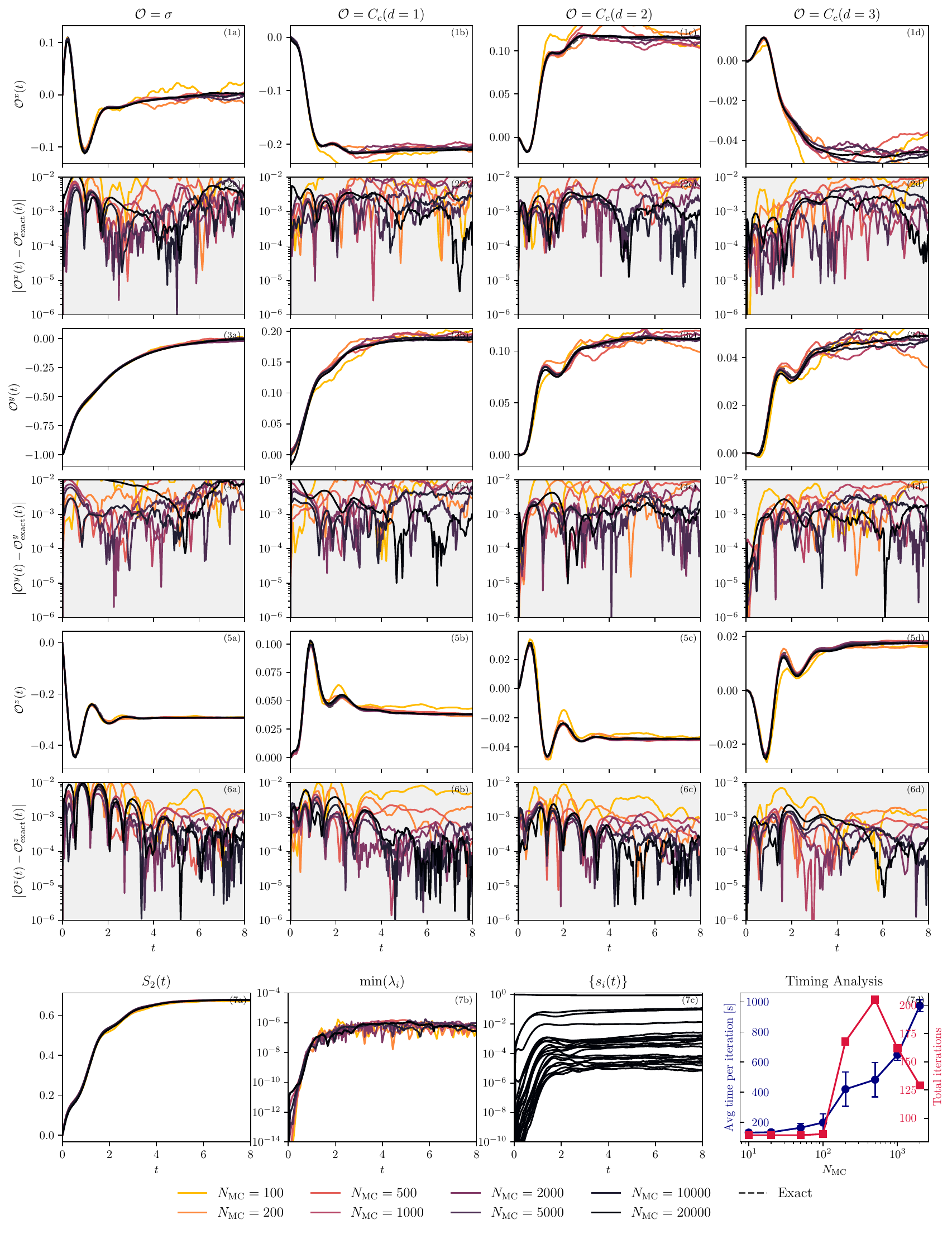}
    \caption{\textbf{Convergence with $N_\text{MC}$. XYZ model with short-range interactions, $N=10$.} For panel descriptions, see Supplementary Note 2 E.}
    \label{fig: XYZ_N10_samples}
\end{figure}

\begin{figure}[H]
    \includegraphics[width=1.0\textwidth]{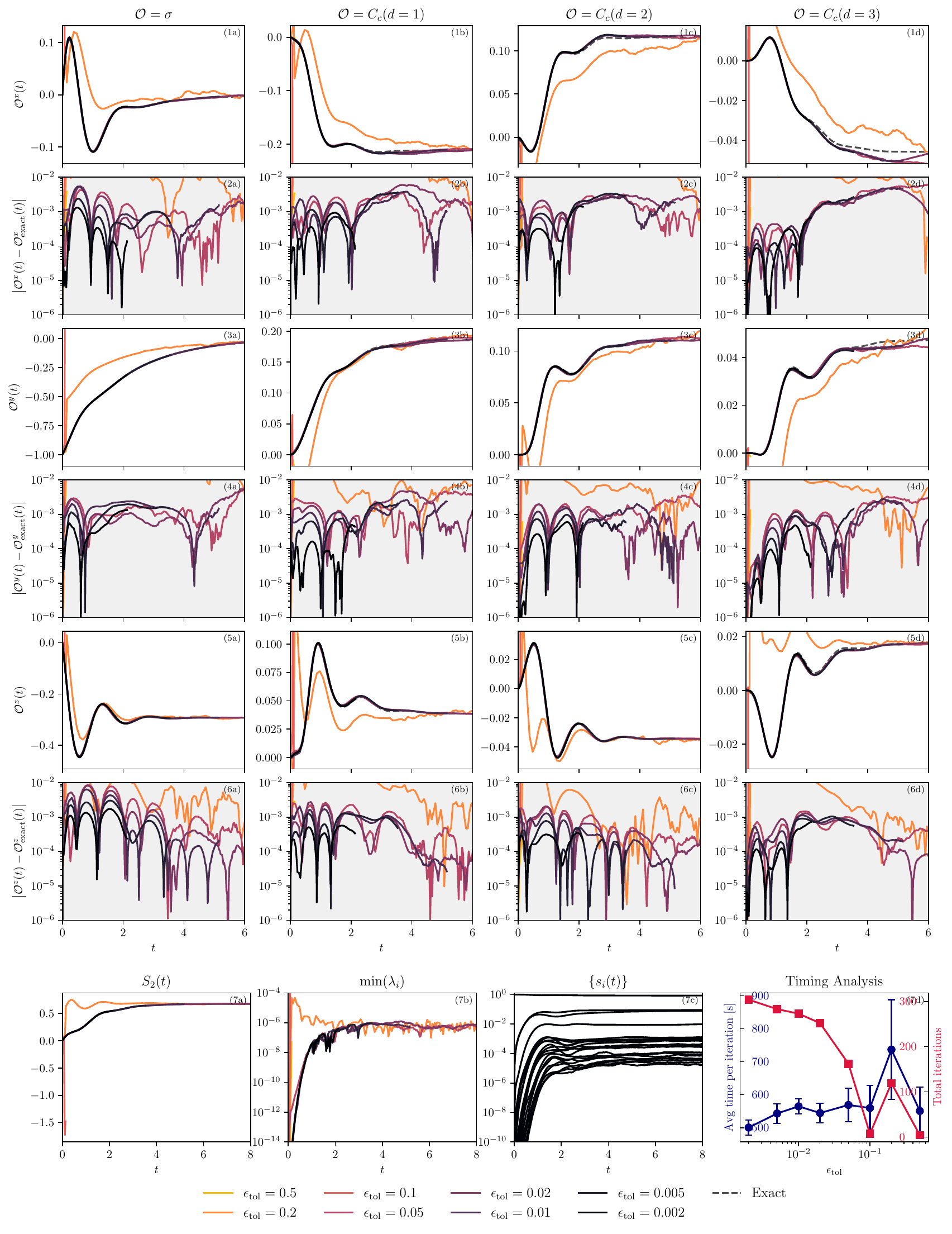}
    \caption{\textbf{Convergence with $\epsilon_\text{tol}$. XYZ model with short-range interactions, $N=10$.} For panel descriptions, see Supplementary Note 2 E.}
    \label{fig: XYZ_N10_eps_tol}
\end{figure}

\begin{figure}[H]
    \includegraphics[width=1.0\textwidth]{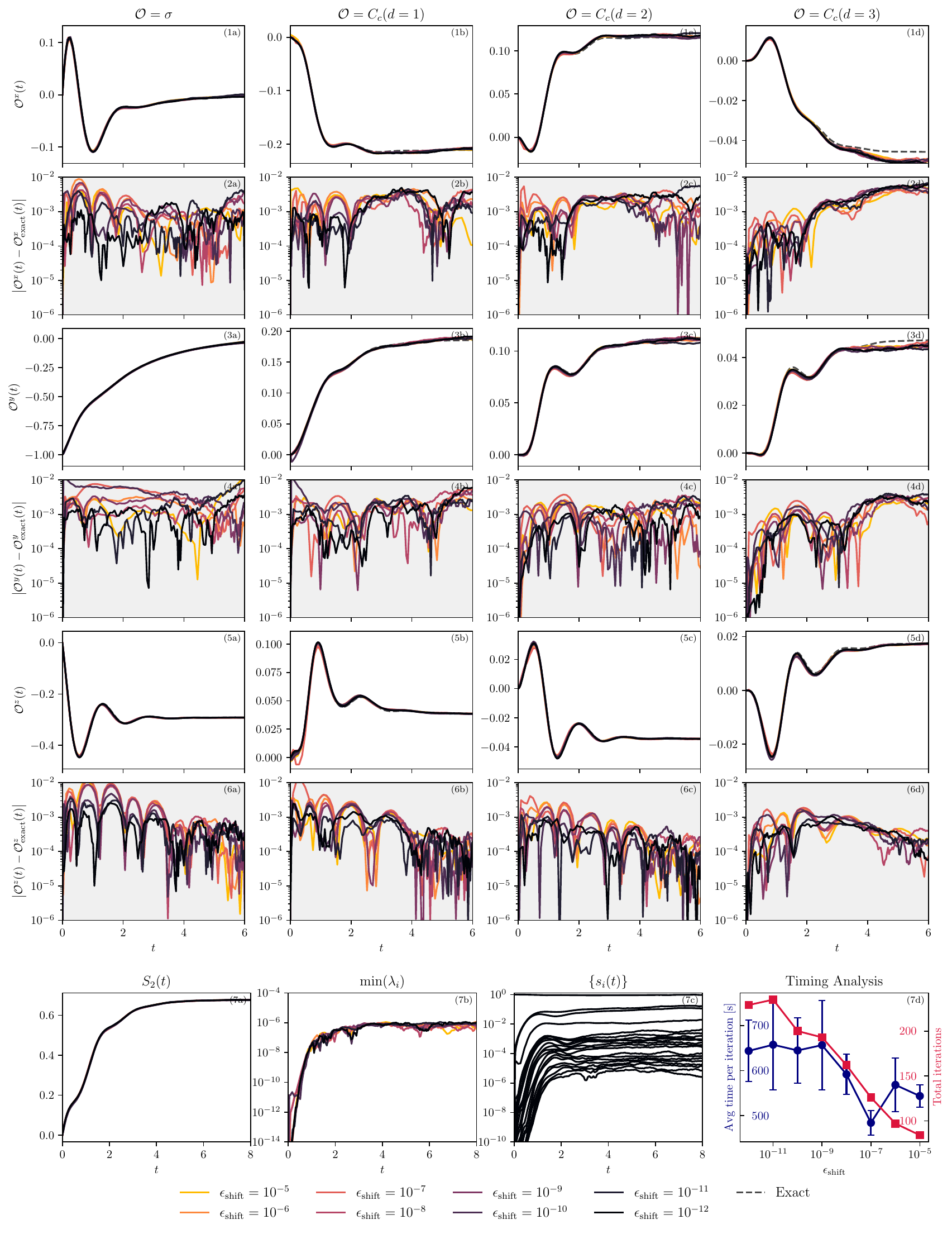}
    \caption{\textbf{Convergence with $\epsilon_\text{shift}$. XYZ model with short-range interactions, $N=10$.} For panel descriptions, see Supplementary Note 2 E.}
    \label{fig: XYZ_N10_eps_shift}
\end{figure}

\begin{figure}[H]
    \includegraphics[width=1.0\textwidth]{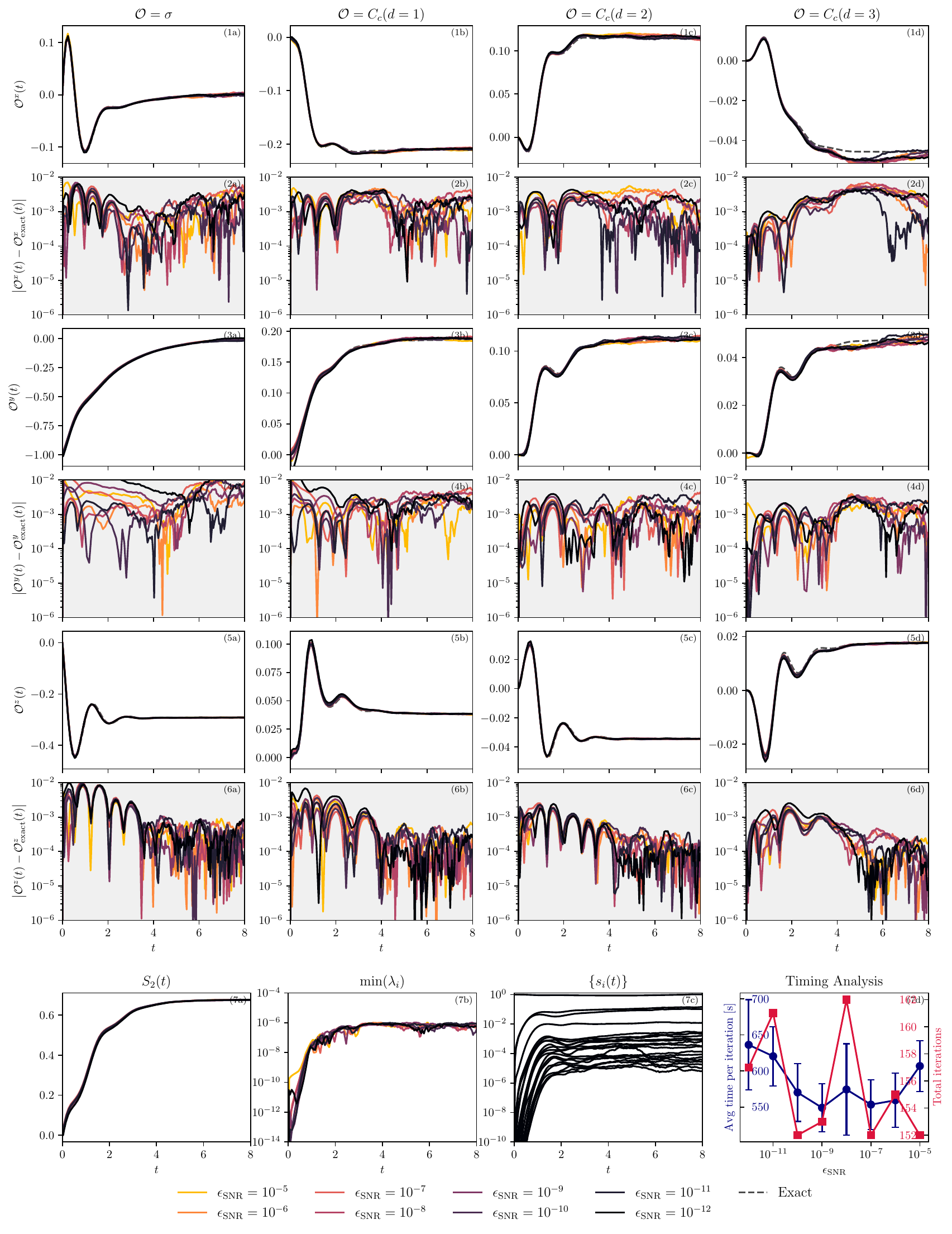}
    \caption{\textbf{Convergence with $\epsilon_\text{SNR}$. XYZ model with short-range interactions, $N=10$.} For panel descriptions, see Supplementary Note 2 E.}
    \label{fig: XYZ_N10_eps_snr}
\end{figure}

\newpage
\subsection{XYZ model with short-range interactions, $N=50$}

\begin{figure}[H]
    \includegraphics[width=1.0\textwidth]{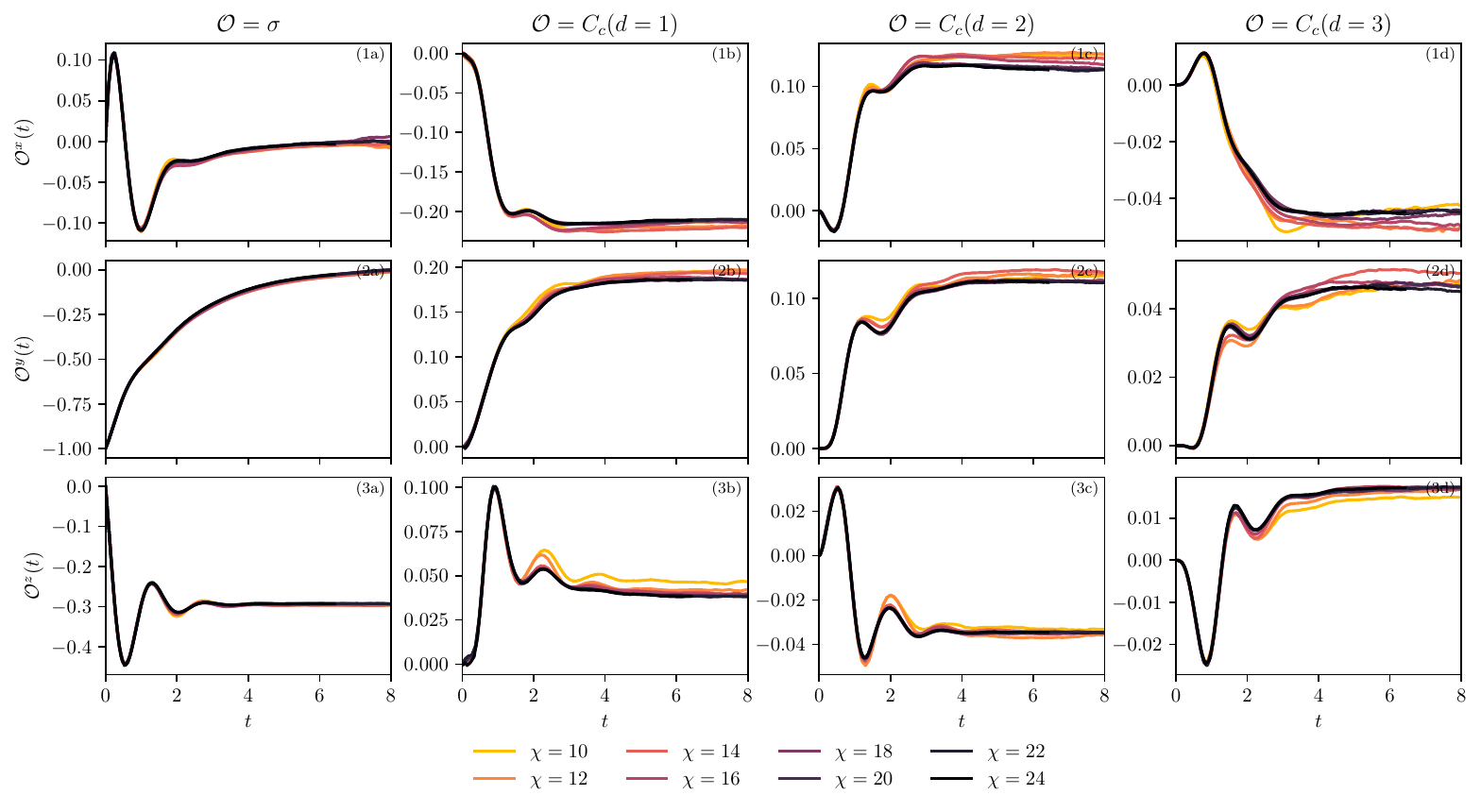}
    \caption{\textbf{Convergence with $\chi$. XYZ model with short-range interactions, $N=50$.} For panel descriptions, see Supplementary Note 2 F.}
    \label{fig: XYZ_N50_chi}
\end{figure}

\begin{figure}[H]
    \includegraphics[width=1.0\textwidth]{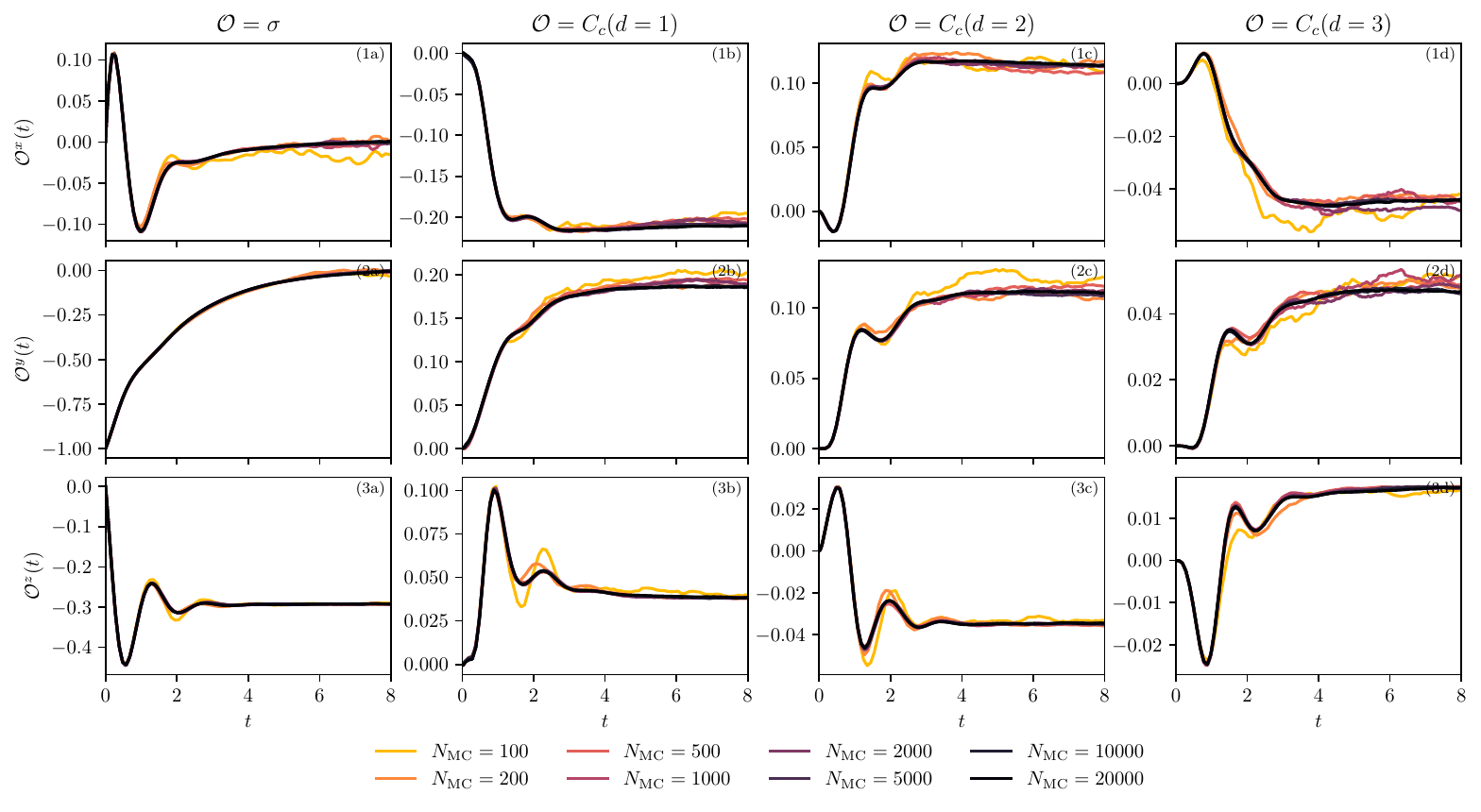}
    \caption{\textbf{Convergence with $N_\text{MC}$. XYZ model with short-range interactions, $N=50$.} For panel descriptions, see Supplementary Note 2 F.}
    \label{fig: XYZ_N50_samples}
\end{figure}

\begin{figure}[H]
    \includegraphics[width=1.0\textwidth]{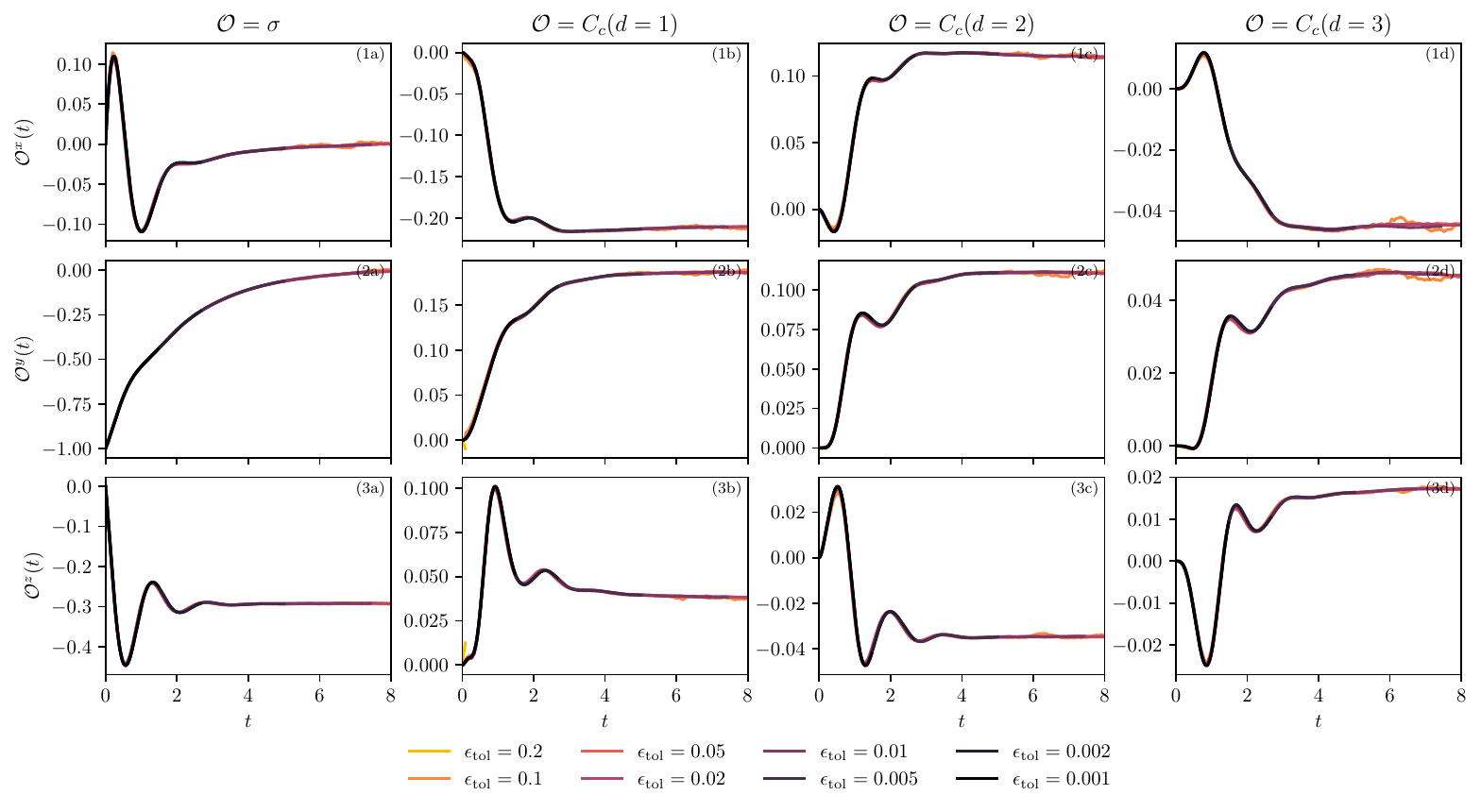}
    \caption{\textbf{Convergence with $\epsilon_\text{tol}$. XYZ model with short-range interactions, $N=50$.} For panel descriptions, see Supplementary Note 2 F.}
    \label{fig: XYZ_N50_eps_tol}
\end{figure}

\begin{figure}[H]
    \includegraphics[width=1.0\textwidth]{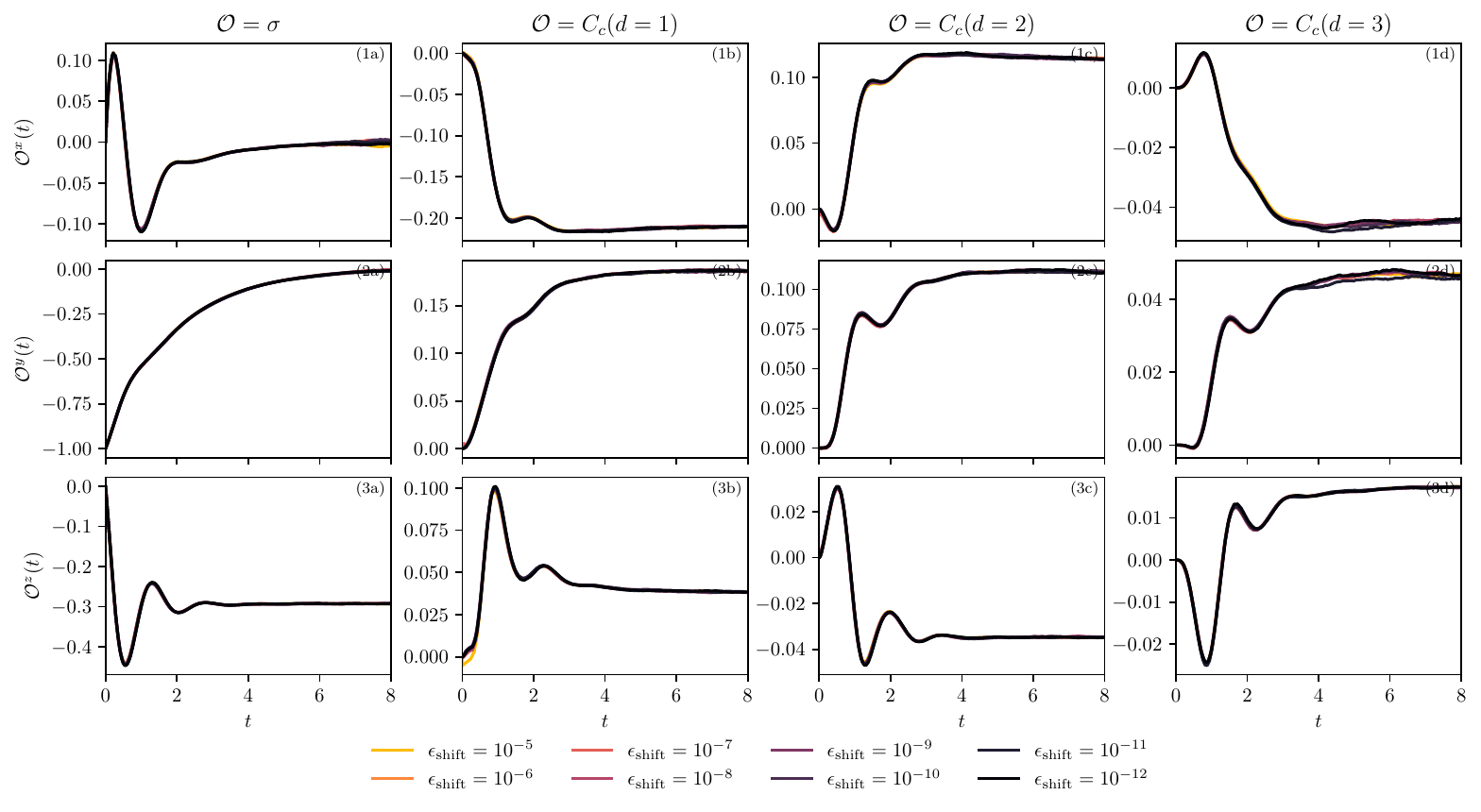}
    \caption{\textbf{Convergence with $\epsilon_\text{shift}$. XYZ model with short-range interactions, $N=50$.} For panel descriptions, see Supplementary Note 2 F.}
    \label{fig: XYZ_N50_eps_shift}
\end{figure}

\begin{figure}[H]
    \includegraphics[width=1.0\textwidth]{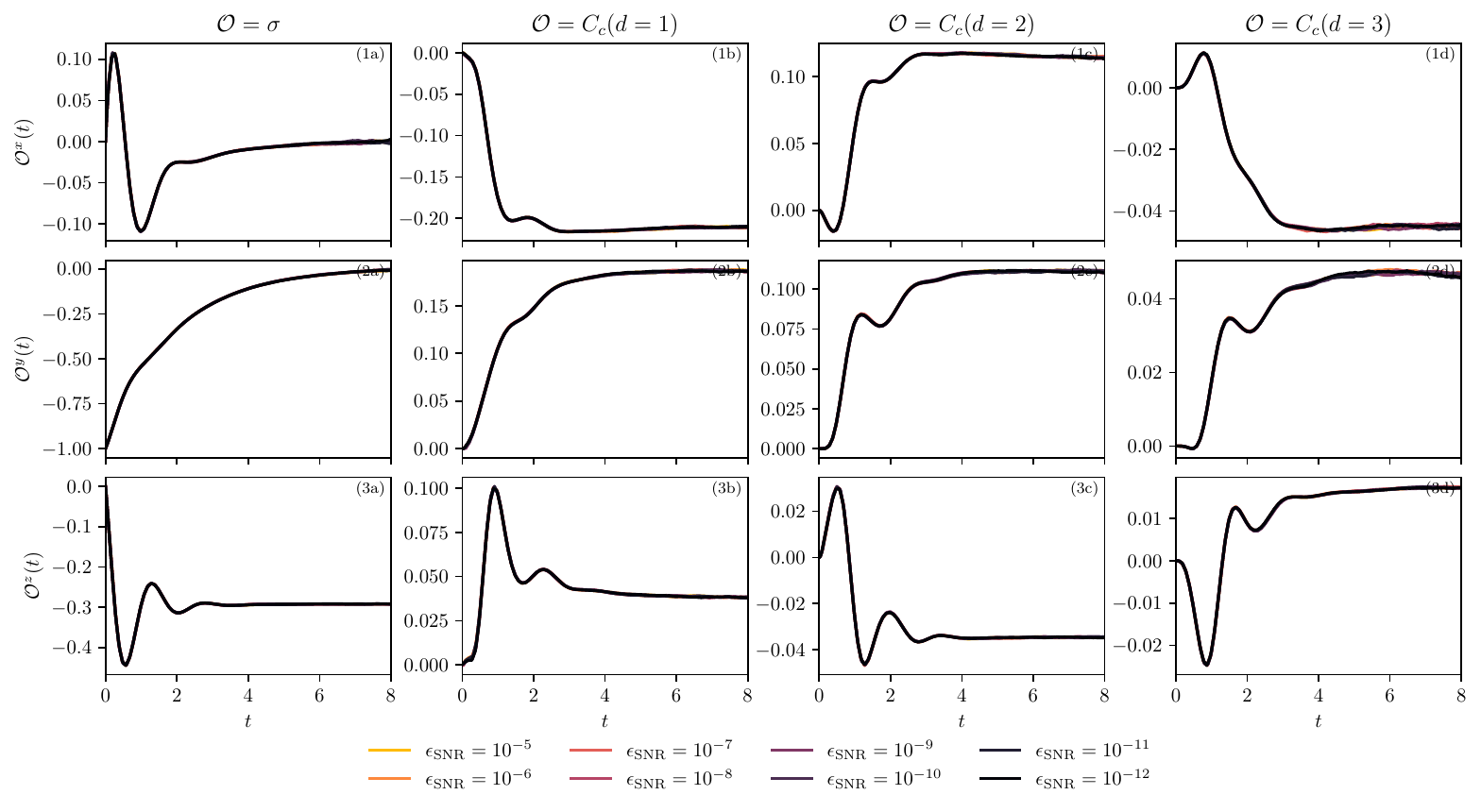}
    \caption{\textbf{Convergence with $\epsilon_\text{SNR}$. XYZ model with short-range interactions, $N=50$.} For panel descriptions, see Supplementary Note 2 F.}
    \label{fig: XYZ_N50_eps_snr}
\end{figure}

\newpage
\subsection{XYZ model with long-range competing interactions, $N=10$}

\begin{figure}[H]
    \includegraphics[width=1.0\textwidth]{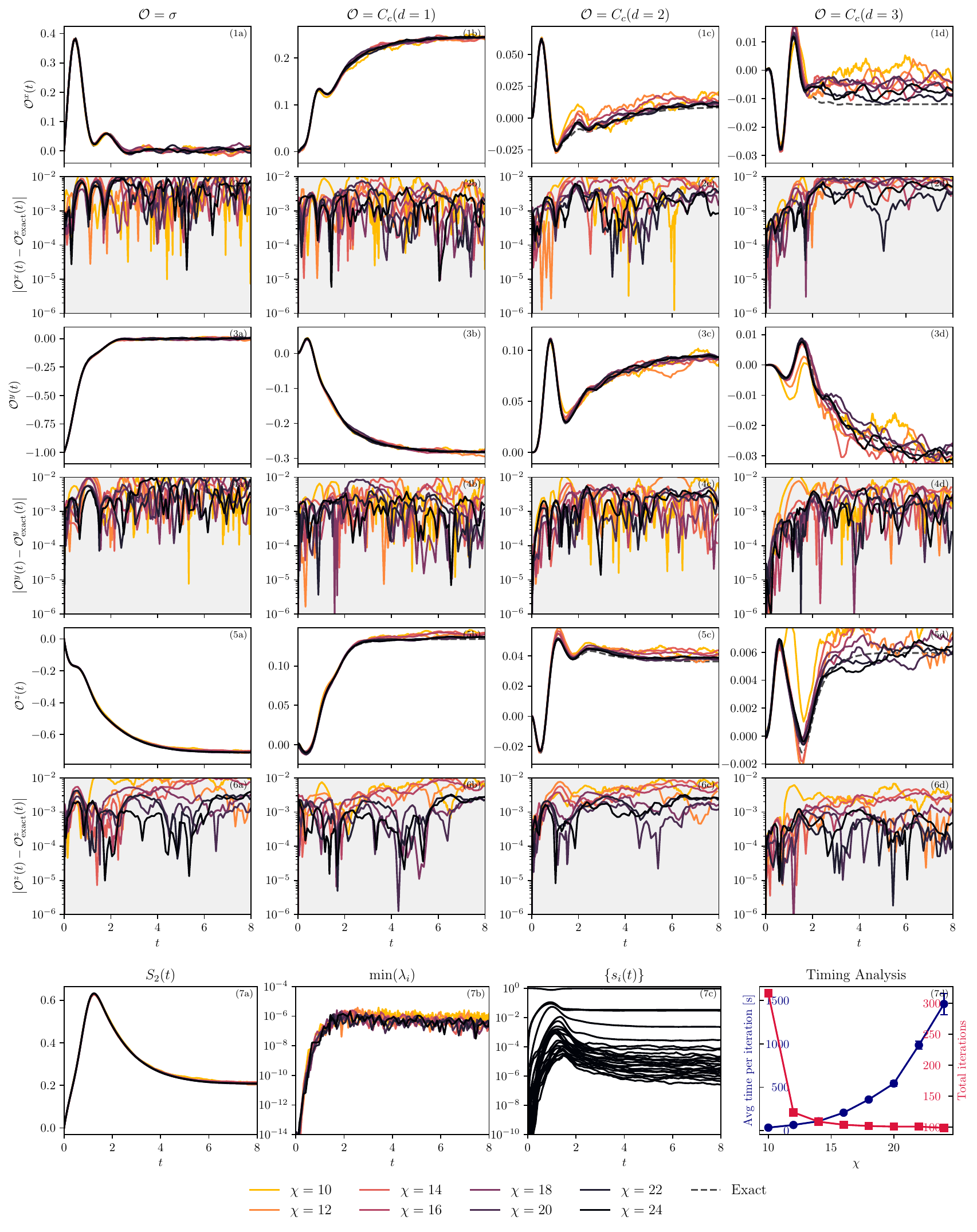}
    \caption{\textbf{Convergence with $\chi$. XYZ model with super-long-range competing interactions, $N=10$.} For panel descriptions, see Supplementary Note 2 E.}
    \label{fig: CXYZ_N10_chi}
\end{figure}

\begin{figure}[H]
    \includegraphics[width=1.0\textwidth]{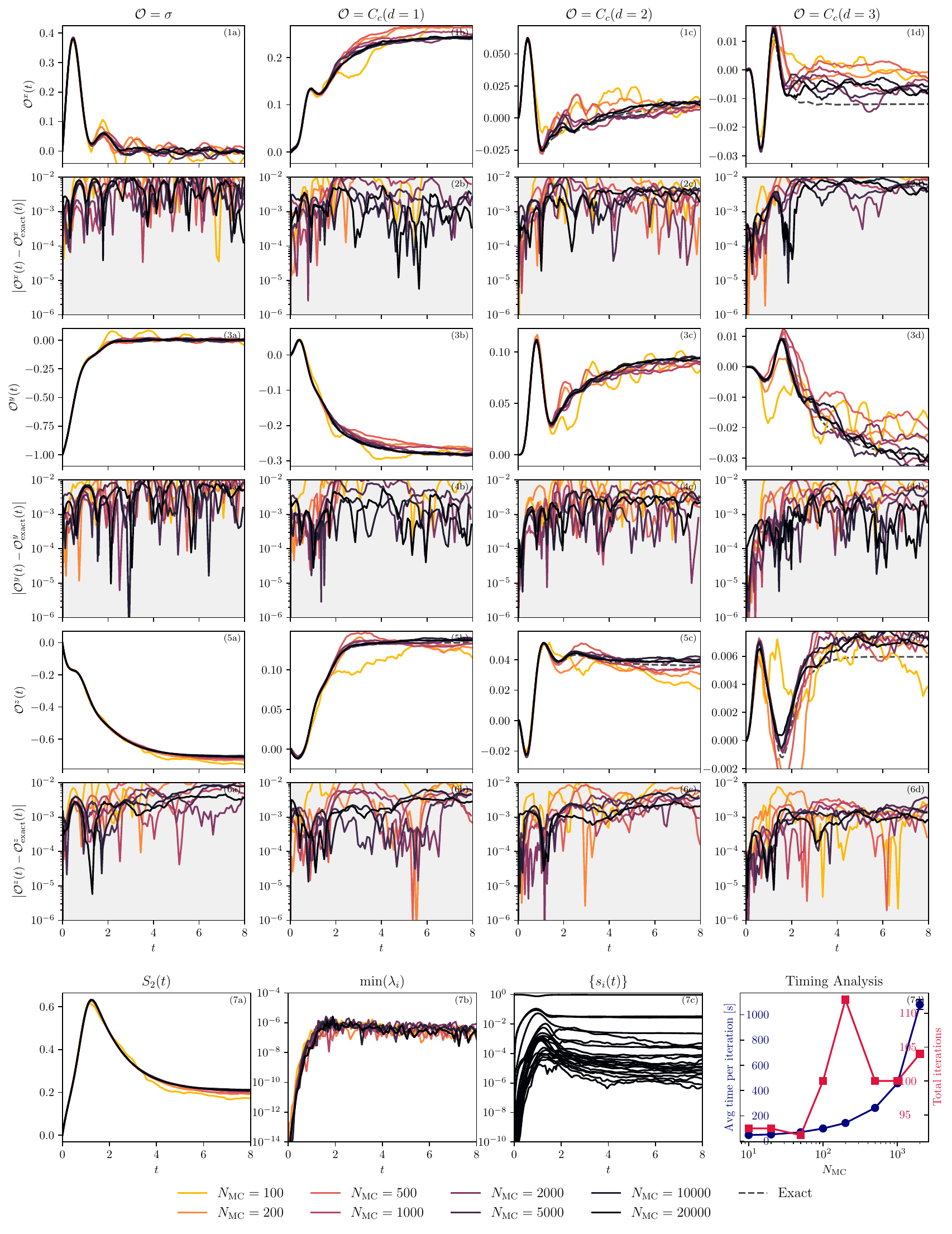}
    \caption{\textbf{Convergence with $N_\text{MC}$. XYZ model with long-range competing interactions, $N=10$.} For panel descriptions, see Supplementary Note 2 E.}
    \label{fig: CXYZ_N10_samples}
\end{figure}

\begin{figure}[H]
    \includegraphics[width=1.0\textwidth]{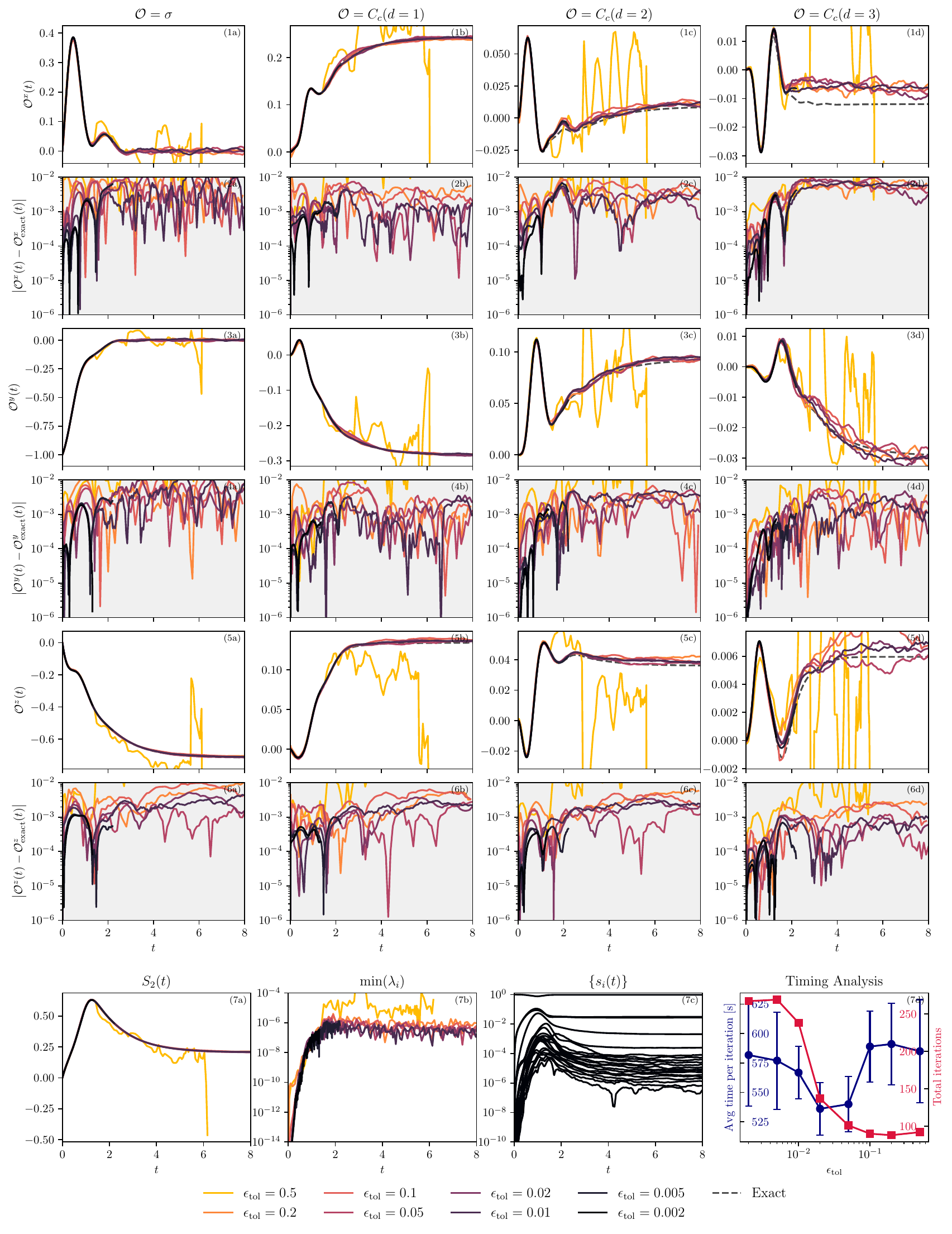}
    \caption{\textbf{Convergence with $\epsilon_\text{tol}$. XYZ model with long-range competing interactions, $N=10$.} For panel descriptions, see Supplementary Note 2 E.}
    \label{fig: CXYZ_N10_eps_tol}
\end{figure}

\begin{figure}[H]
    \includegraphics[width=1.0\textwidth]{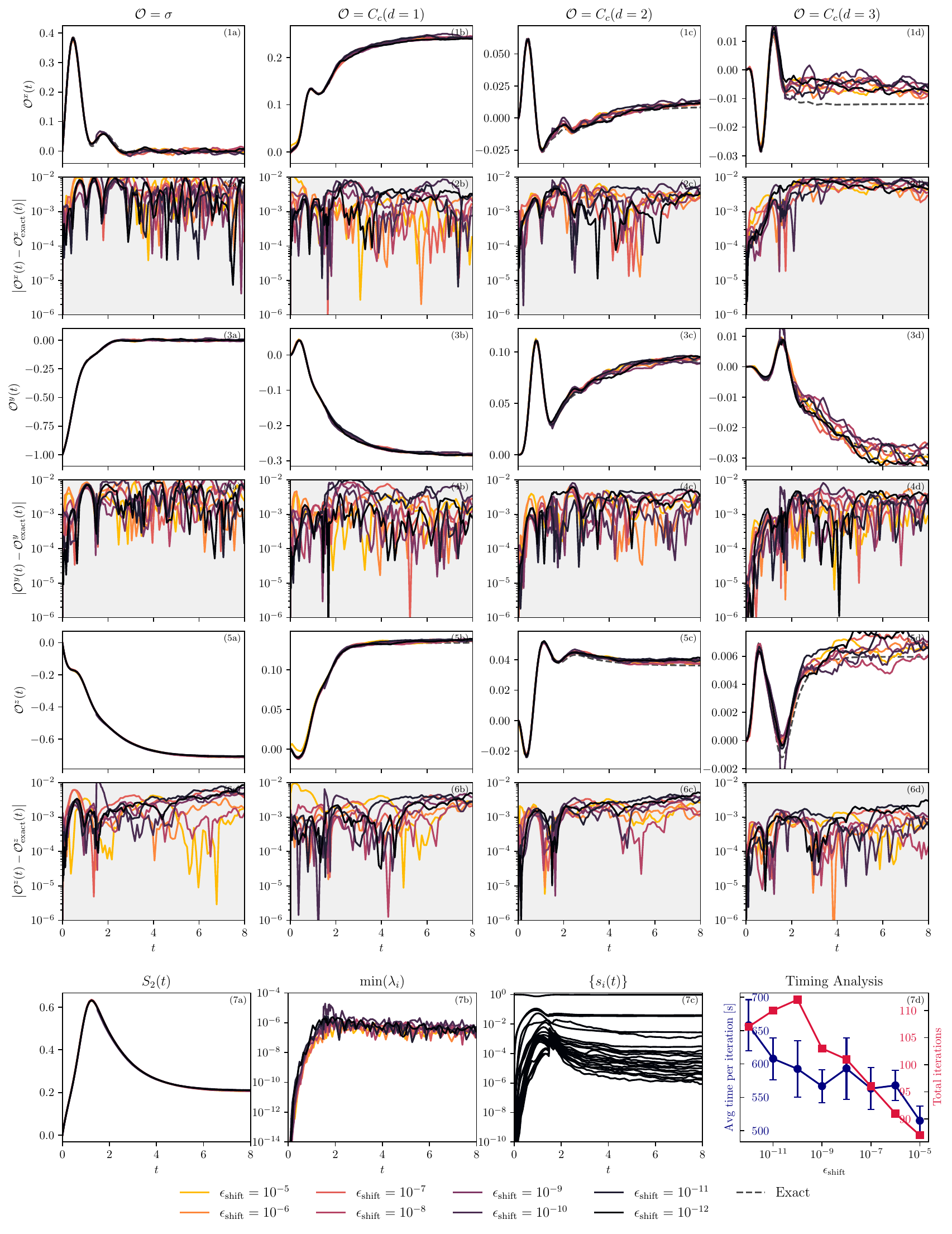}
    \caption{\textbf{Convergence with $\epsilon_\text{shift}$. XYZ model with long-range competing interactions, $N=10$.} For panel descriptions, see Supplementary Note 2 E.}
    \label{fig: CXYZ_N10_eps_shift}
\end{figure}

\begin{figure}[H]
    \includegraphics[width=1.0\textwidth]{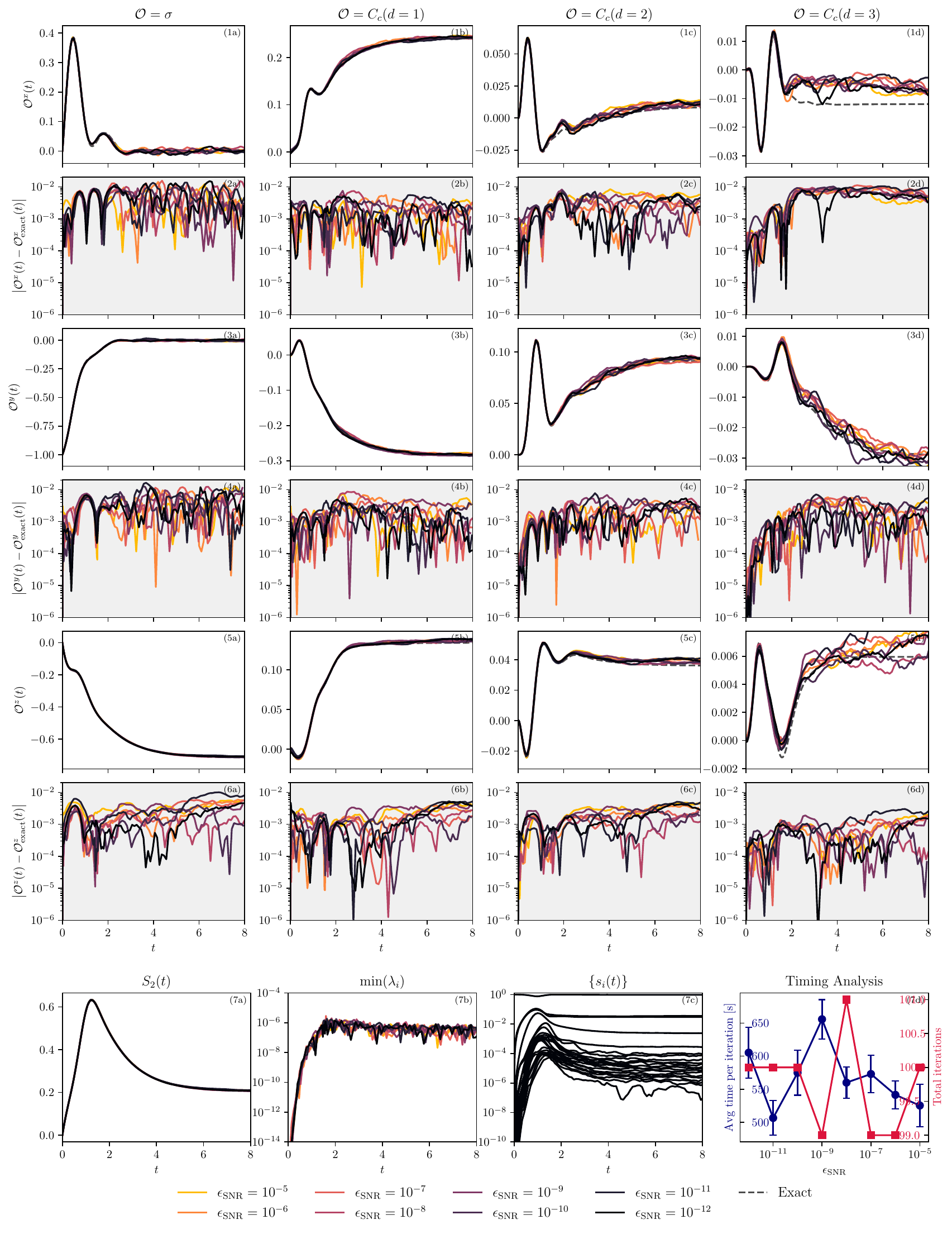}
    \caption{\textbf{Convergence with $\epsilon_\text{SNR}$. XYZ model with long-range competing interactions, $N=10$.} For panel descriptions, see Supplementary Note 2 E.}
    \label{fig: CXYZ_N10_eps_snr}
\end{figure}

\newpage
\subsection{XYZ model with long-range competing interactions, $N=50$}

\begin{figure}[H]
    \includegraphics[width=1.0\textwidth]{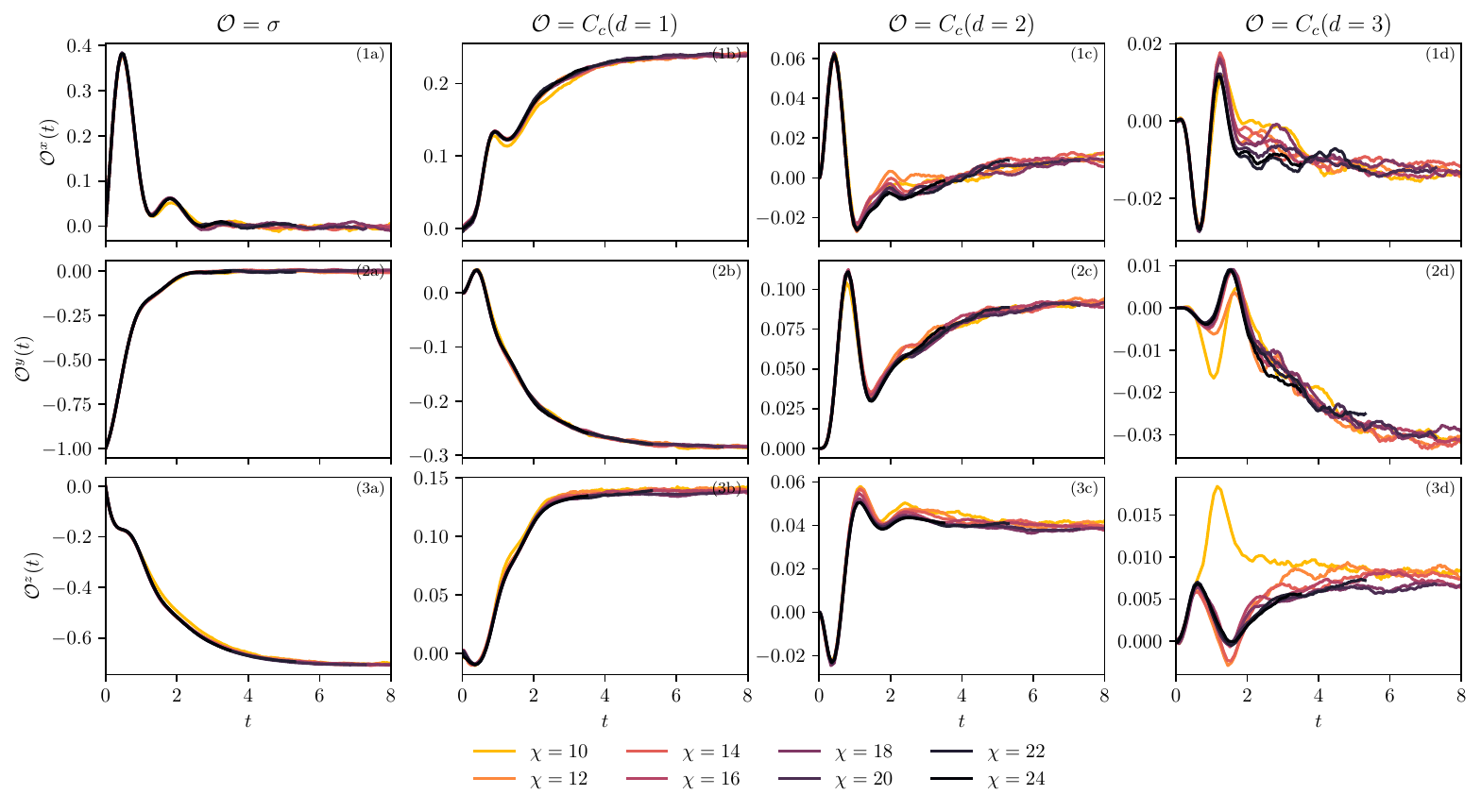}
    \caption{\textbf{Convergence with $\chi$. XYZ model with long-range competing interactions, $N=50$.} For panel descriptions, see Supplementary Note 2 F.}
    \label{fig: CXYZ_N50_chi}
\end{figure}

\begin{figure}[H]
    \includegraphics[width=1.0\textwidth]{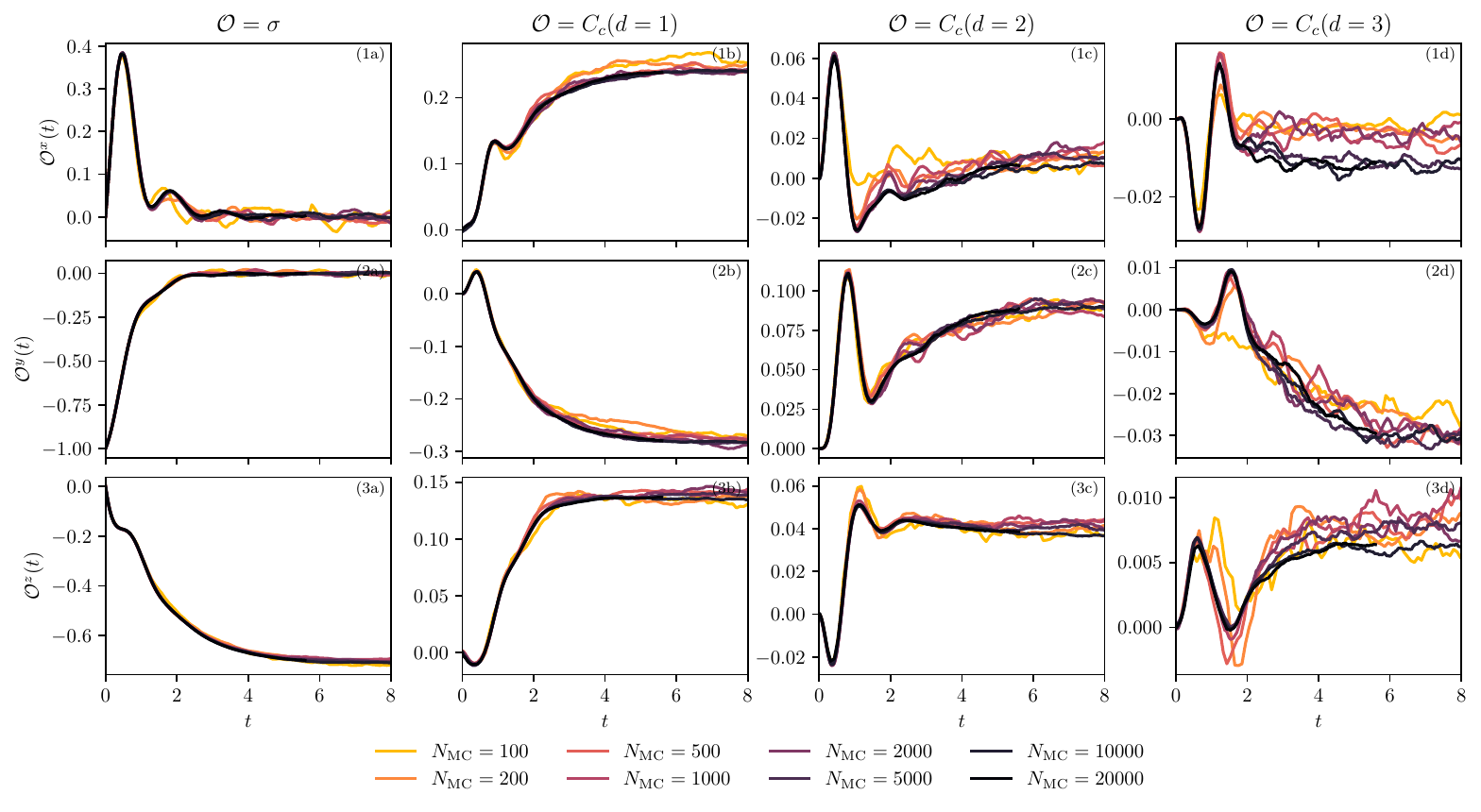}
    \caption{\textbf{Convergence with $N_\text{MC}$. XYZ model with long-range competing interactions, $N=50$.} For panel descriptions, see Supplementary Note 2 F.}
    \label{fig: CXYZ_N50_samples}
\end{figure}

\begin{figure}[H]
    \includegraphics[width=1.0\textwidth]{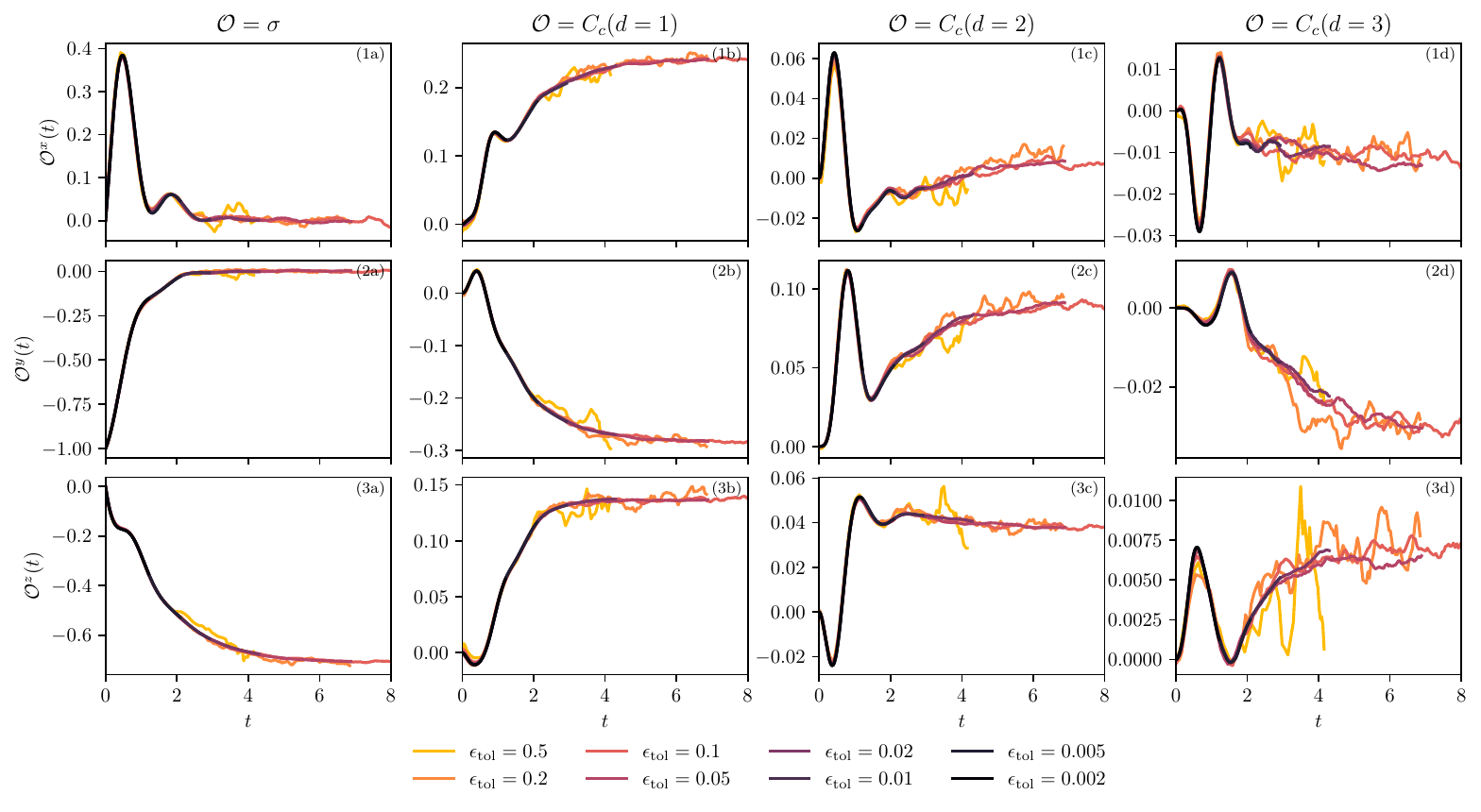}
    \caption{\textbf{Convergence with $\epsilon_\text{tol}$. XYZ model with long-range competing interactions, $N=50$.} For panel descriptions, see Supplementary Note 2 F.}
    \label{fig: CXYZ_N50_eps_tol}
\end{figure}

\begin{figure}[H]
    \includegraphics[width=1.0\textwidth]{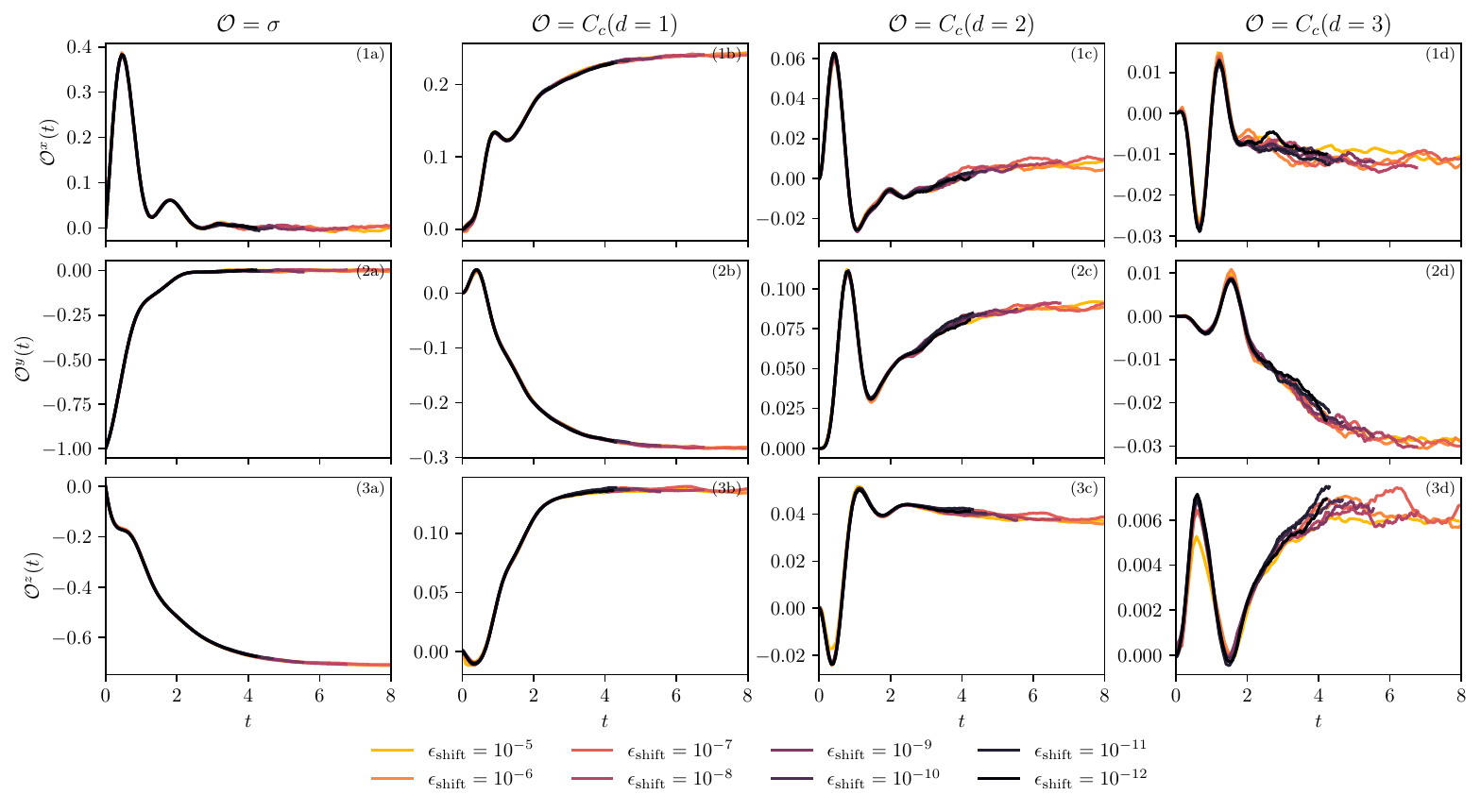}
    \caption{\textbf{Convergence with $\epsilon_\text{shift}$. XYZ model with long-range competing interactions, $N=50$.} For panel descriptions, see Supplementary Note 2 F.}
    \label{fig: CXYZ_N50_eps_shift}
\end{figure}

\begin{figure}[H]
    \includegraphics[width=1.0\textwidth]{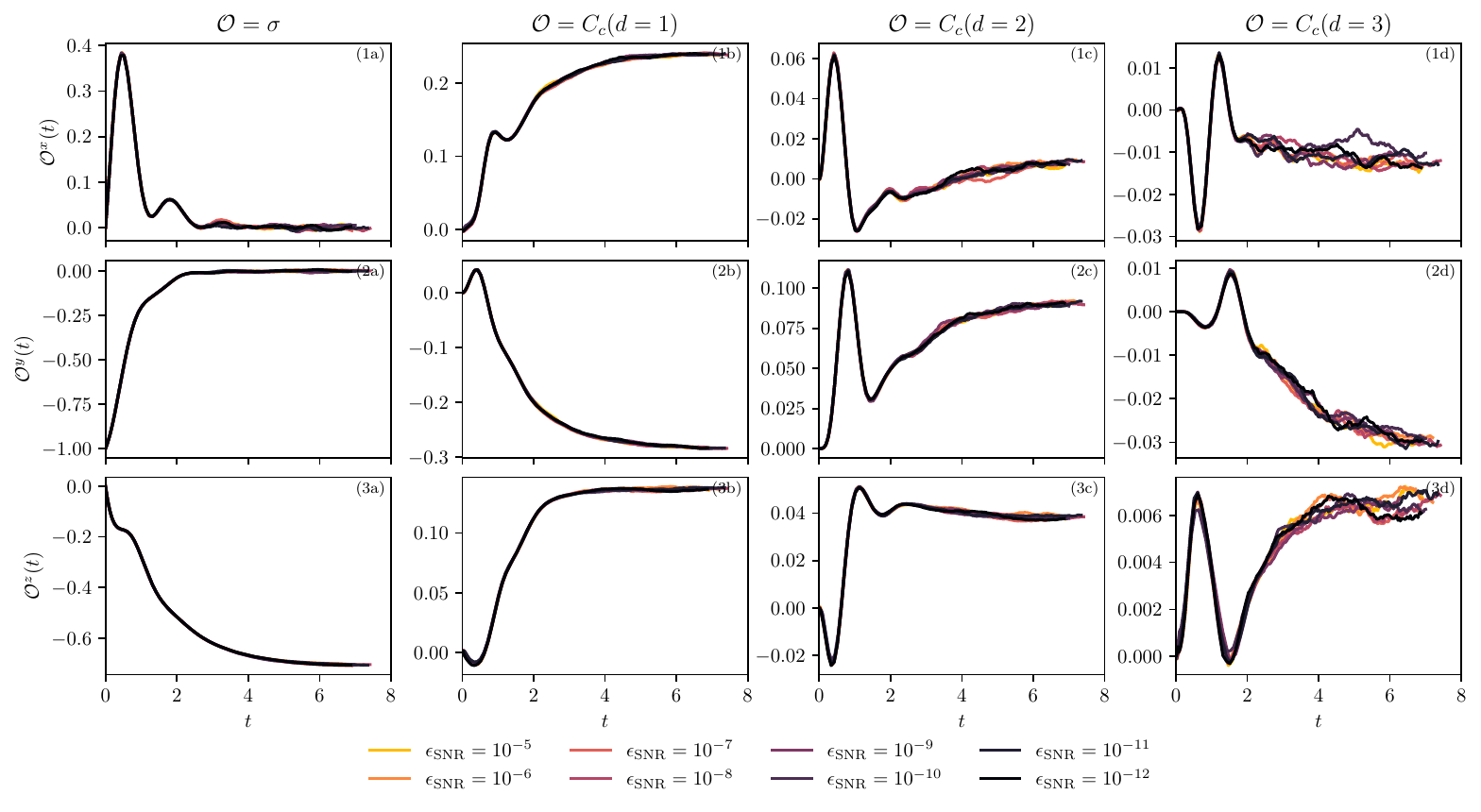}
    \caption{\textbf{Convergence with $\epsilon_\text{SNR}$. XYZ model with long-range competing interactions, $N=50$.} For panel descriptions, see Supplementary Note 2 F.}
    \label{fig: CXYZ_N50_eps_snr}
\end{figure}

\newpage
\subsection{Structure factor convergence}

\begin{figure}[h]
    \includegraphics[width=0.9\textwidth]{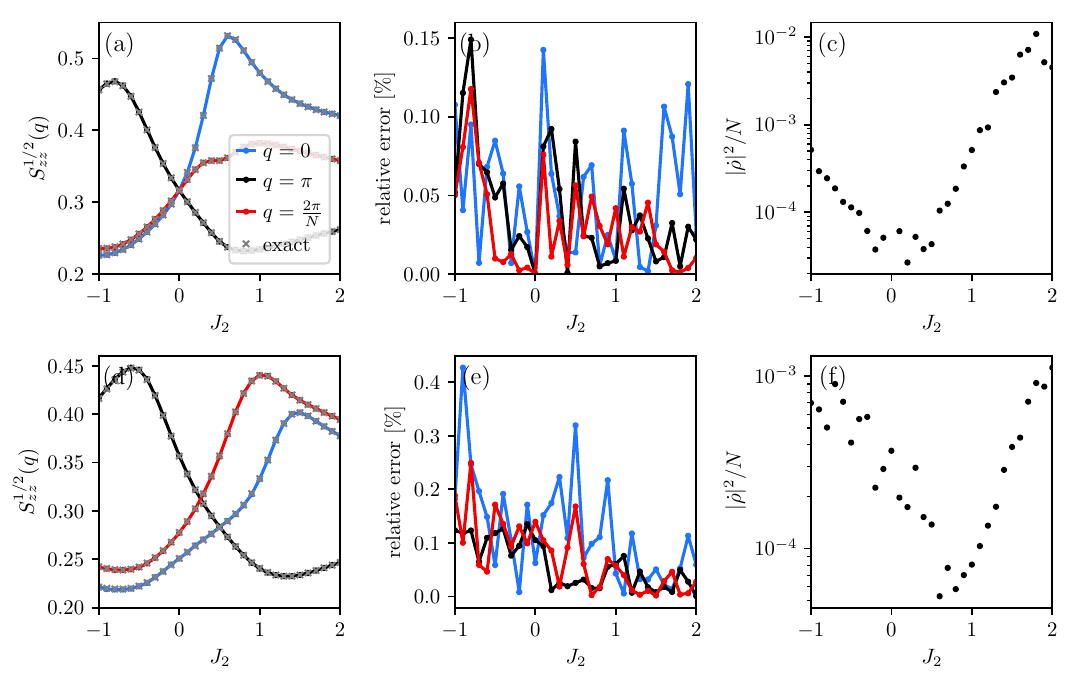}
    \caption{\textbf{Convergence analysis of steady-state structure factor phase diagrams for spin chain with long-range dipolar Ising interactions of $N=10$ sites}. In the bottom row, additional competing long-range Coulomb Ising interactions are present, following Fig. 5 in the main text.
    (a) and (d) steady state structure factors compared to exact results (grey crosses).
    (b) and (e) relative errors in the variational estimate of the structure factor in (a) and (d).
    (c) and (f) value of $|\dot{\rho}|^2/N$ at the steady-state.}
    \label{fig: NESS convergence}
\end{figure}

In Fig. \ref{fig: NESS convergence} we analyse the convergence of t-VMC+MPO in the computation of the steady-state structure factor phase diagrams for $N=10$. The small number of sites allows for a direct comparison with exact results to be made. In (a) and (d) the steady-state structure factors are computed via t-VMC+MPO and exact diagonalization. 
Excellent agreement can be seen, with mean relative errors of 0.08\% and 0.04\% with and without the competing Coulomb interaction, respectively.
We also consider the value of $|\dot{\rho}|^2/N$ at the steady state (subplots (c) and (f)), which acts as a cost function in variational minimization approaches, and which is identically 0 at the true steady state \cite{Hryniuk2024tensornetworkbased}. Small values of less than $10^{-3}$ can be observed in the competing regime.  

\begin{table}
\begin{center}
    \renewcommand{\tabcolsep}{3pt}
    \begin{tabular}{|c|c|c|c|c|c|c|c|c|c|c|c|c|c|c|c|} 
     \hline
     Figure & $N$ & $h$ & $\bm{J}$ & $\alpha$ & $\chi$ & $D$ & method & $N_\text{MC}$& $N_\text{MPI}$ & $\epsilon_\text{shift}$ & $\epsilon_\text{SNR}$ & $\epsilon_\text{tol}/\tau_\text{Euler}$ \\
     \hline 
     3 (a-c) & 200 & -1.0 & \makecell{$(J_x, J_y, J_z)=$\\$(-1.0,-0.9,-1.2)$} & --- & 20 & 1 & adaptive Heun & 200 & 160 & $10^{-8}$ & $10^{-4}$ & 0.1 \\
     3 (d) & 10 & -1.0 & $J=-0.5$ & --- & 20 & 1 & Euler & 5000 & 8 & $10^{-5}$ & $10^{-3}$ & 0.05 \\
     4 (a-b) & 200 & 0.5 & \makecell{$(J_1, J_2)=(-0.5,1.0)$}& \makecell{$\alpha_1=3$\\$\alpha_2=6$} & 10 & 1 & adaptive Heun & 2000 & 4 & $10^{-6}$ & $10^{-4}$ & 0.05 \\
     4 (c-d) & $4\times 4$ & 0.5 & \makecell{$(J_1, J_2)=(-0.25,0.5)$}& \makecell{$\alpha_1=3$\\$\alpha_2=6$} & 10 & 4 & adaptive Heun & 2000 & 80 & $10^{-9}$ & $10^{-4}$ & 0.1 \\
     5 (a-b) & 20 & 0.5 & \makecell{$J_1=0$\\$-1.0\leq J_2\leq 2.0$}& \makecell{$\alpha_1=1$\\$\alpha_2=3$} & 20 & 1 & adaptive Heun & 3000 & 4 & $10^{-4}$ & $10^{-4}$ & 0.05 \\
     5 (c-d) & 20-50 & 0.5 & \makecell{$J_1=1.0$\\$-1.0\leq J_2\leq 2.0$}& \makecell{$\alpha_1=1$\\$\alpha_2=3$} & 30 & 1 & adaptive Heun & 3000 & 4 & $10^{-4}$ & $10^{-4}$ & 0.05 \\
     \hline
    \end{tabular}
\end{center}
\caption{List of parameter and hyperparameter values used in the simulations presented in Figs. 3-5 in the main text. The columns list, in order, total number of sites $N$, local field strength $h$, interaction strength vector $\bm{J}$, interaction decay length $\alpha$, bond dimension $\chi$, unit cell size $D$, numerical integration method, number of Monte Carlo samples per Markov chain $N_\text{MC}$, number of Markov chains $N_\text{MPI}$ (each running on a separate MPI process), regularization hyperparameters $\epsilon_\text{shift}$ and $\epsilon_\text{SNR}$, integration tolerance or step size $\epsilon_\text{tol}/\tau_\text{Euler}$.}
\end{table}

\begin{table}
\begin{center}
    \renewcommand{\tabcolsep}{3pt}
    \begin{tabular}{|c|c|c|c|c|c|c|c|c|c|c|c|c|c|c|c|} 
     \hline
     Figure & $N$ & $h$ & $J$ $(J_1, J_2)$ & $\alpha_1, \alpha_2$ & $\chi$ & $N_\text{MC}$ & $\epsilon_\text{tol}$ & $\epsilon_\text{shift}$ & $\epsilon_\text{SNR}$ \\
     \hline 
     \ref{fig: I_N10_chi} & 10 & 1.0 & 0.5 & --- & var. & $10^5$ & 0.01 & $10^{-8}$ & $10^{-8}$ \\
     \ref{fig: I_N10_samples} & 10 & 1.0 & 0.5 & --- & 20 & var. & 0.01 & $10^{-8}$ & $10^{-8}$ \\
     \ref{fig: I_N10_eps_tol} & 10 & 1.0 & 0.5 & --- & 20 & $10^5$ & var. & $10^{-8}$ & $10^{-8}$ \\
     \ref{fig: I_N10_eps_shift} & 10 & 1.0 & 0.5 & --- & 20 & $10^5$ & 0.01 & var. & $10^{-8}$ \\
     \ref{fig: I_N10_eps_snr} & 10 & 1.0 & 0.5 & --- & 20 & $10^5$ & 0.01 & $10^{-8}$ & var. \\

     \ref{fig: I_N200_chi} & 200 & 1.0 & 0.5 & --- & var. & $10^5$ & 0.01 & $10^{-8}$ & $10^{-8}$ \\
     \ref{fig: I_N200_samples} & 200 & 1.0 & 0.5 & --- & 20 & var. & 0.01 & $10^{-8}$ & $10^{-8}$ \\
     \ref{fig: I_N200_eps_tol} & 200 & 1.0 & 0.5 & --- & 20 & $10^5$ & var. & $10^{-8}$ & $10^{-8}$ \\
     \ref{fig: I_N200_eps_shift} & 200 & 1.0 & 0.5 & --- & 20 & $10^5$ & 0.01 & var. & $10^{-8}$ \\
     \ref{fig: I_N200_eps_snr} & 200 & 1.0 & 0.5 & --- & 20 & $10^5$ & 0.01 & $10^{-8}$ & var. \\

     \ref{fig: CI_N10_chi} & 10 & 1.0 & $0.5, -1.0$ & 3, 6 & var. & $10^5$ & 0.01 & $10^{-8}$ & $10^{-8}$ \\
     \ref{fig: CI_N10_samples} & 10 & 1.0 & $0.5, -1.0$ & 3, 6 & 20 & var. & 0.01 & $10^{-8}$ & $10^{-8}$ \\
     \ref{fig: CI_N10_eps_tol} & 10 & 1.0 & $0.5, -1.0$ & 3, 6 & 20 & $10^5$ & var. & $10^{-8}$ & $10^{-8}$ \\
     \ref{fig: CI_N10_eps_shift} & 10 & 1.0 & $0.5, -1.0$ & 3, 6 & 20 & $10^5$ & 0.01 & var. & $10^{-8}$ \\
     \ref{fig: CI_N10_eps_snr} & 10 & 1.0 & $0.5, -1.0$ & 3, 6 & 20 & $10^5$ & 0.01 & $10^{-8}$ & var. \\

     \ref{fig: CI_N200_chi} & 200 & 1.0 & $0.5, -1.0$ & 3, 6 & var. & $10^5$ & 0.01 & $10^{-8}$ & $10^{-8}$ \\
     \ref{fig: CI_N200_samples} & 200 & 1.0 & $0.5, -1.0$ & 3, 6 & 20 & var. & 0.01 & $10^{-8}$ & $10^{-8}$ \\
     \ref{fig: CI_N200_eps_tol} & 200 & 1.0 & $0.5, -1.0$ & 3, 6 & 20 & $10^5$ & var. & $10^{-8}$ & $10^{-8}$ \\
     \ref{fig: CI_N200_eps_shift} & 200 & 1.0 & $0.5, -1.0$ & 3, 6 & 20 & $10^5$ & 0.01 & var. & $10^{-8}$ \\
     \ref{fig: CI_N200_eps_snr} & 200 & 1.0 & $0.5, -1.0$ & 3, 6 & 20 & $10^5$ & 0.01 & $10^{-8}$ & var. \\

     \ref{fig: CI_N10_strong_chi} & 10 & 1.0 & $1.0, -2.0$ & 3, 6 & var. & $10^5$ & 0.01 & $10^{-8}$ & $10^{-8}$ \\
     \ref{fig: CI_N10_strong_samples} & 10 & 1.0 & $1.0, -2.0$ & 3, 6 & 20 & var. & 0.01 & $10^{-8}$ & $10^{-8}$ \\
     \ref{fig: CI_N10_strong_eps_tol} & 10 & 1.0 & $1.0, -2.0$ & 3, 6 & 20 & $10^5$ & var. & $10^{-8}$ & $10^{-8}$ \\
     \ref{fig: CI_N10_strong_eps_shift} & 10 & 1.0 & $1.0, -2.0$ & 3, 6 & 20 & $10^5$ & 0.01 & var. & $10^{-8}$ \\
     \ref{fig: CI_N10_strong_eps_snr} & 10 & 1.0 & $1.0, -2.0$ & 3, 6 & 20 & $10^5$ & 0.01 & $10^{-8}$ & var. \\

     \ref{fig: CI_N200_strong_chi} & 200 & 1.0 & $1.0, -2.0$ & 3, 6 & var. & $10^5$ & 0.01 & $10^{-8}$ & $10^{-8}$ \\
     \ref{fig: CI_N200_strong_samples} & 200 & 1.0 & $1.0, -2.0$ & 3, 6 & 20 & var. & 0.01 & $10^{-8}$ & $10^{-8}$ \\
     \ref{fig: CI_N200_strong_eps_tol} & 200 & 1.0 & $1.0, -2.0$ & 3, 6 & 20 & $10^5$ & var. & $10^{-8}$ & $10^{-8}$ \\
     \ref{fig: CI_N200_strong_eps_shift} & 200 & 1.0 & $1.0, -2.0$ & 3, 6 & 20 & $10^5$ & 0.01 & var. & $10^{-8}$ \\
     \ref{fig: CI_N200_strong_eps_snr} & 200 & 1.0 & $1.0, -2.0$ & 3, 6 & 20 & $10^5$ & 0.01 & $10^{-8}$ & var. \\

     \ref{fig: CI_N10_far_chi} & 10 & 1.0 & $0.5, -1.0$ & 2, 4 & var. & $10^5$ & 0.01 & $10^{-8}$ & $10^{-8}$ \\
     \ref{fig: CI_N10_far_samples} & 10 & 1.0 & $0.5, -1.0$ & 2, 4 & 20 & var. & 0.01 & $10^{-8}$ & $10^{-8}$ \\
     \ref{fig: CI_N10_far_eps_tol} & 10 & 1.0 & $0.5, -1.0$ & 2, 4 & 20 & $10^5$ & var. & $10^{-8}$ & $10^{-8}$ \\
     \ref{fig: CI_N10_far_eps_shift} & 10 & 1.0 & $0.5, -1.0$ & 2, 4 & 20 & $10^5$ & 0.01 & var. & $10^{-8}$ \\
     \ref{fig: CI_N10_far_eps_snr} & 10 & 1.0 & $0.5, -1.0$ & 2, 4 & 20 & $10^5$ & 0.01 & $10^{-8}$ & var. \\

     \ref{fig: CI_N200_far_chi} & 200 & 1.0 & $0.5, -1.0$ & 2, 4 & var. & $10^5$ & 0.01 & $10^{-8}$ & $10^{-8}$ \\
     \ref{fig: CI_N200_far_samples} & 200 & 1.0 & $0.5, -1.0$ & 2, 4 & 20 & var. & 0.01 & $10^{-8}$ & $10^{-8}$ \\
     \ref{fig: CI_N200_far_eps_tol} & 200 & 1.0 & $0.5, -1.0$ & 2, 4 & 20 & $10^5$ & var. & $10^{-8}$ & $10^{-8}$ \\
     \ref{fig: CI_N200_far_eps_shift} & 200 & 1.0 & $0.5, -1.0$ & 2, 4 & 20 & $10^5$ & 0.01 & var. & $10^{-8}$ \\
     \ref{fig: CI_N200_far_eps_snr} & 200 & 1.0 & $0.5, -1.0$ & 2, 4 & 20 & $10^5$ & 0.01 & $10^{-8}$ & var. \\
     \hline
    \end{tabular}
\end{center}
\caption{List of parameter and hyperparameter values used in the simulations of variants of the dissipative Ising chain presented in Figs. S2-S41 in the Supplementary Material. The columns list, in order, total number of sites $N$, local field strength $h$, interaction strength(s) $J\, (J_1, J_2)$, interaction decay lengths $\alpha_1, \alpha_2$, bond dimension $\chi$, total number of Monte Carlo samples per Markov chain $N_\text{MC}$, integration tolerance $\epsilon_\text{tol}$, regularization hyperparameters $\epsilon_\text{shift}$ and $\epsilon_\text{SNR}$. The $N_\text{MC}$ column lists the total number of Monte Carlo samples used (where $N_\text{MPI}=10$ has been used throughout).}
\end{table}

\begin{table}
\begin{center}
    \renewcommand{\tabcolsep}{3pt}
    \begin{tabular}{|c|c|c|c|c|c|c|c|c|c|c|c|c|c|c|c|} 
     \hline
     Figure & $N$ & $h$ & $\bm{J}$ $(\bm{J}_1, \bm{J}_2)$ & $\alpha_1, \alpha_2$ & $\chi$ & $N_\text{MC}$ & $\epsilon_\text{tol}$ & $\epsilon_\text{shift}$ & $\epsilon_\text{SNR}$ \\
     \hline 

     \ref{fig: XYZ_N10_chi} & 10 & 0.5 & (0.6, -0.5, 0.4) & --- & var. & $10^5$ & 0.05 & $10^{-8}$ & $10^{-8}$ \\
     \ref{fig: XYZ_N10_samples} & 10 & 0.5 & (0.6, -0.5, 0.4) & --- & 20 & var. & 0.05 & $10^{-8}$ & $10^{-8}$ \\
     \ref{fig: XYZ_N10_eps_tol} & 10 & 0.5 & (0.6, -0.5, 0.4) & --- & 20 & $10^5$ & var. & $10^{-8}$ & $10^{-8}$ \\
     \ref{fig: XYZ_N10_eps_shift} & 10 & 0.5 & (0.6, -0.5, 0.4) & --- & 20 & $10^5$ & 0.05 & var. & $10^{-8}$ \\
     \ref{fig: XYZ_N10_eps_snr} & 10 & 0.5 & (0.6, -0.5, 0.4) & --- & 20 & $10^5$ & 0.05 & $10^{-8}$ & var. \\

     \ref{fig: XYZ_N50_chi} & 50 & 0.5 & (0.6, -0.5, 0.4) & --- & var. & $10^5$ & 0.05 & $10^{-8}$ & $10^{-8}$ \\
     \ref{fig: XYZ_N50_samples} & 50 & 0.5 & (0.6, -0.5, 0.4) & --- & 20 & var. & 0.05 & $10^{-8}$ & $10^{-8}$ \\
     \ref{fig: XYZ_N50_eps_tol} & 50 & 0.5 & (0.6, -0.5, 0.4) & --- & 20 & $10^5$ & var. & $10^{-8}$ & $10^{-8}$ \\
     \ref{fig: XYZ_N50_eps_shift} & 50 & 0.5 & (0.6, -0.5, 0.4) & --- & 20 & $10^5$ & 0.05 & var. & $10^{-8}$ \\
     \ref{fig: XYZ_N50_eps_snr} & 50 & 0.5 & (0.6, -0.5, 0.4) & --- & 20 & $10^5$ & 0.05 & $10^{-8}$ & var. \\

     \ref{fig: CXYZ_N10_chi} & 10 & 0.5 & (0.6, -0.5, 0.4), (-0.9, 1.0, -1.1) & 3, 6 & var. & $10^5$ & 0.05 & $10^{-8}$ & $10^{-8}$ \\
     \ref{fig: CXYZ_N10_samples} & 10 & 0.5 & (0.6, -0.5, 0.4), (-0.9, 1.0, -1.1) & 3, 6 & 20 & var. & 0.05 & $10^{-8}$ & $10^{-8}$ \\
     \ref{fig: CXYZ_N10_eps_tol} & 10 & 0.5 & (0.6, -0.5, 0.4), (-0.9, 1.0, -1.1) & 3, 6 & 20 & $10^5$ & var. & $10^{-8}$ & $10^{-8}$ \\
     \ref{fig: CXYZ_N10_eps_shift} & 10 & 0.5 & (0.6, -0.5, 0.4), (-0.9, 1.0, -1.1) & 3, 6 & 20 & $10^5$ & 0.05 & var. & $10^{-8}$ \\
     \ref{fig: CXYZ_N10_eps_snr} & 10 & 0.5 & (0.6, -0.5, 0.4), (-0.9, 1.0, -1.1) & 3, 6 & 20 & $10^5$ & 0.05 & $10^{-8}$ & var. \\

     \ref{fig: CXYZ_N50_chi} & 50 & 0.5 & (0.6, -0.5, 0.4), (-0.9, 1.0, -1.1) & 3, 6 & var. & $10^5$ & 0.05 & $10^{-8}$ & $10^{-8}$ \\
     \ref{fig: CXYZ_N50_samples} & 50 & 0.5 & (0.6, -0.5, 0.4), (-0.9, 1.0, -1.1) & 3, 6 & 20 & var. & 0.05 & $10^{-8}$ & $10^{-8}$ \\
     \ref{fig: CXYZ_N50_eps_tol} & 50 & 0.5 & (0.6, -0.5, 0.4), (-0.9, 1.0, -1.1) & 3, 6 & 20 & $10^5$ & var. & $10^{-8}$ & $10^{-8}$ \\
     \ref{fig: CXYZ_N50_eps_shift} & 50 & 0.5 & (0.6, -0.5, 0.4), (-0.9, 1.0, -1.1) & 3, 6 & 20 & $10^5$ & 0.05 & var. & $10^{-8}$ \\
     \ref{fig: CXYZ_N50_eps_snr} & 50 & 0.5 & (0.6, -0.5, 0.4), (-0.9, 1.0, -1.1) & 3, 6 & 20 & $10^5$ & 0.05 & $10^{-8}$ & var. \\
     \hline
    \end{tabular}
\end{center}
\caption{List of parameter and hyperparameter values used in the simulations of variants of the dissipative XYZ chain presented in Figs. S42-S61 in the Supplementary Material. The columns list, in order, total number of sites $N$, local field strength $h$, interaction strength(s) $\bm{J}\, (\bm{J}_1, \bm{J}_2)$, interaction decay lengths $\alpha_1, \alpha_2$, bond dimension $\chi$, total number of Monte Carlo samples per Markov chain $N_\text{MC}$, integration tolerance $\epsilon_\text{tol}$, regularization hyperparameters $\epsilon_\text{shift}$ and $\epsilon_\text{SNR}$. The $N_\text{MC}$ column lists the total number of Monte Carlo samples used (where $N_\text{MPI}=10$ has been used throughout).}
\end{table}

\bibliography{references.bib}